%% file: main.tex
\documentclass[acmlarge]{acmart}
\usepackage{booktabs} % For formal tables

\usepackage{balance}
\usepackage{amsmath}
\usepackage{graphicx}
\usepackage{wrapfig}
\usepackage{subcaption}
\usepackage[ruled,vlined]{algorithm2e}

\usepackage{algorithmic}
\usepackage{color,soul}
\usepackage{hyperref}
\usepackage{gensymb}
\usepackage{enumerate}

\usepackage{soul}
\usepackage[export]{adjustbox}
\usepackage{pdfpages}

\newcommand {\majorRev}[1]{\ignorespaces}
\newcommand{\majorRevText}[1]{\textcolor{black}{#1}}
\newcommand{\minorRevText}[1]{\textcolor{black}{#1}}

\usepackage{microtype}
\def\ie{{i.e.},~}
\def\eg{{e.g.},~}

\AtBeginDocument{%
	\providecommand\BibTeX{{%
			\normalfont B\kern-0.5em{\scshape i\kern-0.25em b}\kern-0.8em\TeX}}}

\setcopyright{acmcopyright}
\acmJournal{IMWUT}
\acmYear{2021} 
\acmDOI{10.1145/3448085}

\def \sys {\textit{$S^3$}}

\begin{document}
	
	\title{S{\textsuperscript{3}}: Side-Channel Attack on Stylus Pencil through Sensors}
	
	\author{Habiba Farrukh}
    \email{hfarrukh@purdue.edu}
    \affiliation{%
        \institution{Purdue University}
        \streetaddress{305 N. University St}
        \city{West Lafayette}
        \state{IN}
        \country{USA}
        \postcode{47907}
    }
    
    \author{Tinghan Yang}
    \email{yang1683@mpurdue.edu}
    \affiliation{%
        \institution{Purdue University}
        \streetaddress{305 N. University St}
        \city{West Lafayette}
        \state{IN}
        \country{USA}
        \postcode{47907}
    }

    \author{Hanwen Xu}
    \email{xuhw20@mails.tsinghua.edu.cn}
    \affiliation{%
      \institution{Tsinghua University}
      \city{Beijing}
      \country{China}}

    \author{Yuxuan Yin}
    \email{yin-yx16@tsinghua.org.cn}
    \affiliation{%
      \institution{Tsinghua University}
      \city{Beijing}
      \country{China}
    }

    \author{He Wang}
    \email{hw@purdue.edu}
    \affiliation{%
        \institution{Purdue University}
        \streetaddress{305 N. University St}
        \city{West Lafayette}
        \state{IN}
        \country{USA}
        \postcode{47907}
    }
    
    \author{Z. Berkay Celik}
    \email{zcelik@purdue.edu}
    \affiliation{%
        \institution{Purdue University}
        \streetaddress{305 N. University St}
        \city{West Lafayette}
        \state{IN}
        \country{USA}
        \postcode{47907}
    }

    \renewcommand{\shortauthors}{Farrukh et al.}

	
	\begin{abstract}
		With smart devices being an essential part of our everyday lives, unsupervised access to the mobile sensors' data can result in a multitude of side-channel attacks.
		In this paper, we study potential data leaks from Apple Pencil (2\textsuperscript{nd} generation) supported by the Apple iPad Pro, the latest stylus pen which attaches to the iPad body magnetically for charging.
		We observe that the Pencil's body affects the magnetic readings sensed by the iPad's magnetometer when a user is using the Pencil.
		Therefore, we ask: \textit{Can we infer what a user is writing on the iPad screen with the Apple Pencil, given access to only the iPad's motion sensors' data?}
		To answer this question, we present \textbf{S}ide-channel attack on \textbf{S}tylus pencil through \textbf{S}ensors (\sys{}), a system that identifies what a user is writing from motion sensor readings. 
		We first use the sharp fluctuations in the motion sensors' data to determine when a user is writing on the iPad. 
		We then introduce a high-dimensional particle filter to track the location and orientation of the Pencil during usage. 
		Lastly, to guide particles, we build the Pencil's magnetic map serving as a bridge between the measured magnetic data and the Pencil location and orientation. 
		We evaluate \sys{} with $10$ subjects and demonstrate that we correctly identify $93.9$\%, $96$\%, $97.9$\%, and $93.33$\% of the letters, numbers, shapes, and words by only having access to the motion sensors' data.
	\end{abstract}



	\maketitle
	
	\input{introduction}

	\input{threat_model}

	\input{2D_tracking}

	\input{system_architecture}
	
	\input{system_design}
	
	\input{evaluation}

	\input{discussion}
	
	\input{relatedwork}

	\input{conclusion}

	\begin{acks}
		\input{acknowledgements}
	\end{acks}

	\input{groundtruth}
	
	\bibliographystyle{ACM-Reference-Format}
	\bibliography{references}
	
	\appendix
	
	\input{appendix}

% 	\newpage
%  	\input{response}
%   	\input{commentsResponse}
  	
%   	\newpage
% 	\includepdf[pages=-]{S3_Diff.pdf}
\end{document}

%% file: introduction.tex
\section{Introduction}
\label{sec:introduction}
Modern-day smart devices come embedded with various sensors, enabling a vast range of applications in activity recognition, context awareness, mobile health, and productivity.  
Unfortunately, these sensors are also gateways for unintended information leakage about users' activities.
Unauthorized access to users' personal information, habits, behaviors, and preferences through sensor data has long been a major security and privacy concern.
Mobile operating systems such as Android and iOS require users' explicit permission to grant access to sensitive data (\eg GPS locations, microphone, and camera recordings) to an application. 
However, data from more generic sensors such as magnetometer, accelerometer and gyroscope are less regulated, and often can be accessed by applications without users' permission.
Unmonitored access to these sensors' data has recently opened the door to a multitude of side-channel attacks.
Prior efforts have unveiled security and privacy concerns that result from applications having access to motion sensors' data.
Such works propose attacks targeted at inferring users’ privacy-sensitive data such as touch actions and keystrokes~\cite{wang2015mole, MehrnezhadTSH16, Maiti2016},  passwords and PIN codes~\cite{owusu2012accessory, MehrnezhadTSH16}, and application and webpage fingerprinting~\cite{deepMag, matyunin2019magneticspy}.
Unfortunately, side-channel attacks continue to be a major source of concern to users and developers due to frequent software updates and new devices introduced in the market.
\begin{figure}%
	%	\centering
	\begin{subfigure}{0.40\linewidth}
		\centering
		\includegraphics[height=2.0in, width=2.0in]{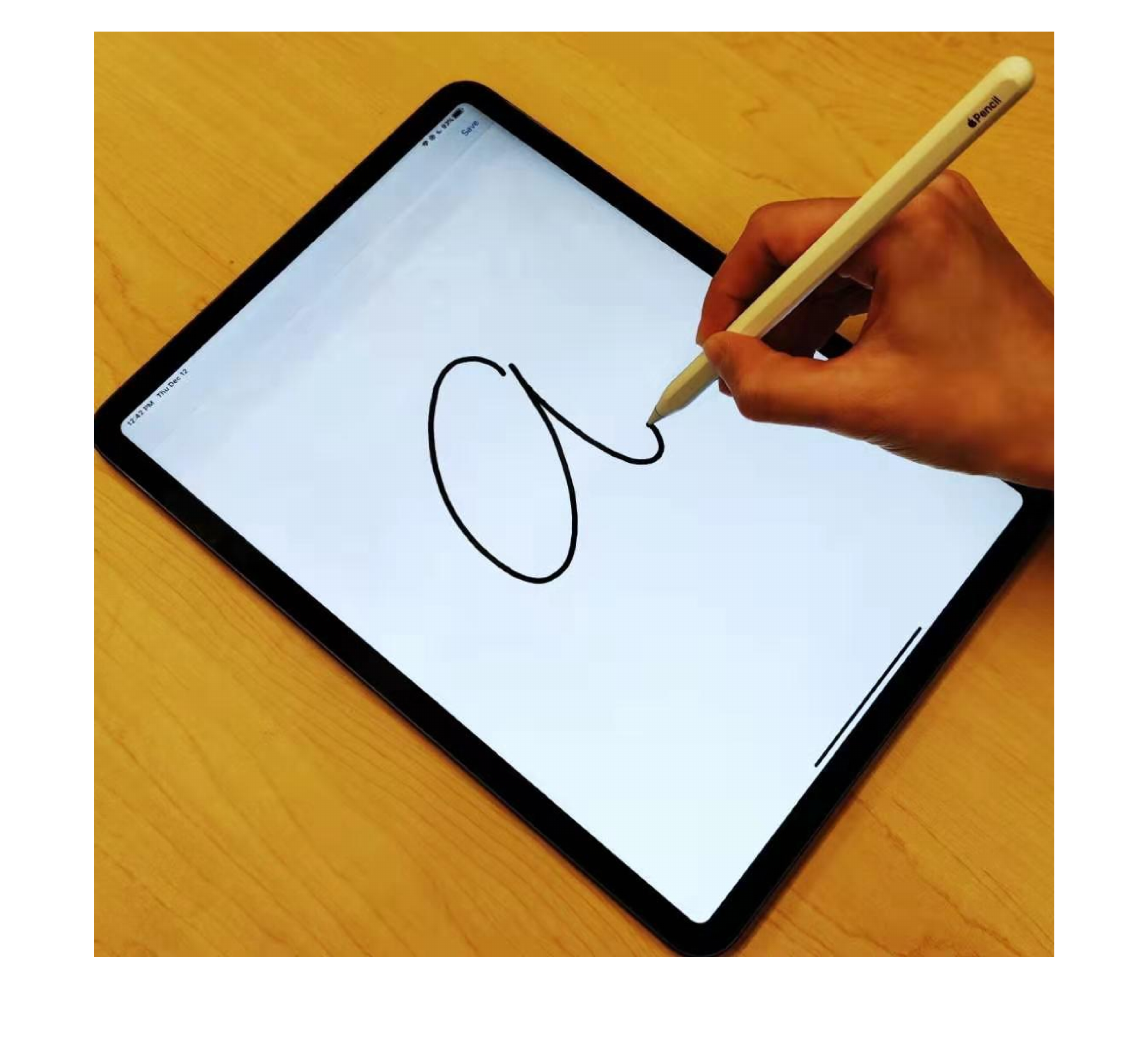}
		\caption{}
		\label{fig:screen_basic}
	\end{subfigure}	
	\begin{subfigure}{0.58\linewidth}
		\centering
		\includegraphics[height=2.0in, width=3.0in]{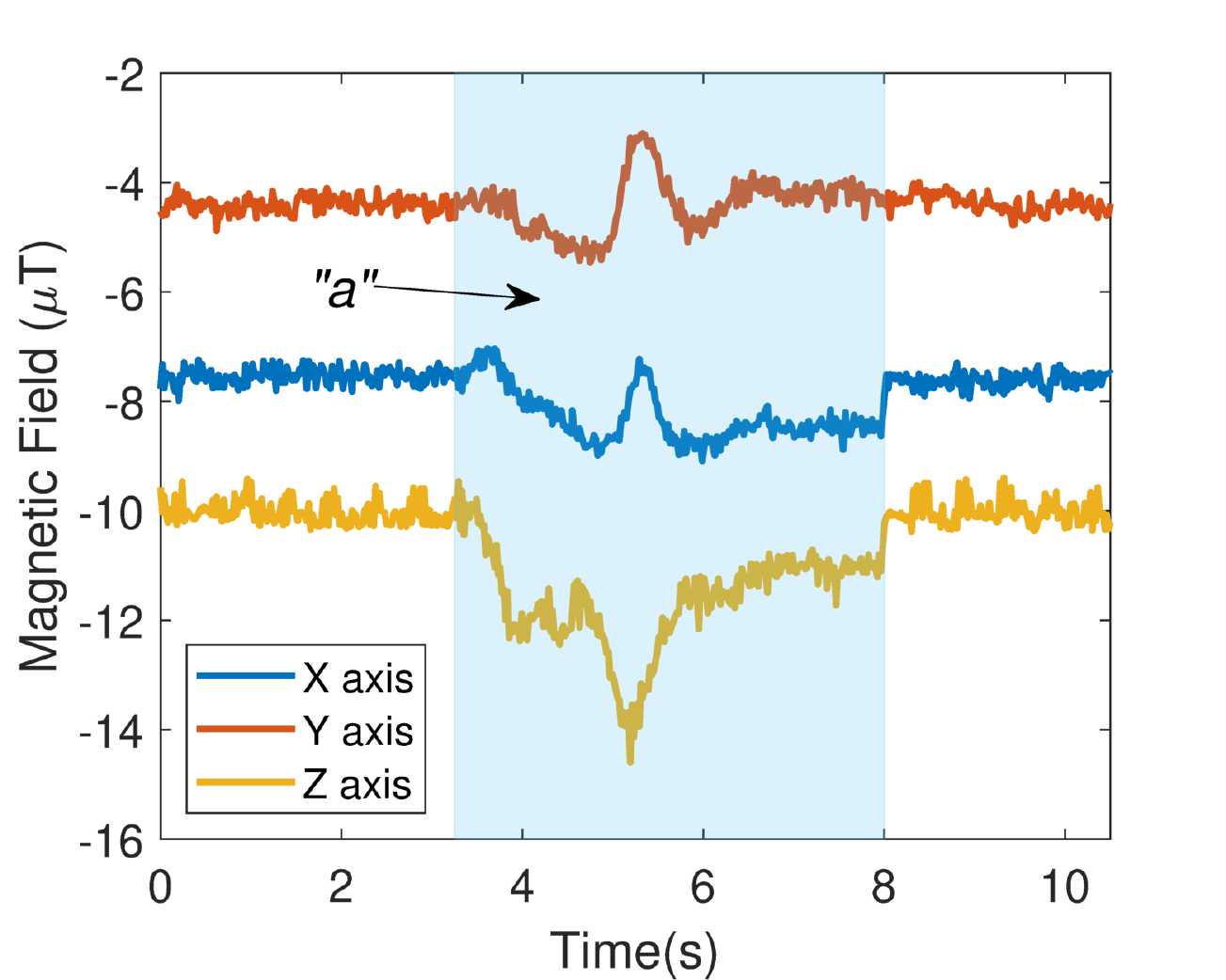}
		\caption{}
		\label{fig:basic_pattern3d}		
	\end{subfigure}
	\caption{Example of magnetometer readings when a user writes on the iPad using Apple Pencil.}
	\label{fig:basic_word}
\end{figure}

With the ever-increasing popularity of large-screen handheld devices, many manufacturers have started equipping their products with \emph{stylus pencils} to improve user experience.
In this paper, we study the recently launched "Apple Pencil" used with one of Apple's most popular devices, the iPad Pro.
The Apple Pencil lets users write, draw, and make selections in a variety of generic and custom-designed applications with ease.
The second generation of this product launched in 2018 offers a convenient new feature where the pencil pairs with the iPad to charge wirelessly.
The pencil's body is embedded with multiple magnets, allowing it to magnetically attach to the iPad's body.
In this paper, we present a novel side-channel attack on the Apple Pencil.
The magnetic readings sensed by the iPad's magnetometer are impacted when the user interacts with the screen using the pencil.
\majorRev{[MR1, C1, C5]}\majorRevText{We show that, given that the Apple Pencil is often used to fill text fields in applications~\cite{pencil20scribble}, an adversary can infer a victim's private data such as passwords, notes, text messages and signatures by only tracking the movement of the pencil through the motion sensors' data.}
%
%Given the huge popularity of the iPad, this side-channel attack has a huge impact on user security and privacy.} 
%%

\begin{figure}[ht!]%
	\begin{subfigure}{0.45\linewidth}
		\centering
		\includegraphics[height=2in, width=1.5in]{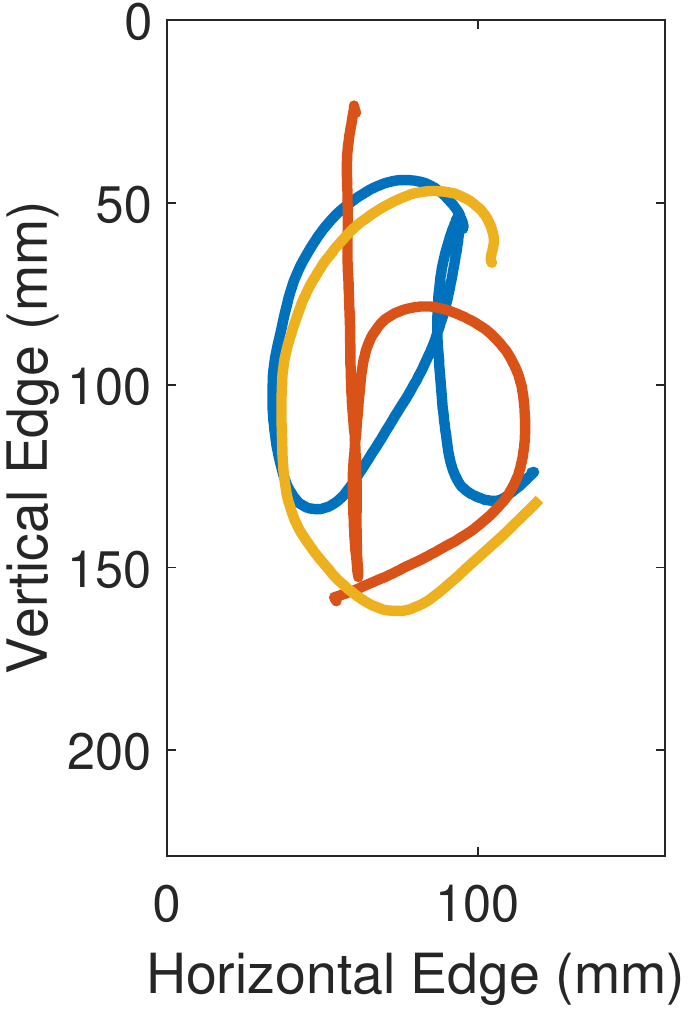}
		\caption{}
		\label{fig:screen_differentletters}
	\end{subfigure}	
	\begin{subfigure}{0.45\linewidth}
		\begin{subfigure}{1\linewidth}
			\centering
			\includegraphics[height=0.66in, width=1.8in]{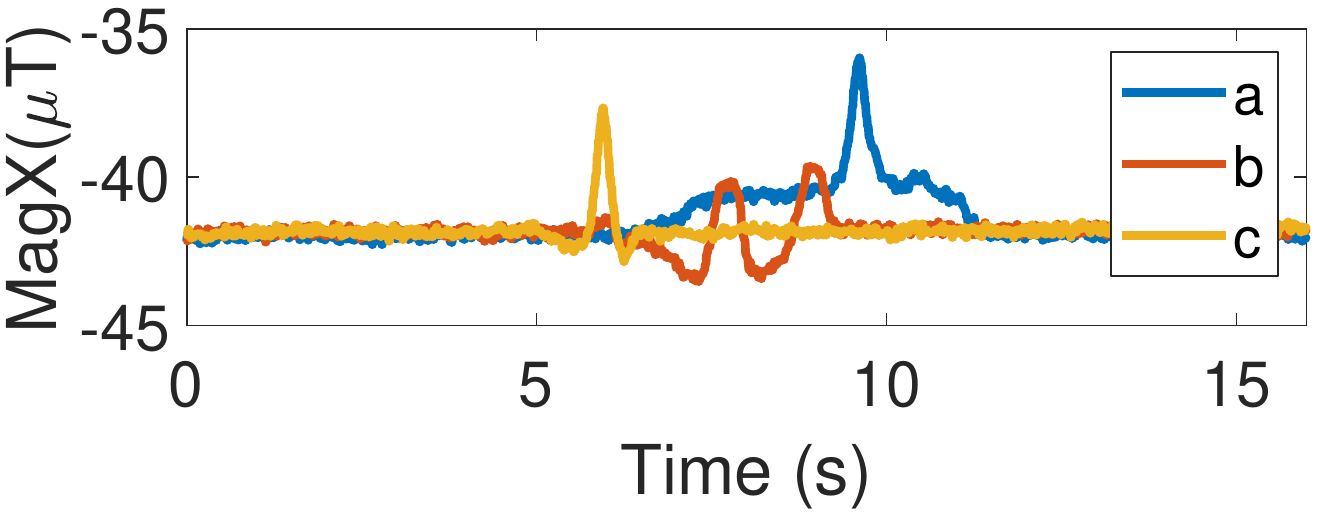}
		\end{subfigure}
		%\label{fig:basic_pattern3d}
		\begin{subfigure}{1\linewidth}
			\centering
			\includegraphics[height=0.66in, width=1.8in]{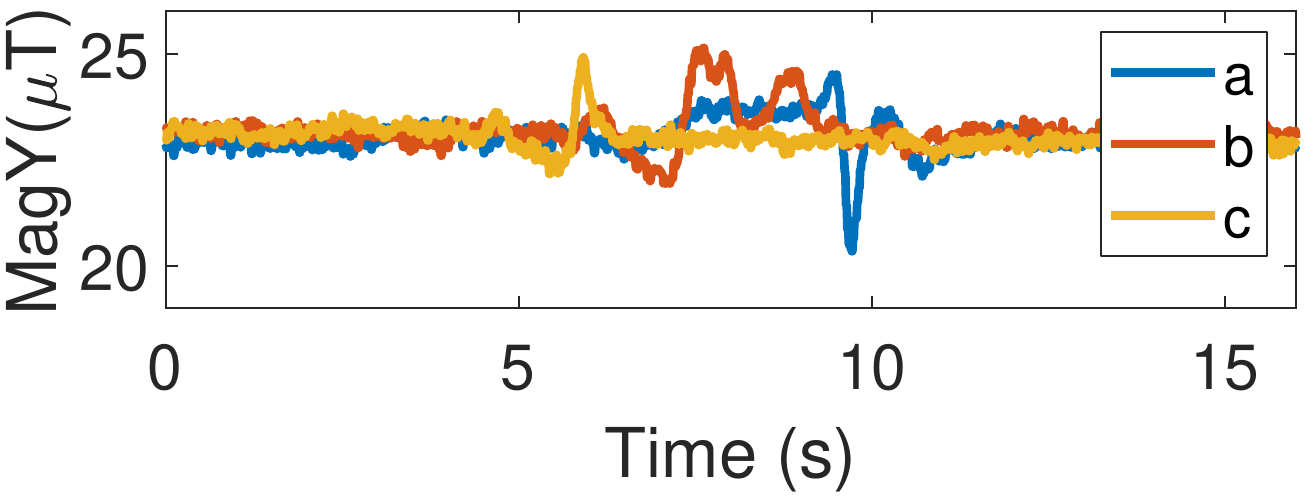}
			%\label{fig:basic_pattern3d}
		\end{subfigure}		
		\begin{subfigure}{1\linewidth}
			\centering
			\includegraphics[height=0.66in, width=1.8in]{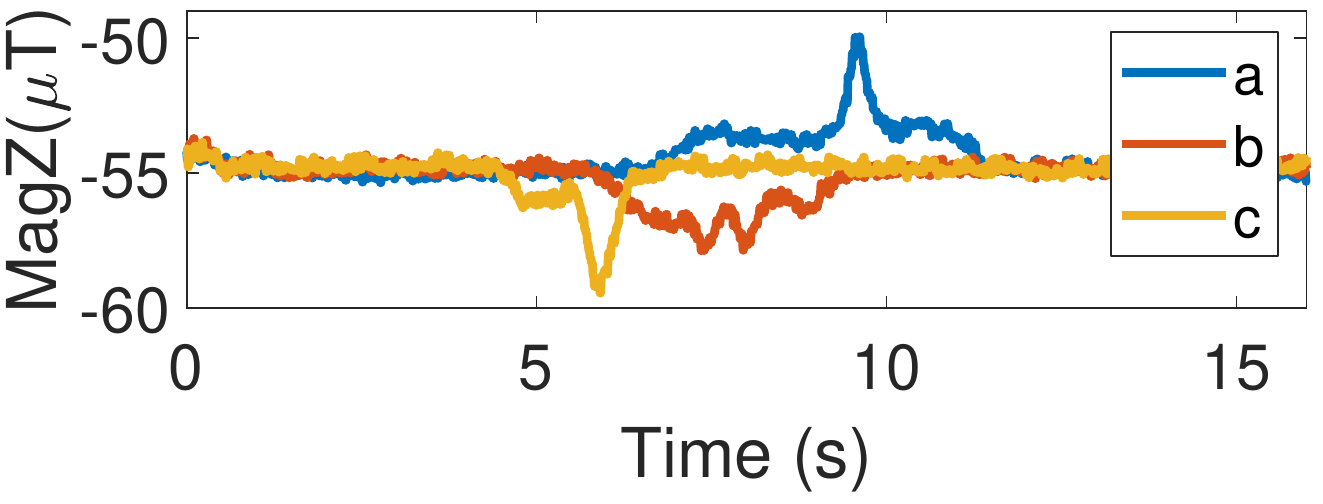}
			%\label{fig:basic_pattern3d}
		\end{subfigure}
		\caption{}
		\label{fig:mag_differentletters}
	\end{subfigure}
	\caption{The magnetometer readings change when a user writes different characters on the screen.}
	\label{fig:different_letters}
\end{figure}

To illustrate our idea, Figure~\ref{fig:mag_differentletters} presents the recorded magnetometer data in X, Y, and Z directions when a user writes the character `a' on the iPad with the Pencil, as shown in Figure~\ref{fig:screen_basic}.
We show that an adversary can infer what users are writing on the iPad's screen by observing the fluctuations in the magnetic readings by \emph{only} having access to motion sensors' data.
We note that the attack does not rely on any touch information (\ie the pencil tip position) since iOS does not allow third-party applications running in the background to access this information.

As observed in Figure~\ref{fig:basic_word}, we can readily determine when the pencil is used to write on the screen from the magnetic data.
However, to infer the \emph{contents} written on the screen, the magnetic readings require detailed analysis.
In our preliminary experiments, we observe various subtle fluctuations in magnetic readings characterized by different letters and words written on the screen.
To illustrate, we present the recorded magnetometer data in Figure~\ref{fig:different_letters} when the letters `a', `b' and `c' are written on the iPad's screen (Figure~\ref{fig:screen_differentletters}).
Figure~\ref{fig:mag_differentletters} shows the magnetometer readings in three axes, where fluctuations in the magnetometer readings are different for each of the three characters. 
One intuitive approach to extract information from these readings is to train a learning model.
This model can then be used to identify useful patterns (\eg letters, numbers, shapes) from similarities between the magnetic data over time for a given set of characters.
Yet, this approach works when the pencil's magnetic impact is consistent when it is held at different locations or with different orientations on the screen for a given character. 
\begin{figure}[t!]%
	\begin{subfigure}{0.45\linewidth}
		\centering
		\includegraphics[height=2in, width=1.5in]{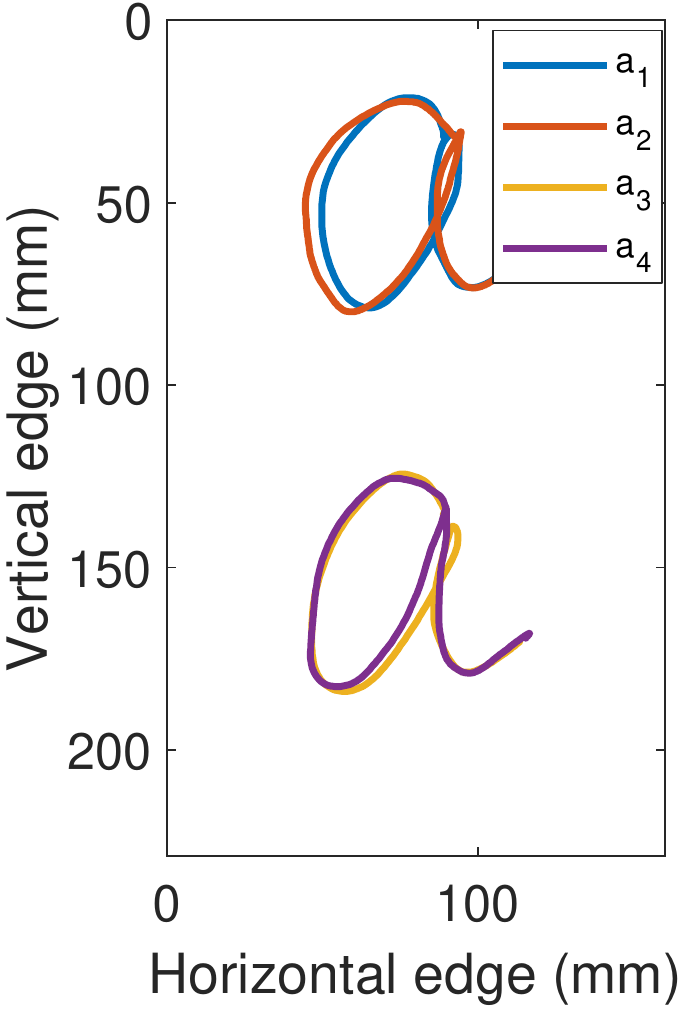}
		\caption{}
		\label{fig:screen_different_locations_orientations}
	\end{subfigure}	
	\begin{subfigure}{0.45\linewidth}
		\begin{subfigure}{1\linewidth}
			\centering
			\includegraphics[height=0.66in, width=1.8in]{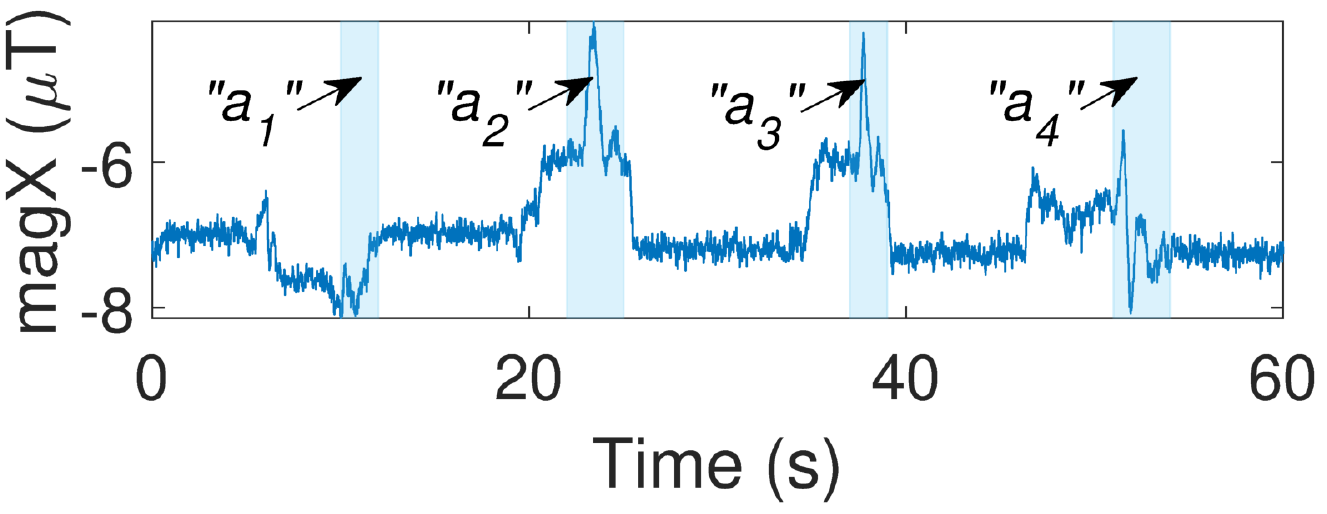}
		\end{subfigure}
		%\label{fig:basic_pattern3d}
		\begin{subfigure}{1\linewidth}
			\centering
			\includegraphics[height=0.66in, width=1.8in]{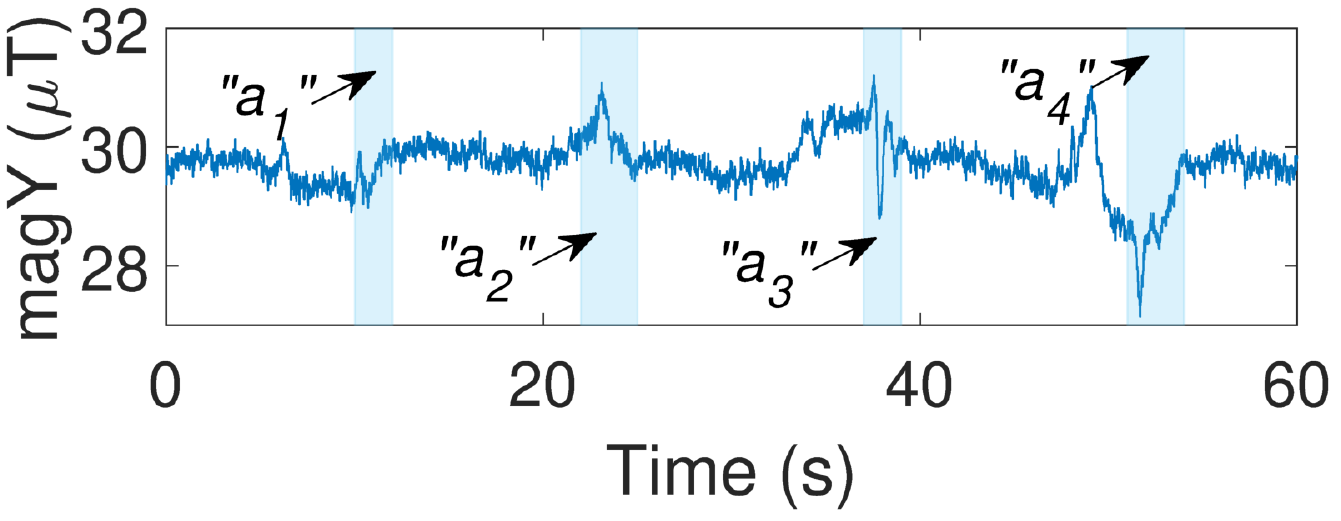}
			%\label{fig:basic_pattern3d}
		\end{subfigure}
		
		\begin{subfigure}{1\linewidth}
			\centering
			\includegraphics[height=0.66in, width=1.8in]{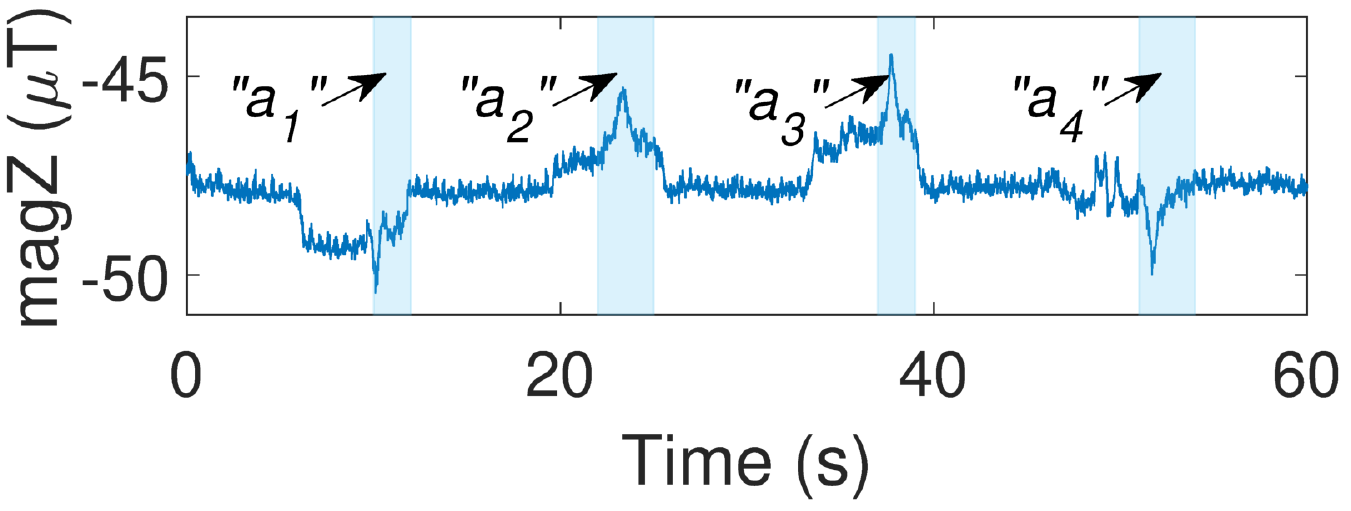}
			%\label{fig:basic_pattern3d}
		\end{subfigure}
		\caption{}
	\end{subfigure}
	\caption{The magnetometer readings change when a user writes the same character on the screen.}
	\label{fig:different_locations_orientations}
\end{figure}
To determine whether this is the case for Apple Pencil and iPad Pro, we wrote the character `a' with the pencil at different locations and orientations on the iPad's screen.
Figure~\ref{fig:screen_different_locations_orientations} illustrates two letters written at different locations ($a_1$ and $a_3$)  and two letters in the same location when the pencil is rolled by $180^\circ$ ($a_3$ and $a_4$). 
We observe that magnetometer readings vary significantly, although the traces of characters are visually the same.

Based on these observations,  we design a tracking algorithm to track the pencil's tip movement using the magnetic field data to identify users' writing.
Building such an algorithm requires finding the correct mapping between a given magnetic reading and the pencil tip's location and its orientation with respect to the screen.
A common method for finding such mappings is through war-driving.
Unfortunately, the mapping space is five-dimensional since we want to track the pencil's 2D location on the screen \emph{while} keeping a record of its 3D orientation.
War-driving for all possible locations and orientations of the pencil on the iPad leads to a huge search space and necessitates a huge amount of human effort.

To address this, we reduce the war-driving space by building a 3D magnetic field map \emph{around the pencil} instead of determining the magnetic field for the screen itself.
We collect magnetometer data while writing with different orientations of the pencil at different locations on the screen.
We employ a computer vision-based approach for pose-estimation to track the pencil's 3D orientation since iOS's touch API does not provide full information about the pencil's 3D orientation.
We apply Kalman filtering~\cite{zarchan2000fundamentals} and smoothing to the estimated pencil orientations to remove noisy data.
We then use this orientation along with the pencil tip location and magnetic data to build the magnetic map with respect to the pencil.
Lastly, to further improve our magnetic field map precision, we adopt a magnetic vector field reconstruction technique~\cite{vectorfield2012}, which uses the divergence and curl properties~\cite{chow2006introduction} of the magnetic fields for optimizing the reconstruction process.
This map generation is conducted offline and does not require information collected from the target user.

Using the designed magnetic field map, we build a multi-dimensional \emph{particle filter}~\cite{particleFilter} to track the status of the pencil on the iPad's screen, which solely operates with the data collected from motion sensors.
The particles' state includes the pencil tip's location and the 3D orientation of the pencil represented as quaternion vectors~\cite{diebel2006representing}.
We use the human writing behavior to guide the particles’ initialization and transition (components of our particle filter).
%
% Further, we use KLD resampling~\cite{fox2002kld} to improve efficiency,  incorporate particles' history to handle scattered particle clusters that may exist at the beginning of the tracking, and the variance of magnetometer, accelerometer, and gyroscope readings to estimate when a user begins and ends writing on the screen.
Additionally, we used KLD resampling~\cite{fox2002kld} to improve efficiency and incorporated particles' history to handle scattered particle clusters that may exist at the beginning of the tracking process.
%Additionally, we leveraged the human writing behavior to guide the
%particles' initialization and transition (components of our particle filter) to improve our tracking accuracy.
%
For an end-to-end attack, we also analyzed the variance in the magnetometer, accelerometer, and gyroscope readings to estimate when the user begins and ends writing on the screen.

We implement our system, \sys{} (\textbf{S}ide-channel attack on \textbf{S}tylus pencil through \textbf{S}ensors), on Apple 11" iPad Pro running iOS 12.0.
We evaluate \sys{} with $10$ subjects, where the subjects write different letters, numbers, shapes, and words at different locations of the screen.
We show that an adversary (randomly chosen from the subjects) is able to correctly identify $93.9$\%, $96$\%, and $97.9$\% of the written letters, numbers, and shapes, and English words with an accuracy of $93.33$\%. 
%In this work, we make the following contributions:

The main contributions of this paper are summarized as follows:
\begin{itemize}
% \item We introduce potential private data leaks through motion sensor data in  tablet devices from the embedded magnets in stylus pencils. 
% %
% \item We present \sys{}, a novel system that infers users' writing from motion sensors’ data. \sys{} includes a tracking algorithm to track the pencil's tip movement and a multi-dimensional particle filter algorithm to precisely infer the user's writing. 
% %
% \item We evaluate \sys{} on an iPad Pro with Apple Pencil with $10$ human subjects and demonstrate its high accuracy in uncovering letters, numbers, shapes, and words without significant runtime overhead.
	\item We unveil a privacy leakage through motion sensor data in modern tablet devices due to the introduction of stylus pencils with embedded magnets. 
	\item We design a novel tracking algorithm by constructing a magnetic map of the pencil and building a sophisticated multi-dimensional particle filter that can track the user's writing. 
	\item We implemented our system on an iPad Pro with Apple Pencil. A thorough evaluation with $10$ human subjects demonstrates its high accuracy in uncovering letters, numbers, shapes, and words in a variety of scenarios, without significant overhead.
\end{itemize}

%% file: threat_model.tex
\section{Threat Model}
\label{sec:threat_model}
We use Apple's iPad and Pencil (2nd generation) to demonstrate a proof-of-concept of inferring information about what the user is writing from motion sensors data. However, our attack can be a threat to any mobile device with a stylus pen support using embedded magnets.
\majorRev{[MR1, C1, C5]}\majorRevText{The sensitive information being leaked out from user writings can be an unprecedented threat to the confidentiality of user data processed by the writing apps; an adversary can infer passwords, text messages, secure access codes, and other secrets of the user.}

\majorRev{[MR5, C8, C9]}\majorRevText{We consider an adversary that controls a malicious native application installed on the victim's device, which has access to the motion sensors data and runs in the background during data collection. 
The native application does not have any information about other running applications and does not have access to system resources.
To detail, the iOS applications, by default, have access to the motion sensor data, and only web applications from iOS $12.2$ and onwards require explicit user permission.
Starting from iOS $5$, an application can stay active in the background if it performs specific tasks such as playing audio, receiving updates from a server, and using location services. 
Given these facts, an adversary can easily mimic a benign legitimate app such as a fitness and activity tracker to stay active in the background to collect motion data.
The application periodically logs the motion sensors' data, including accelerometer, gyroscope, and magnetometer readings.
The recorded sensor data is stored locally and sent to a remote server for processing. These processes incur minimal energy and computation overhead in order to remain undetected (Detailed in Section~\ref{sec:evaluation}).}

\majorRev{[MR5, C2, C6, C10, C11]}\majorRevText{We additionally assume that the adversary can obtain or learn a model to capture the users' writing behavior, which is used to predict how the Pencil's movement changes over time through Pencil's previous positions.
The adversary could learn such a model by collecting handwriting samples (various letters, numbers, shapes, etc.) from a small number of people (up to 3 users is sufficient for the attack to succeed as detailed in our evaluation in Section~\ref{sec:evaluation}).
To build this model, an adversary uses a computer vision-based technique (See Section~\ref{sec:map_generation}) to track the Pencil's orientation while logging the Pencil's location and motion sensors' data. 
We note that the handwriting samples \emph{are not collected from the potential attack victims} and are solely obtained from the adversary and their accomplices.
After the model is learned, the adversary uses the model to infer the free-form handwriting of unaware users \emph{solely by collecting motion sensor data}.
We emphasize that the Pencil location and orientation data are collected only during the training process. An adversary does not need to access this information at attack time to infer user writings.}

%% file: 2D_tracking.tex
\section{Pencil Tracking: A First Look}
\label{sec:first_look}

In our quest for inferring the user's writing from the sensors' data, we begin by exploring the simplest case:
\textit{assume that the user always holds the Pencil in a fixed orientation, i.e., vertically such that the Pencil altitude remains $90^\circ$.}
%as shown in Figure~\ref{fig:vertical_pencil}.
%
%More specifically, we assume the Pencil is held vertically with an altitude of $90^\circ$.
%
In this case, to track how the Pencil moves on the screen, we first determine a mapping between the magnetic readings and the location of the Pencil tip. %since, the orientation stays the same.
We define this mapping as the \textit{2D magnetic map}.

\subsection{2D Magnetic Map}
%In order to build the magnetic field map, we have to determine the variation in the magnetic field caused by the Pencil at each location on the iPad screen.
%
The 2D magnetic field map would be a function, $f$, that returns the magnetic readings corresponding to a given location on the screen i.e.
\begin{equation}
(m_x, m_y, m_z) = f(x, y)
\label{eq:2dmap}
\end{equation}
where $x$ and $y$ are the coordinates of the Pencil tip location and $m_x$, $m_y$ and $m_z$ are the magnetic reading in $x$, $y$ and $z$ directions. We consider the shorter edge of the iPad as $x$-axis and the longer edge as $y$-axis.

\begin{figure}[t!]
	\begin{subfigure}{0.32\linewidth}
		\centering
		\includegraphics[height=1.8in]{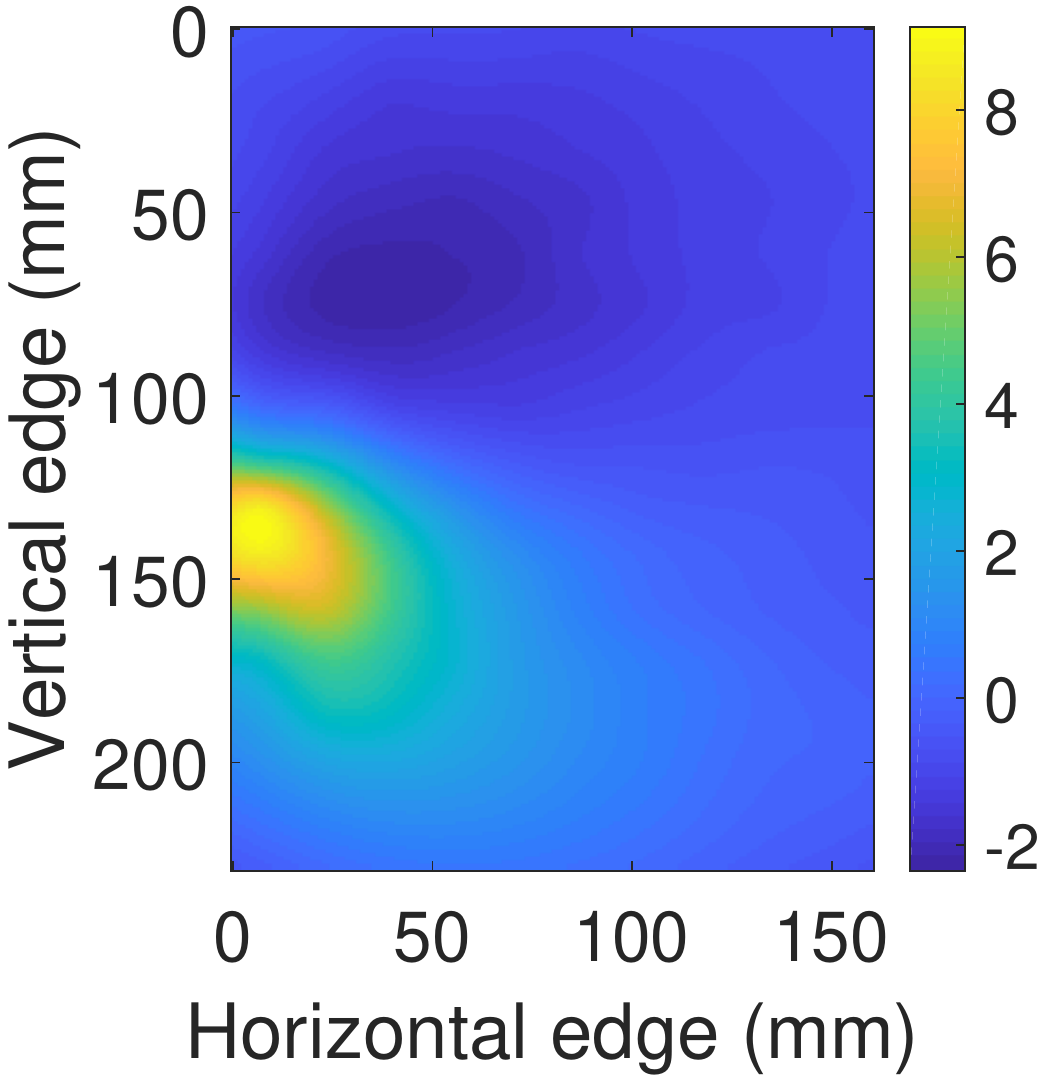}
		\caption{}
		\label{fig:2dmap_x}
	\end{subfigure}
	\begin{subfigure}{0.32\linewidth}
		\centering
		\includegraphics[height=1.8in]{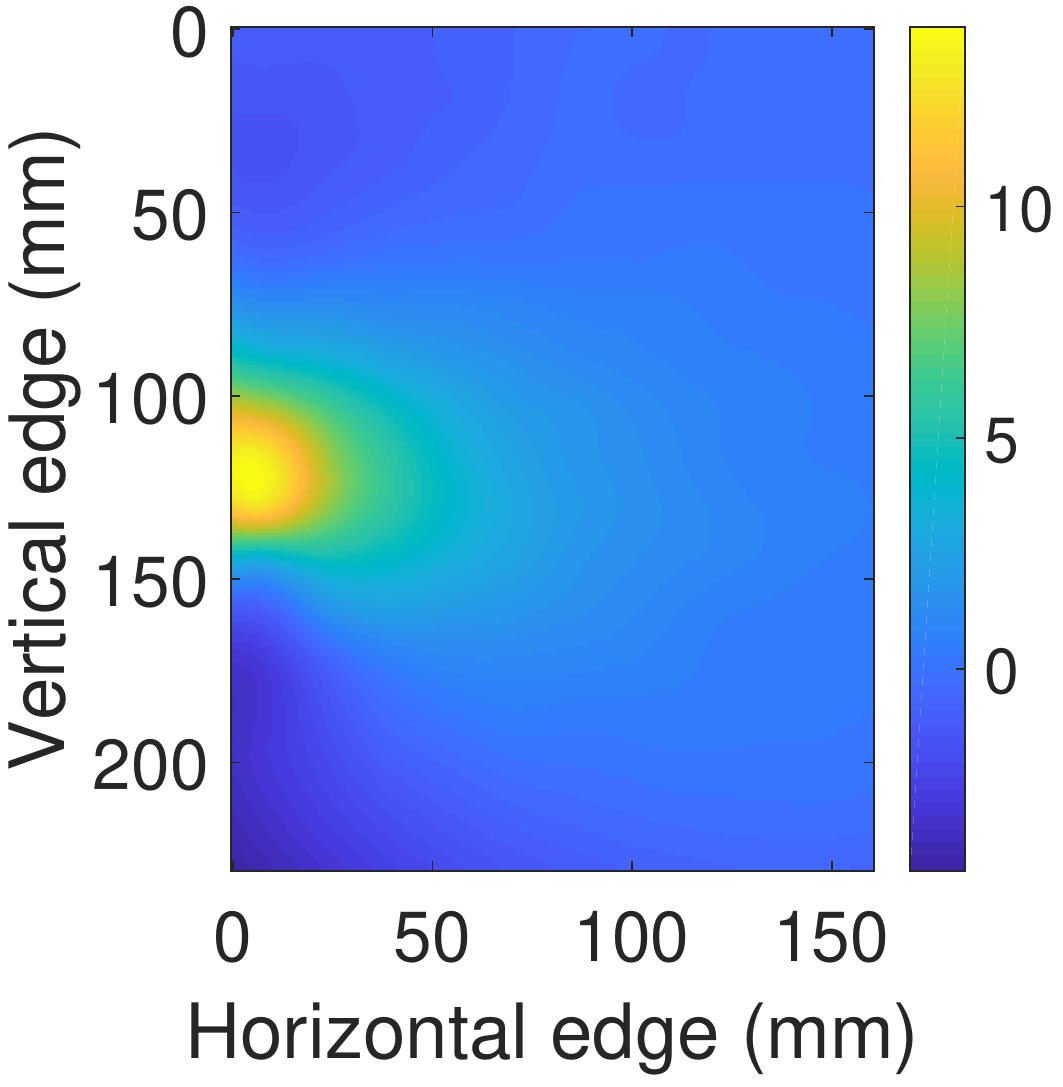}
		\caption{}
		\label{fig:2dmap_y}
	\end{subfigure}
	\begin{subfigure}{0.32\linewidth}
		\centering
		\includegraphics[height=1.8in]{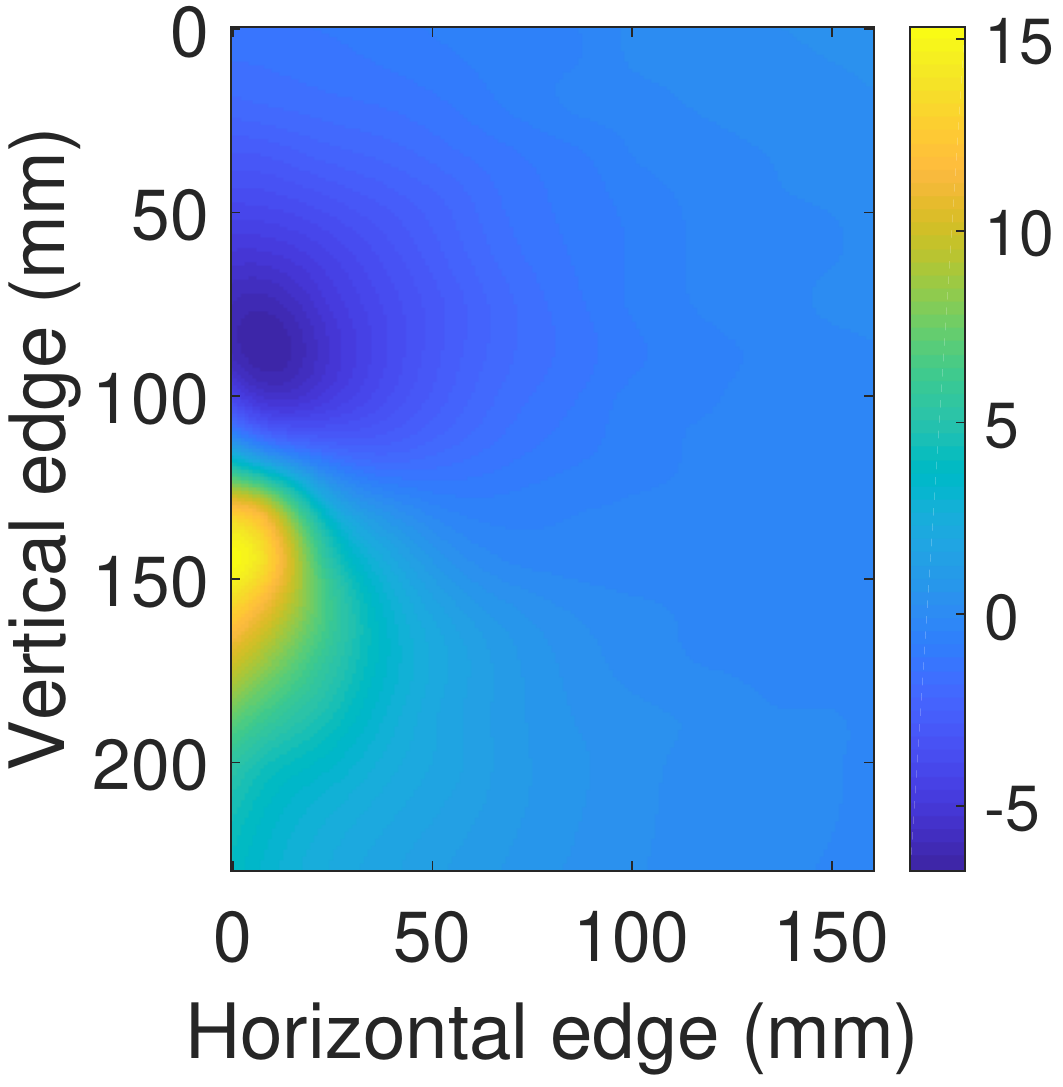}
		\caption{}
		\label{fig:2dmap_z}
	\end{subfigure}
	\caption{2D magnetic map: the magnetic impact of the Pencil at different locations on the screen in (a) $X$, (b) $Y$, and (c) $Z$ dimensions.}
	\label{fig:2dmap}
\end{figure}

To build this map, we draw on the screen using the Pencil such that we cover the entire screen.
We use a custom drawing application that logs the $x$ and $y$ coordinates of the Pencil on the screen using iOS touch API, at a rate of $120$Hz.
%
%OS's API provides the location of the touch ($x$ and $y$ coordinates of the screen in pixels) of the Pencil.
%
%The API scans for a touch at a rate of $120$ times per second.
%
The magnetic data is recorded in the background at a frequency of $50$Hz.
The timestamps recorded with the magnetic and touch data are used to align the magnetic readings with each of the touch samples. 
%
%The magnetic data is recorded at a frequency of $50$Hz.
%  
%For a 12.9" iPad Pro, the screen is $155.8mm$ wide and $222.6mm$ in length, with a pixel resolution of ??.
%
%Based on iOS API's touch coordinates, 
We start collecting the magnetic data a few minutes before drawing (when the Pencil is detached from the iPad and is not on the screen) to sense the ambient magnetic field.
The average of the ambient magnetic data is subtracted from the collected data to eliminate the magnetic impact of the surroundings.
Figure~\ref{fig:2dmap} shows the 2D magnetic field map in the $x$, $y$ and $z$ directions. 
%
%The axes for the map are measured in millimeters, representing the width and height of the drawable region of the iPad screen.
\subsection{2D Tracking}
\label{sec:2dtracking}

Once we have built the 2D magnetic map, we ask the question: \textit{how can we track the Pencil location using only the magnetic data?}
Here we borrow a standard tracking algorithm, \textit{particle filter}, to track the Pencil's location.
Particle filtering is a probabilistic approximation algorithm for state estimation problems~\cite{particleFilter}.
In particular, for object location estimation, particle filter operates by maintaining a probability distribution for the location estimate at time $t$, which is represented by a set of weighted samples called particles. 
The state of these particles is updated based on a motion model for the object, and the weights for these particles are assigned at each timestamp by a likelihood model for the recorded sensor data. 
The particles are then resampled at each timestamp using importance sampling to choose a new set of particles according to the weights of the prior samples.
%Here we realize that we can benefit from the existing approaches for object tracking in robotics and signal processing.
%
%We consider \textit{particle filter}, a popular state estimation algorithm, to track the Pencil's location.

\begin{figure}[t!]
	\begin{subfigure}{0.32\linewidth}
		\centering
		\includegraphics[width=1.5in]{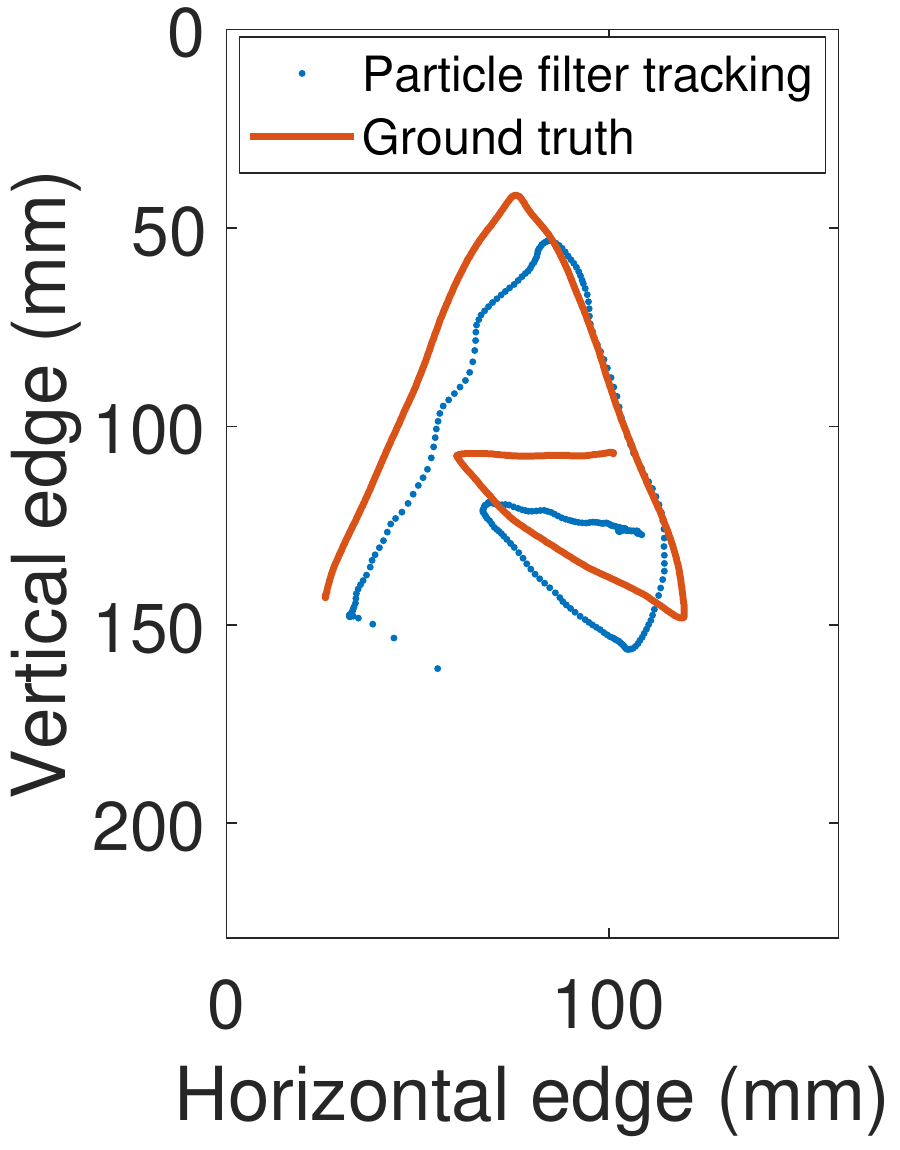}
		\caption{}
		\label{fig:2dpf_a}
	\end{subfigure}
	\begin{subfigure}{0.32\linewidth}
		\centering
		\includegraphics[width=1.5in]{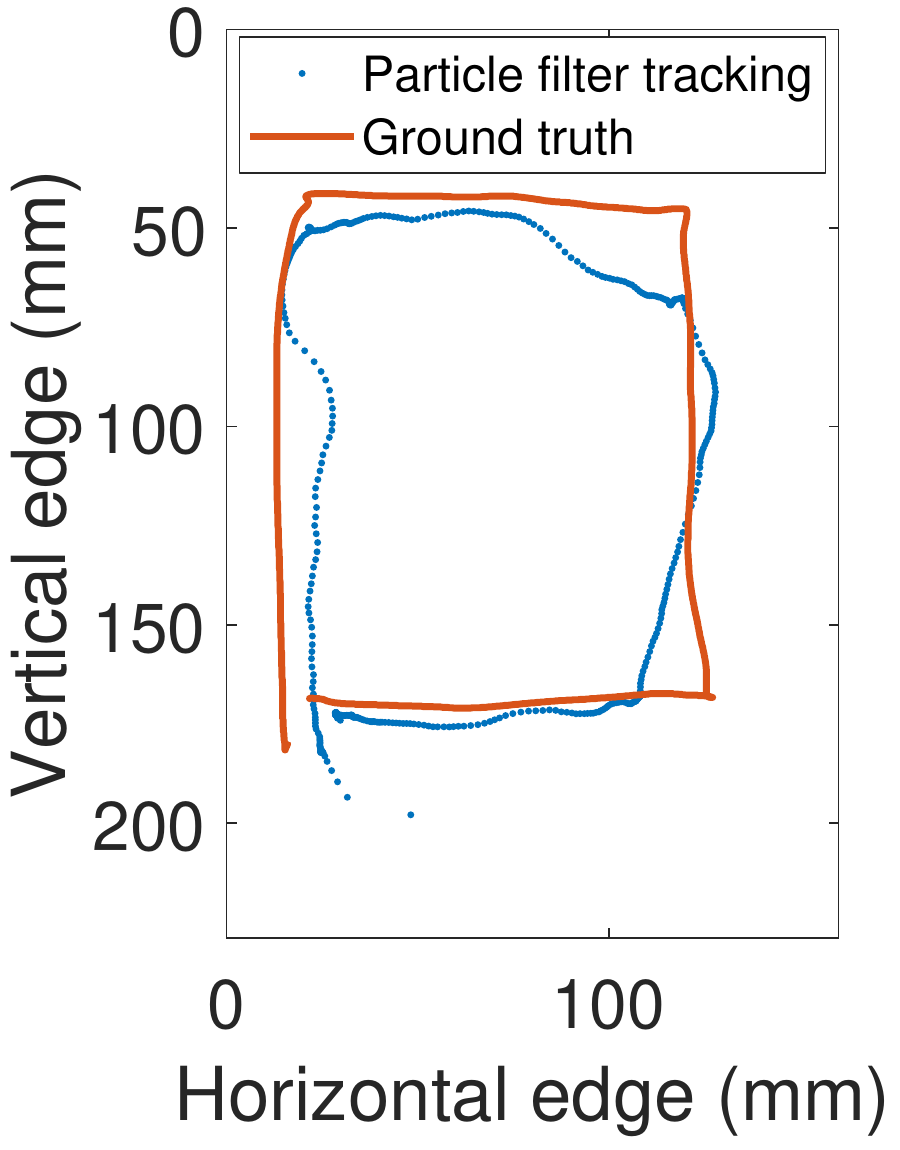}
		\caption{}
		\label{fig:2dpf_shape}
	\end{subfigure}
	\begin{subfigure}{0.32\linewidth}
		\centering
		\includegraphics[width=1.5in]{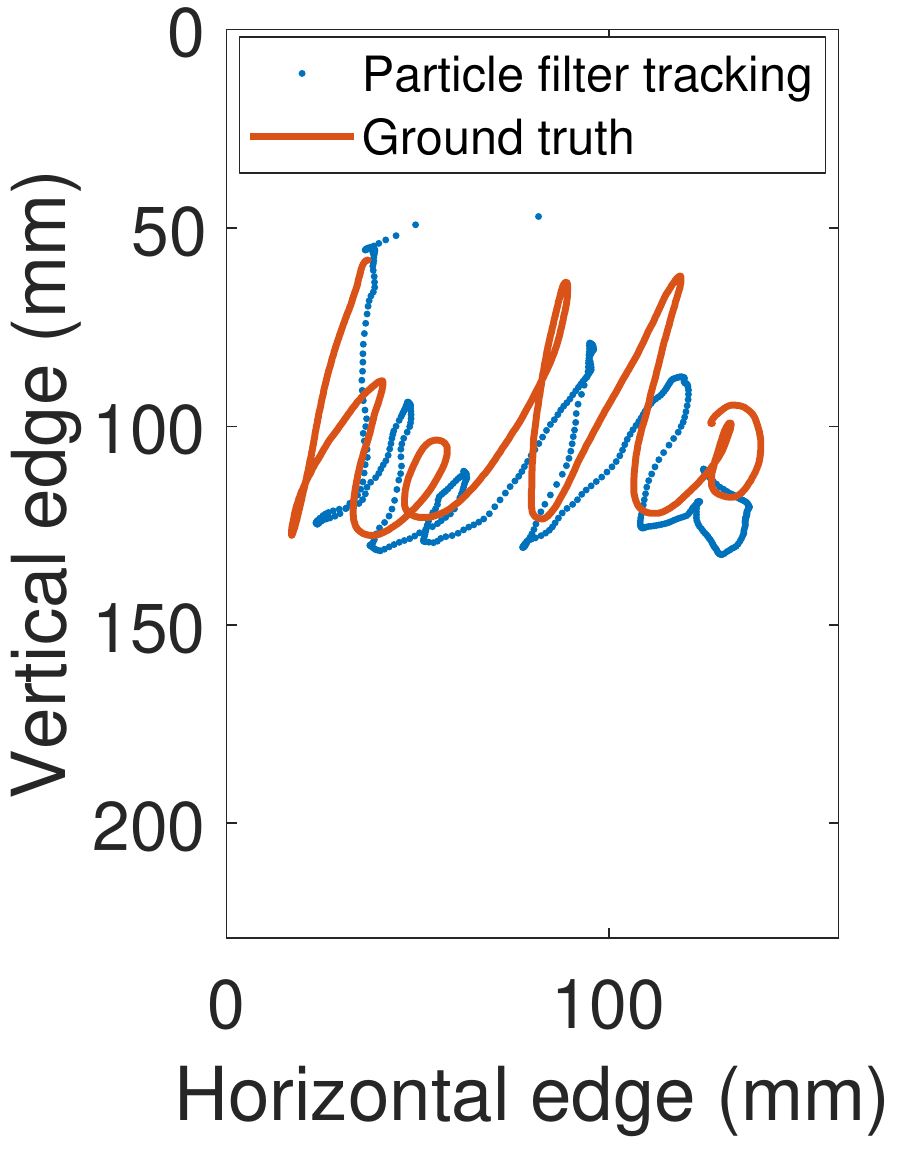}
		\caption{}
		\label{fig:2dpf_hello}
	\end{subfigure}
	\caption{Tracking results from a simple implementation of particle filter.}
	\label{fig:2dparticle_filter}
\end{figure}

We define the state vector for the particle filter as $s_t$, representing the Pencil tip location.
Here, $t$ represents the timestamp, which is updated at a frequency of $50$Hz.
For the simple  case, we define the Pencil movement model as:
\begin{equation}
s_{t+1} = s_t+ b_t
\label{eq:2dparticle_transition}
\end{equation}
%where $s_t$ represents the state vector which is the estimated location of the Pencil tip ($\langle x, y \rangle$ coordinates in millimeter) at time $t$ in our case.
%
where $b_t$ is a random perturbation added to the state, with a Gaussian distribution.

Weights are assigned to each particle at each timestamp with the function:
\begin{equation}
%w_t^i = w_{t-1}^i p(u_t|s_t^i) =  w_{t-1}^i p_{e_t}(dist(u_t, f(s_t^i)))
w_t^i = \exp\left( \frac{(m_t - r_t)^2}{\sigma}\right)
\label{eq:2dparticle_weight}
\end{equation}
Here $r_t$ is the magnetic readings obtained from the magnetometer at time $t$ and $m_t$ is the queried magnetic readings given state is $s_t$, using Equation~\ref{eq:2dmap}. 
The location of the Pencil tip at each timestamp is the weighted average of the particles.
Figure~\ref{fig:2dparticle_filter} shows the results of this simple implementation of particle filter when used to track a letter (\textit{`A'} - Figure~\ref{fig:2dpf_a}), a shape (a square - Figure~\ref{fig:2dpf_shape}) and a word (\textit{`hello'} - Figure~\ref{fig:2dpf_hello}).
%
%The figure shows the ground truth and the tracking results from the particle filter.
%
Even though we assumed that the Pencil orientation is fixed for this case, the tracking results show promise that we can infer what the user is writing based on sensors' data. 
%Here, even though we do not use any trend information in the particle transition function, we are able to track the Pencil movement pretty well.
%
%Thus, the usage of particle shows promise for tracking what the user is writing.

%% file: system_architecture.tex
\section{System Architecture}
\begin{figure}[t!]
	\centering
	\includegraphics[width=4.5in]{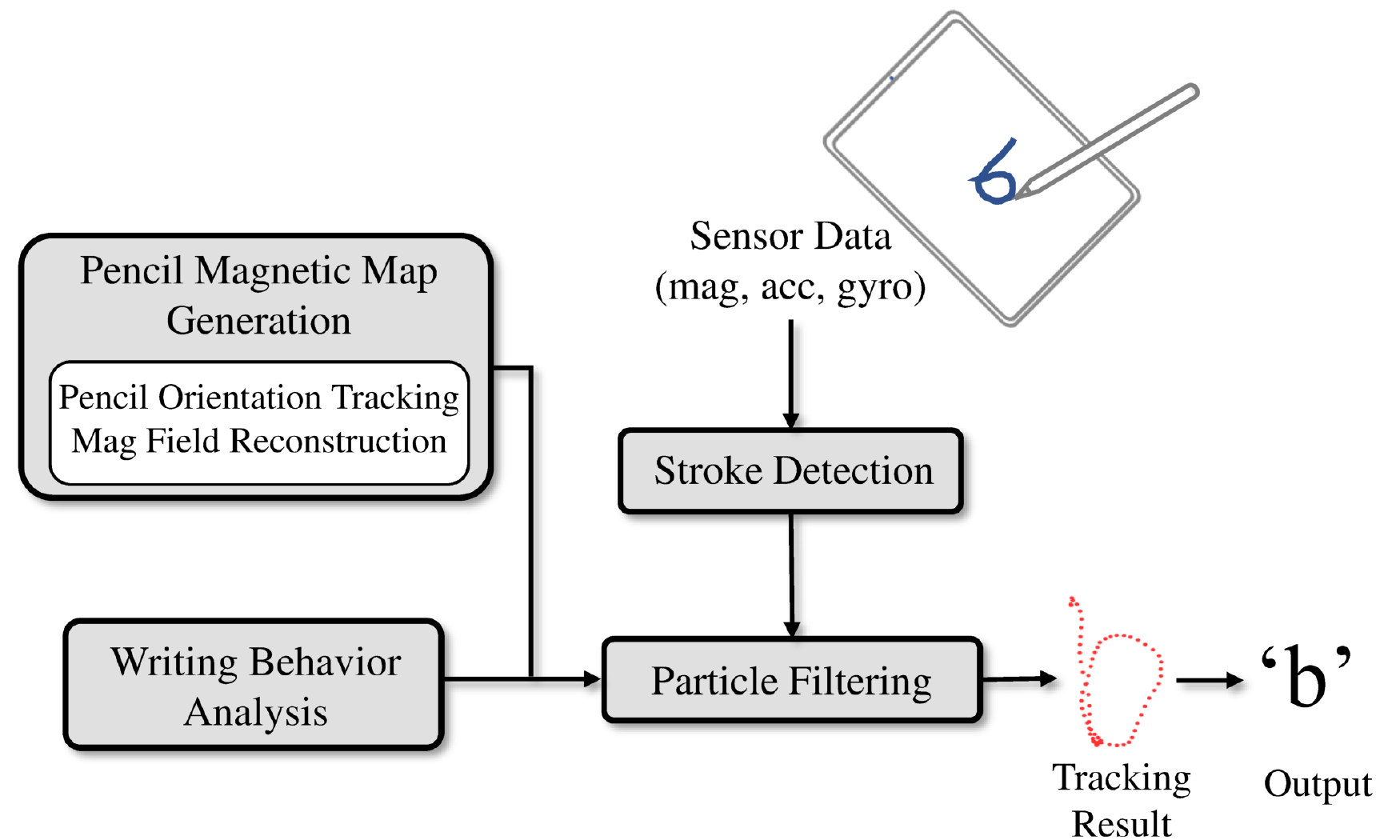}
	\caption{System architecture.}
	\label{fig:system_architecture}
\end{figure}
This section presents the functional overview of our system \sys as shown in Figure~\ref{fig:system_architecture}.
We will detail each component of our system in Section~\ref{sec:system_design}.
%Figure~\ref{fig:system_architecture} shows the overview of our system.
%
%Since in realistic attack scenarios, an attacker cannot restrict the orientation of the Pencil movement while a user is writing, solving the simple case is not enough.
%

To launch a realistic attack, our system should find out what the user is writing in their natural handwriting style, unlike the fixed orientation case in the previous section.
For this purpose, we need a mapping between the magnetic data collected and the status of the Pencil, which now includes location and orientation, i.e., $5$ degrees of freedom ($\langle x, y \rangle$ location and 3D orientation of the Pencil).
%In this paper, we develop a system that only uses the data collected by the magnetometer when a user is writing on the iPad with Apple Pencil. 
%
We generate this map through war-driving and reduce the effort involved by building the magnetic map around the Pencil.
%We reduce the effort involved in war driving for building this magnetic mapping by generating the magnetic map around the Pencil.
%the degrees of freedom for the Pencil's movement can be reduced from $5$ to $3$ i.e. the 3D orientation of the Pencil.
%Given the huge space for possible 3D orientations of the Pencil, war driving is not a feasible option.
%
We extend computer vision techniques to track the Pencil's orientation by attaching a checkerboard on top of the Pencil and recording the Pencil movement with a camera.
%Therefore, we design an algorithm to build the magnetic map around the Pencil to reduce our search space.
%
%We assume that the user can hold the Pencil in any way they like.
%
%To build such a system, the first step is to build a magnetic field map that can act as a bridge between the magnetic readings collected and the corresponding position of the Pencil considering its location as well as orientation.
%
%In order to build such a map, we rely on computer vision techniques to accurately track the orientation of the Pencil.
%
%
%Given the huge space for possible 3D orientations of the Pencil, we design an algorithm to build the magnetic map around the Pencil to reduce our search space.
%
%The magnetic and touch data collected for the magnetic map is noisy and discontinuous. 
%
A magnetic field reconstruction technique~\cite{vectorfield2012} is used to remove noise from the collected magnetic data and generate a continuous magnetic field map for the Pencil. 
This magnetic map for a given iPad and Pencil can be built solely by the attacker. It is a one-time effort since the same map is used to track the user's handwriting in different locations and environments.

We also build a model for writing behavior by writing different letters, numbers, and shapes at different locations on the screen multiple times while tracking the Pencil orientation with a camera, as previously mentioned.
This model finds the relationship between orientation and location changes while writing via linear regression.
It predicts how these parameters should change at a given timestamp with respect to the Pencil's previous position.
Similar to the magnetic map, this writing behavior model is also trained on data collected from the attacker and their accomplices and does not involve the victim.

The motion sensors' data (including magnetometer, accelerometer, and gyroscope data) collected by the attacker's application is sent to a back-end server for processing.
We apply stroke detection on this motion data to detect the start and end of a stroke.
The magnetic data corresponding to the detected stroke is then fed into a multi-dimensional particle filter (Section~\ref{sec:pencil_tracking}), together with the writing behavior model.
%To facilitate the stroke separation, we also use accelerometer and gyroscope data to capture the tiny movements in the iPad when the Pencil interacts with the iPad.
%
%Finally, the 3D magnetic map and strokes' magnetic data is fed into the particle filtering module.
%
%We observe that for tracking the free-form writing, at each time instance, we can benefit from the information known about Pencil's position in the previous time steps. 
%
%Essentially, using the history of the particles in particle filtering helps us achieve better and more accurate tracking.
%
%To take the history into account, we use the concept of sequential particle filtering (Sequential Monte Carlo method) to infer what the user is writing.
%
%
The writing behavior model helps in predicting the next state of the particle filter.
The particle filter outputs the tracking traces of the stroke.
An attacker then looks at the tracking result to guess what the user wrote.
%tune the various parameters of particle filter such as the variance in the magnetic data while writing with the Pencil.

%In the following section, we describe the details of each component of our system. 
%%
%We first define the different coordinate systems involved followed by our approach for determining the Pencil's orientation.
%%
%This is followed by details of the map generation and reconstruction process.
%%
%Finally, we describe our algorithm for our sequential particle filtering.  

%% file: system_design.tex
\section{System Design}
\label{sec:system_design}
This section describes the details for tracking when the users write with their natural handwriting style.
To this end, we first describe our approach for building the Pencil's magnetic map.
This is followed by the details of our tracking algorithm and stroke detection.
%Once we have solved the simple case of inferring what is written on the screen with a fixed orientation of the Pencil, we proceed to the more general and realistic case:
%Assume that the user can write with their natural free-form writing style.
%
%This means that we have to relax the constraints on the Pencil's orientation during writing.
%%
%In this case, the magnetic field map needed for tracking the Pencil movement is a function, $f$, mapping the location $\langle x, y \rangle$ of the Pencil tip and its 3D orientation to the magnetic readings in $x$, $y$ and $z$ directions i.e.
%\begin{equation}
%(m_x, m_y, m_z) = f(x, y, orientation)
%\label{eq:5dmap}
%\end{equation}
%%
%
%For building this magnetic map, we first want to introduce the coordinates systems for the iPad screen and the Pencil and clarify the notion of the Pencil's orientation.
\subsection{Pencil Magnetic Map Generation}
\label{sec:map_generation}
The magnetic field map needed for tracking the Pencil movement is a function, $f$, mapping the location of the Pencil tip and its 3D orientation to the magnetic readings in $x$, $y$ and $z$ directions i.e.
\begin{equation}
(m_x, m_y, m_z) = f(x, y, orientation)
\label{eq:5dmap}
\end{equation}
%This means that the map has $5$ degrees of freedom.
%%
%However, if we build the map around the Pencil, the degrees of freedom are reduced to $3$.
%%
%This approach reduces the efforts involved in building the map.
%%
%In this section, we describe the details of the map generation process.
%%
To build this magnetic map, we first want to introduce the coordinate systems for the iPad screen and the Pencil and clarify the notion of the Pencil's orientation.
%\begin{figure}[h!]
%	\centering
%	\includegraphics[width=2in]{figures/designDetails/coordinate_systems.pdf}
%	\caption{Screen and Pencil coordinate systems and the location of magnetometer.}
%	\label{fig:coordinate_systems}
%\end{figure}

\begin{figure}
	\begin{subfigure}{0.32\linewidth}
		\centering
		\includegraphics[height=2in,width=2in]{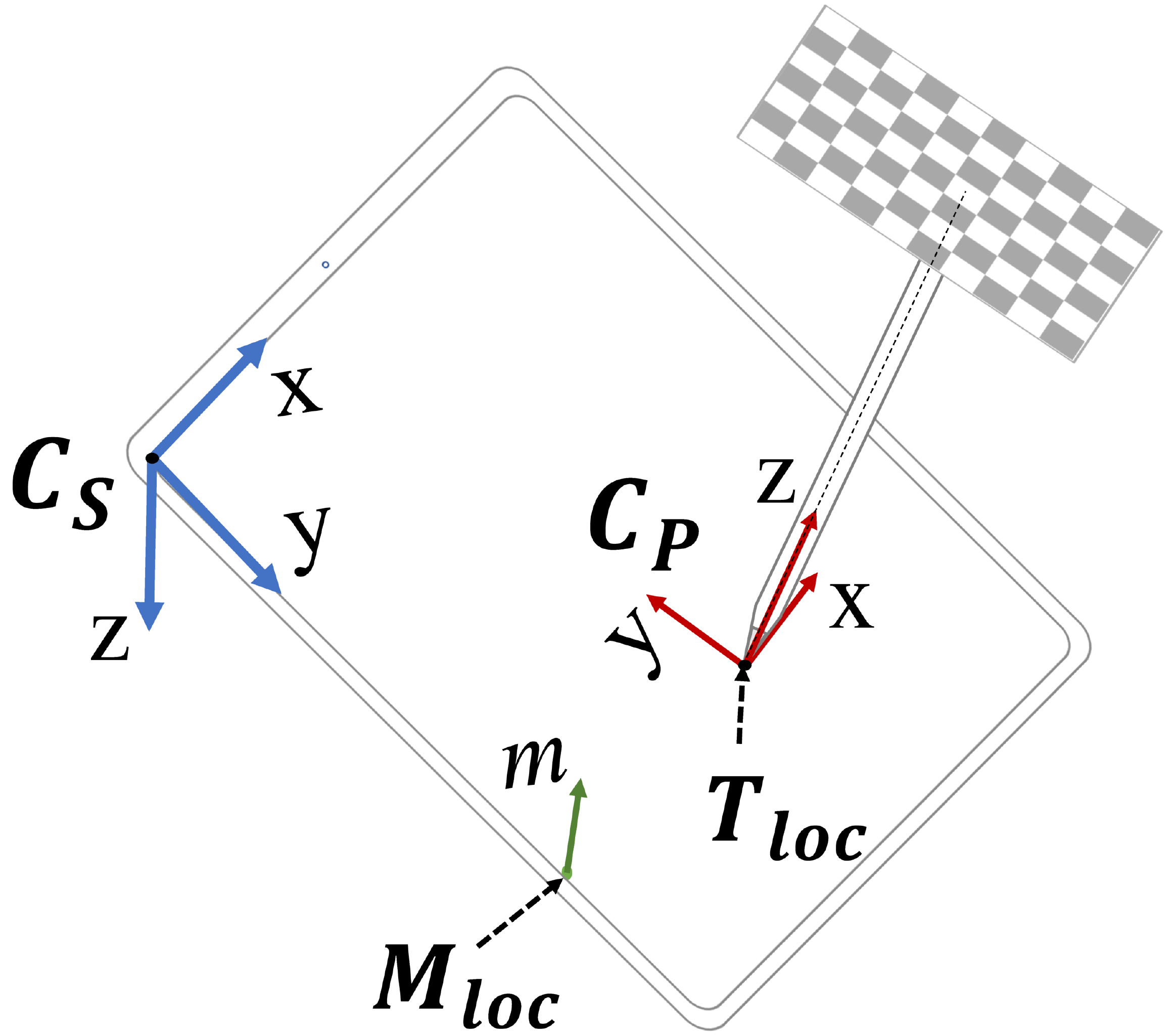}
		\caption{}
		\label{fig:coordinate_systems}
	\end{subfigure}
	\begin{subfigure}{0.32\linewidth}
		\centering
		\includegraphics[height=2in,width=1.8in]{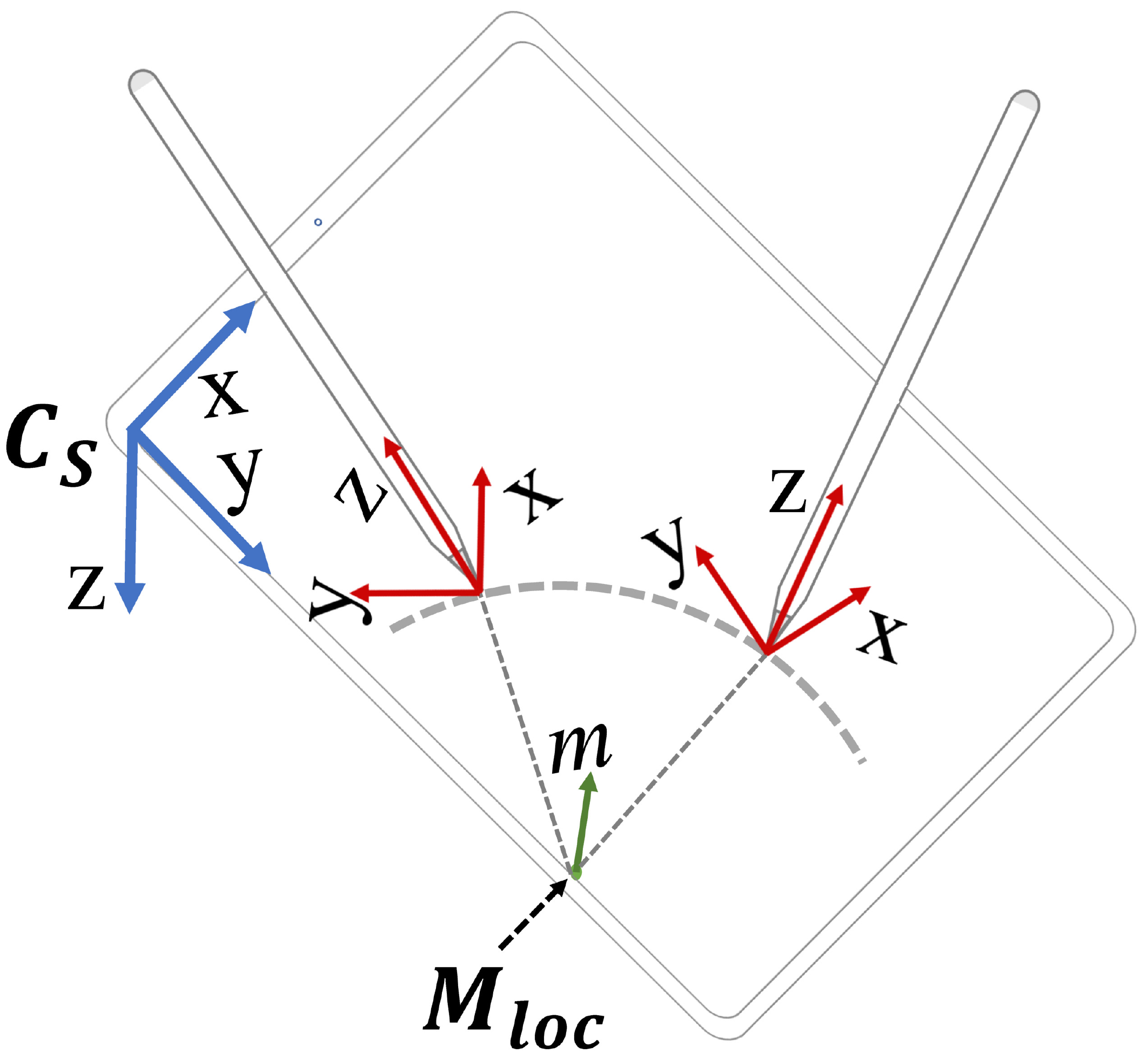}
		\caption{}
		\label{fig:pencil_states}
	\end{subfigure}
	\begin{subfigure}{0.32\linewidth}
		\centering
		\includegraphics[height=2in,width=1.5in]{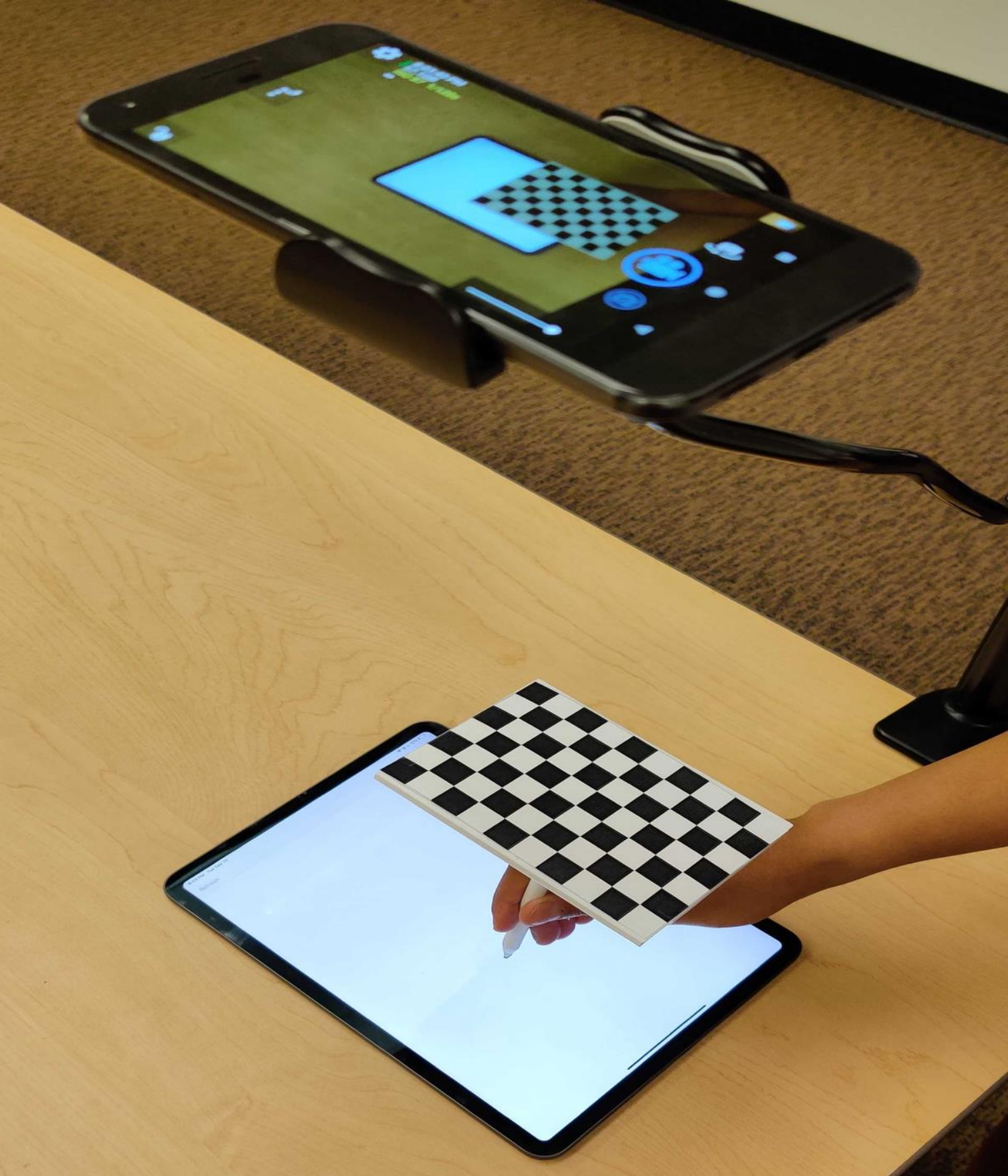}
		\caption{}
		\label{fig:camera_setup}
	\end{subfigure}
	\caption{(a) Screen and Pencil coordinate systems and the location of the magnetometer. (b) An example that the location of the magnetometer in the Pencil coordinate system is the same, even when the Pencil's locations and orientations are different. (c) Setup for tracking the Pencil orientation.}
	\label{fig:screen_pencil}
\end{figure}
%\textbf{iPad Screen and Pencil Coordinate Systems}: 

Figure~\ref{fig:coordinate_systems} shows the coordinate systems for the iPad screen, $C_s$.
The top-left corner of the iPad screen is the origin for $C_s$.
The Pencil tip location reported by the iOS touch API is also in the screen coordinate system.
The $Z$ axis of $C_s$ points downwards based on the right-hand rule.
$M_{loc}$ and $T_{loc}$ are the locations of the magnetometer and Pencil tip in $C_s$, respectively.
We assume that $M_{loc}$ is known beforehand.
Figure~\ref{fig:coordinate_systems} also shows the Pencil's coordinate system, $C_p$.
The $Z$-axis of the Pencil runs vertically through the Pencil's body.
We define the orientation of the Pencil as its axes expressed as vectors in the $C_s$ denoted by $X_{ps}$, $Y_{ps}$ and $Z_{ps}$.

We can now define equation~\ref{eq:5dmap} as
\begin{equation}
(m_x, m_y, m_z) = f(T_{loc}, X_{ps}, Y_{ps}, Z_{ps})
\end{equation}

For the free-form writing case, we need to move the Pencil in all possible locations and orientations to get the magnetic impact for each location and orientation.
%
%Determining the magnetic readings corresponding to every orientation through war driving 
This requires a huge amount of human effort.
Hence, we focus on opportunities to reduce the war-driving space for building the magnetic field map.

%\begin{figure}[h!]
%	\centering
%	\includegraphics[height=2in, width=2in]{figures/designDetails/pencil_states.pdf}
%	\caption{An example that the location of the magnetometer in the Pencil coordinate system is the same even when the Pencil's locations and orientations are different.}
%	\label{fig:pencil_states}
%\end{figure}

%
So far, we have considered building the magnetic map from the iPad's perspective, \ie the Pencil's location and orientation, and the magnetic readings are in $C_s$.
Another way to look at this problem is to build the map from the Pencil's perspective.
Recording the Pencil movement's magnetic impact on the screen is essentially the same as keeping the Pencil stationary and moving the magnetometer around it.
In other words, for any position of the Pencil in $C_s$, we can find the corresponding magnetometer position in $C_p$.
%
%In other words, if we consider Pencil's coordinate system, $C_p$, where the Pencil tip's coordinates are always $[0, 0, 0]$, as the Pencil moves on the screen, the magnetometer moves and rotates in $C_p$.
%
Given that we know the location of the magnetometer, $M_{loc}$, in $C_s$, if the Pencil tip is at location, $T_{loc}$, the 3D position of the magnetometer in $C_p$, $M_{loc}^\prime$ can be represented by 
\begin{equation}
M_{loc}^\prime = P_{s}*(M_{loc}-T_{loc})
\label{eq:mloc}
\end{equation}
where $P_s$ is the tuple $(X_{ps}, Y_{ps}, Z_{ps})$ representing Pencil's axes in screen space.

%\begin{figure}[h!]
%	\centering
%	\includegraphics[width=1.3in]{figures/designDetails/camera_setup.pdf}
%	\caption{Setup for tracking the Pencil orientation.}
%	\label{fig:camera_setup}
%\end{figure}

Similarly, we can also transform the magnetic readings in $C_s$, $m$ to magnetic readings in $C_p$, $m^\prime$ by
\begin{equation}
m^\prime = P_{s}*m
\label{eq:mr}
\end{equation}

%The $M_{loc}^\prime$ can be the same for different location and orientations of the Pencil.
%This means that for any two points on the screen which are at the same distance from $M_{loc}$, their corresponding $M_{loc}^\prime$ will be the same in Pencil's coordinate space. 
%
%two Pencil positions which are different in $C_s$, however from Pencil's perspective, the magnetometer will be at the same location in $C_p$ for these two positions.

Building a map around the Pencil reduces the search space for war-driving
%, since now we only need to cover all the possible 3D locations of the magnetometer in $C_p$.
%
since $M_{loc}^\prime$ can be the same for more than one Pencil position in $C_s$.
This means that if we express equation~\ref{eq:mloc} as
\begin{equation}
T_{loc} = M_{loc} - P_{s}^T*M_{loc}^\prime
\label{eq:mloc_inverse}
\end{equation}
we can show that there are multiple pairs of $T_{loc}$  and $P_{s}$ which would correspond to the same $M_{loc}^\prime$ in $C_p$.
Figure~\ref{fig:pencil_states} shows an example of this case.
Even though the $T_{loc}$ and $P_s$ are different for the two Pencils, $M_{loc}^\prime$ is the same for these two Pencils in $C_p$.
%
%Hence, based on the transformation in equation~\ref{eq:mloc}, there must be two different $P_s$ such that the magnetic reading recorded by the magnetometer for these two Pencil positions is the same in $C_p$. 
%
Due to this redundancy, if we cover one of these Pencil positions while war-driving, we also determine the magnetic impact for a large number of other positions automatically.
This greatly reduces the number of positions we need to cover through war-driving.
Additionally, instead of collecting only one sample for each Pencil location and orientation, we can collect multiple samples for a position in $C_p$ when we move the Pencil at different locations on the screen. 

Considering the benefits mentioned above, we build the magnetic map around the Pencil, which can now be expressed as
\begin{equation}
m^\prime = f(M_{loc}^\prime)
\end{equation}
For this purpose, we use our setup shown in Figure~\ref{fig:camera_setup} to track the orientation of the Pencil while randomly drawing on the screen for a few hours.
We record the Pencil movement with a camera placed at a distance above the iPad and looking down upon the screen.
In this random drawing, we rotate and move the Pencil such that we can cover the entire screen and maximum possible orientations.
%
%This random drawing approach is faster than using a grid-based approach and lets us collect multiple sample points for a given Pencil position.
%
Similar to the procedure for building the 2D magnetic map, we collect the magnetic data in the background at $50$Hz and touch data, $T_{loc}$, from iOS API at $120$Hz.
The touch API only provides information about the azimuth and altitude angles~\cite{celestial} of the Pencil.
However, the Pencil rotation around its $Z$-axis also affects the magnetic readings sensed by the magnetometer.
Hence, the Pencil orientation, $P_s$, cannot be obtained without any ambiguity directly from the touch API.
Therefore, we employ a computer vision approach to track the orientation.

\begin{figure*}[h!]
	\centering
	\begin{subfigure}{0.32\linewidth}
		\centering
		\includegraphics[width=2in]{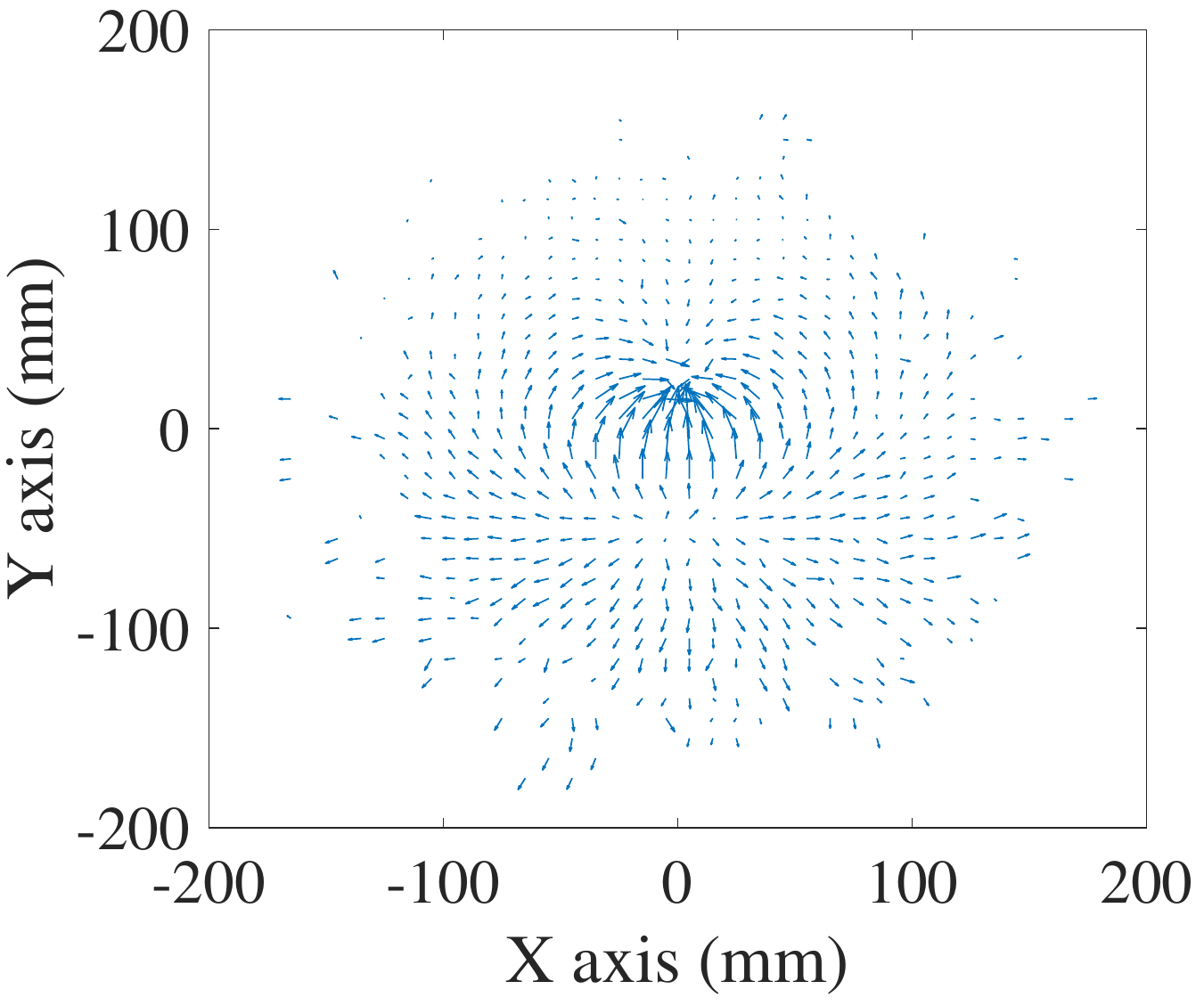}
		\label{fig:pencilMap_before_minus5}
	\end{subfigure}
	\begin{subfigure}{0.32\linewidth}
		\centering
		\includegraphics[width=2in]{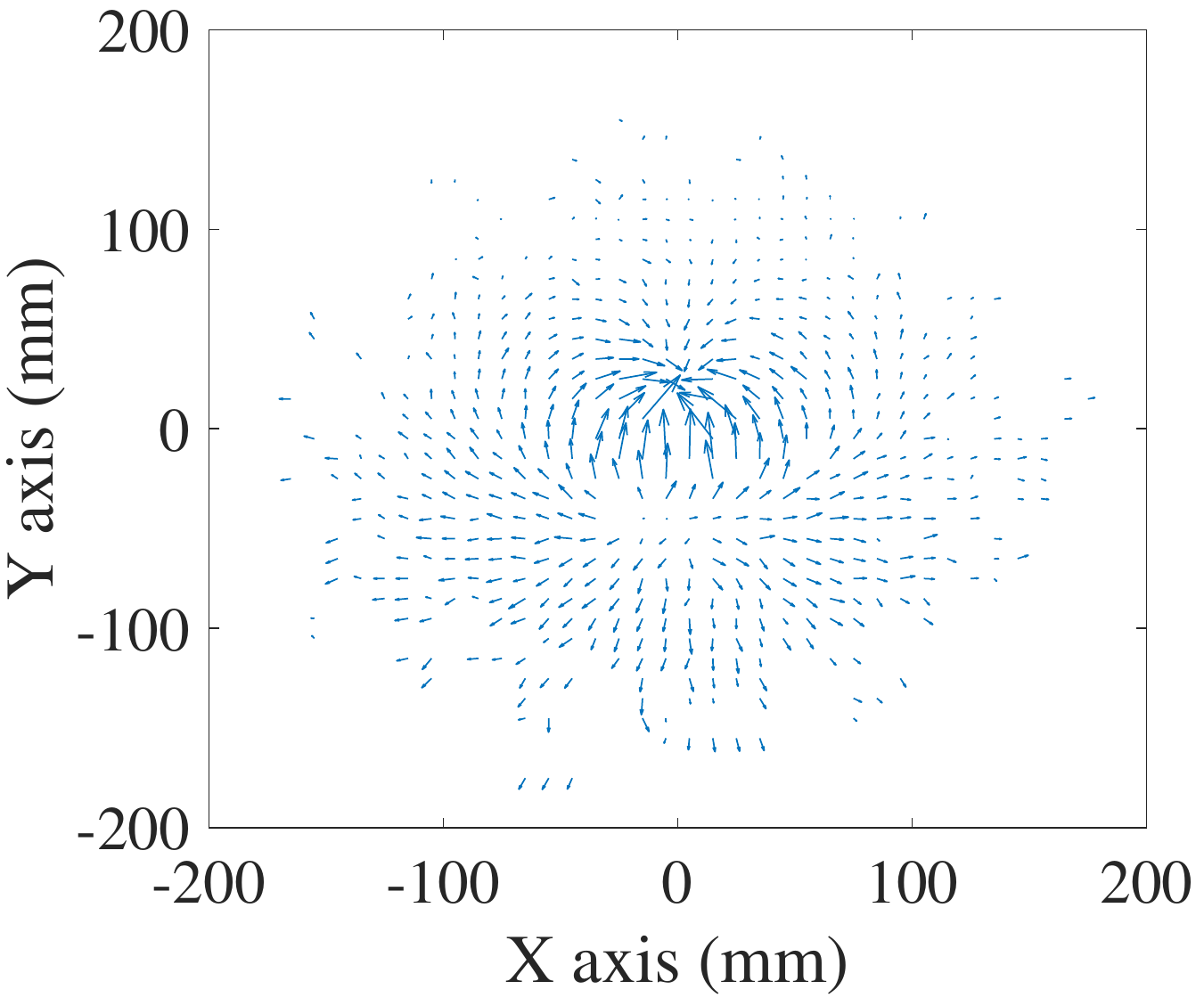}
		\label{fig:pencilMap_before_0}
	\end{subfigure}
	\begin{subfigure}{0.32\linewidth}
		\centering
		\includegraphics[width=2in]{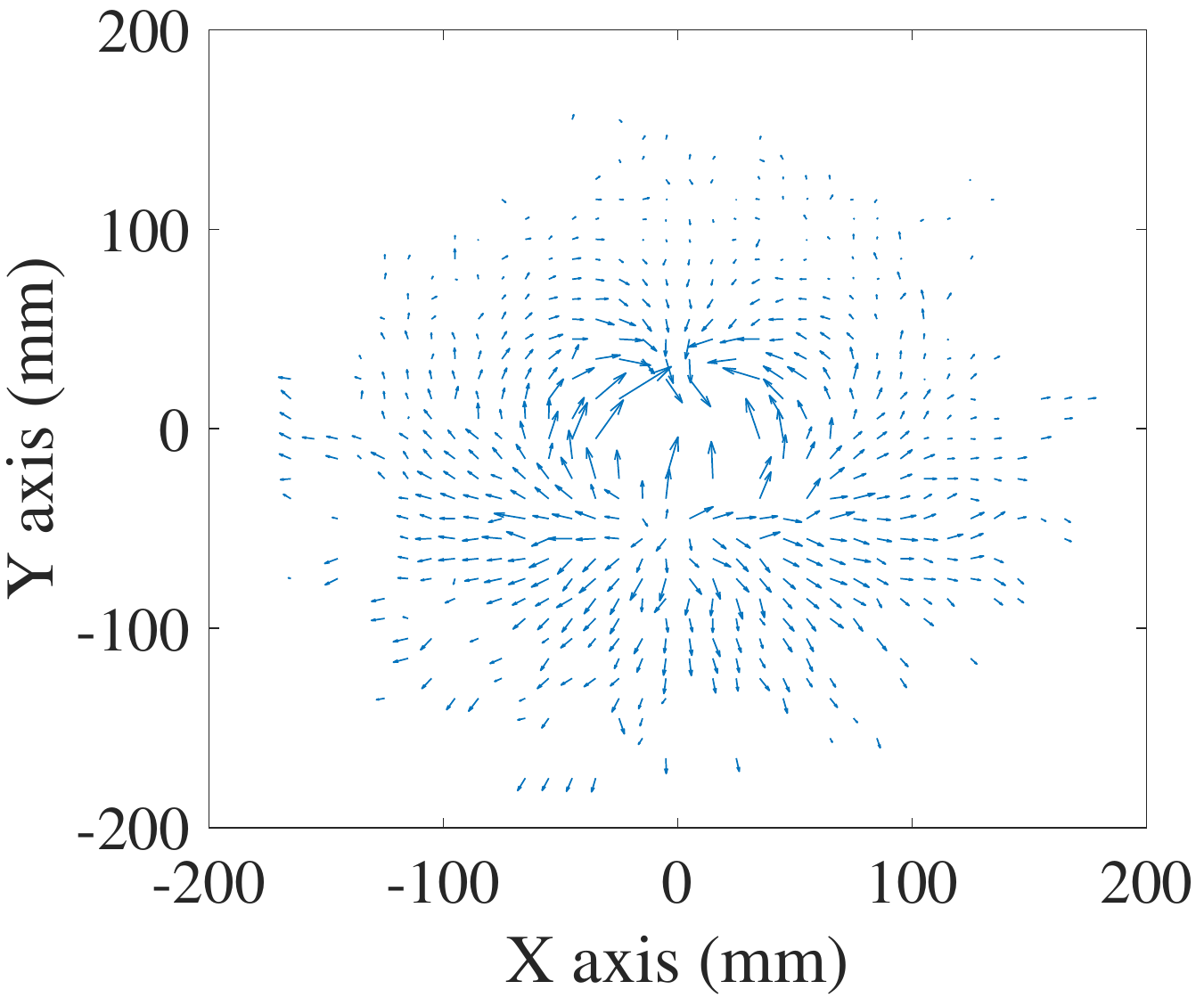}
		\label{fig:pencilMap_before_5}
	\end{subfigure}
	\begin{subfigure}{0.32\linewidth}
		\centering	\includegraphics[width=2in]{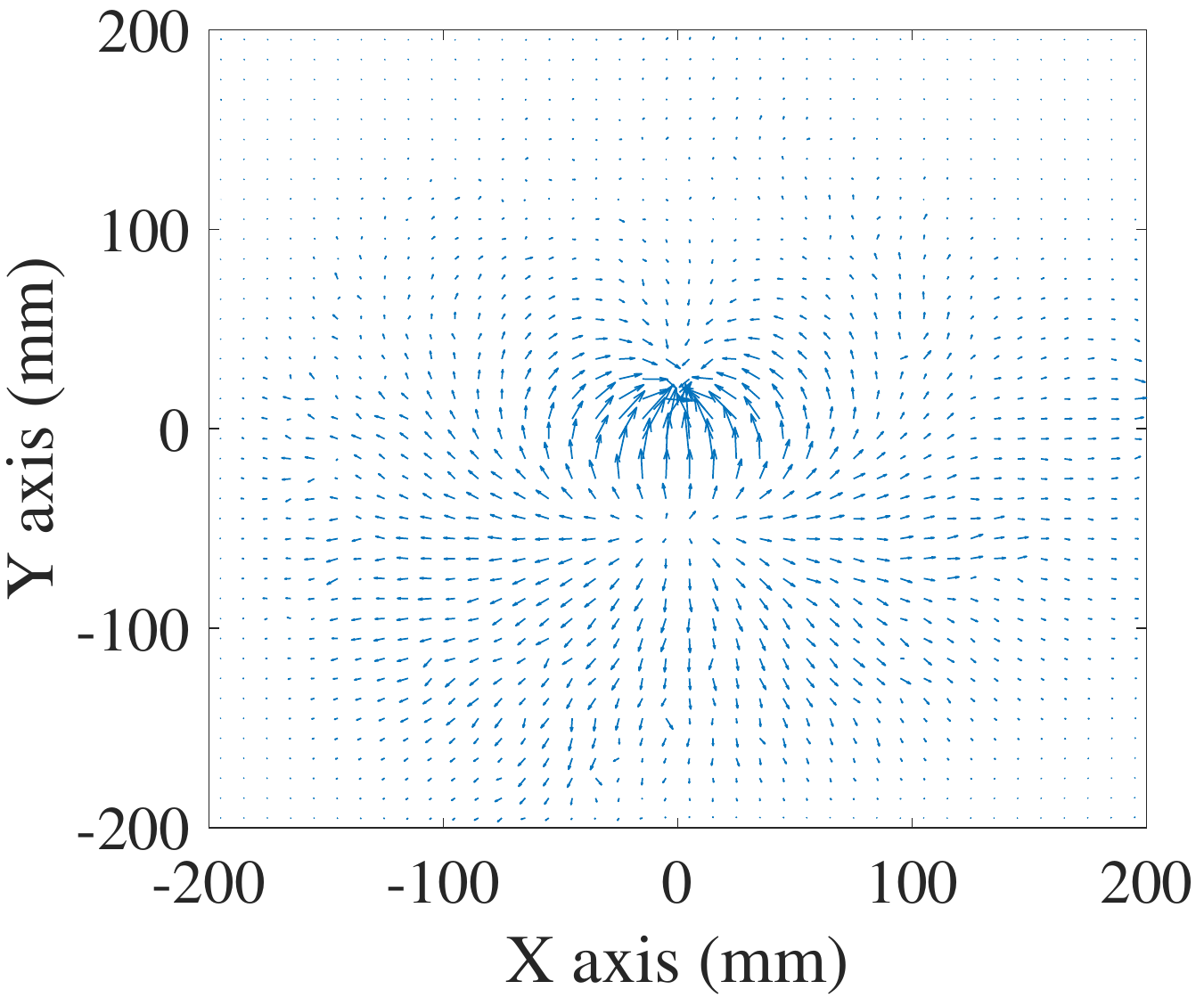}
		\label{fig:pencilMap_after_minus5}
	\end{subfigure}
	\begin{subfigure}{0.32\linewidth}
		\centering	\includegraphics[width=2in]{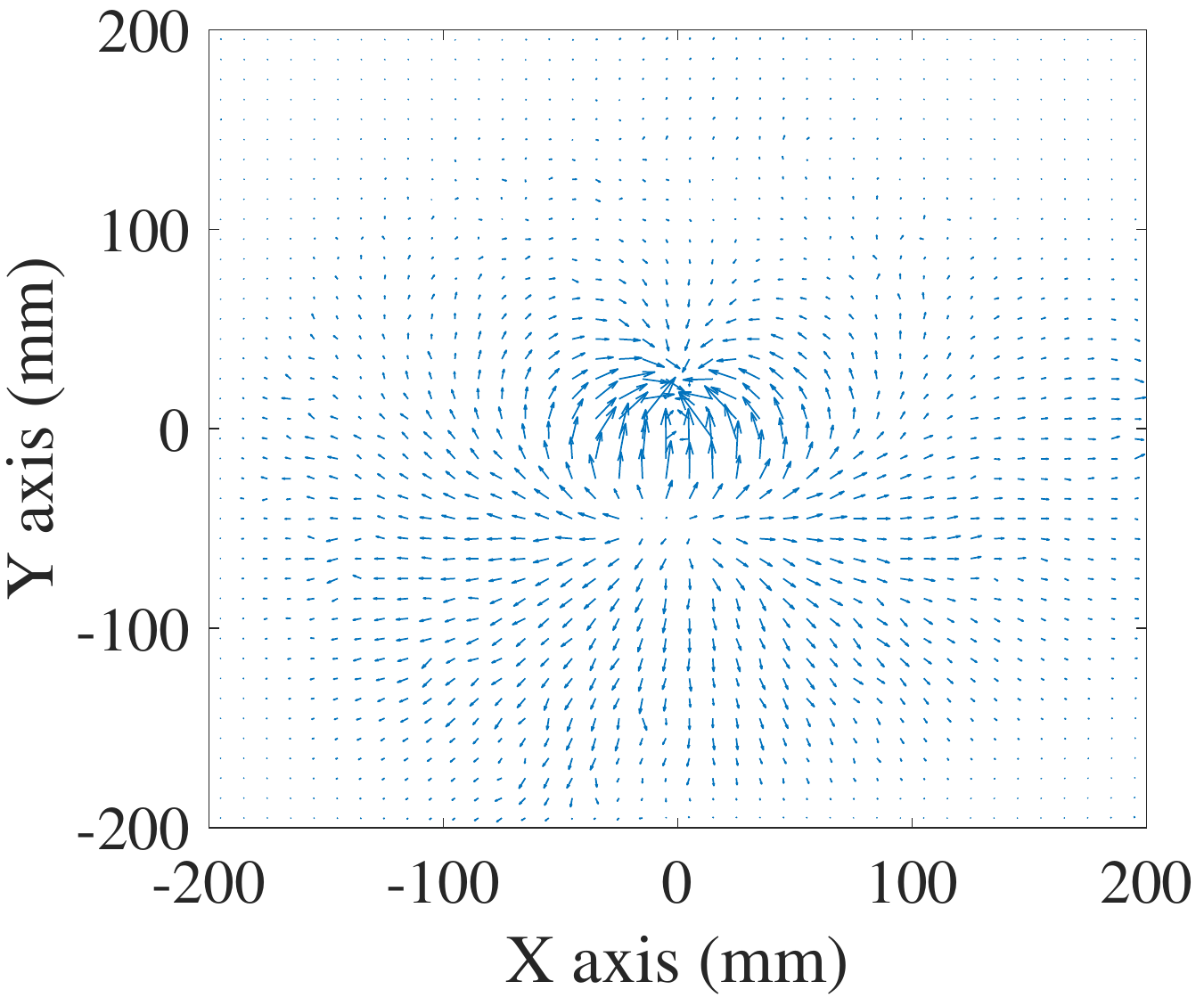}
		\label{fig:pencilMap_after_0}
	\end{subfigure}
	\begin{subfigure}{0.32\linewidth}
		\centering	\includegraphics[width=2in]{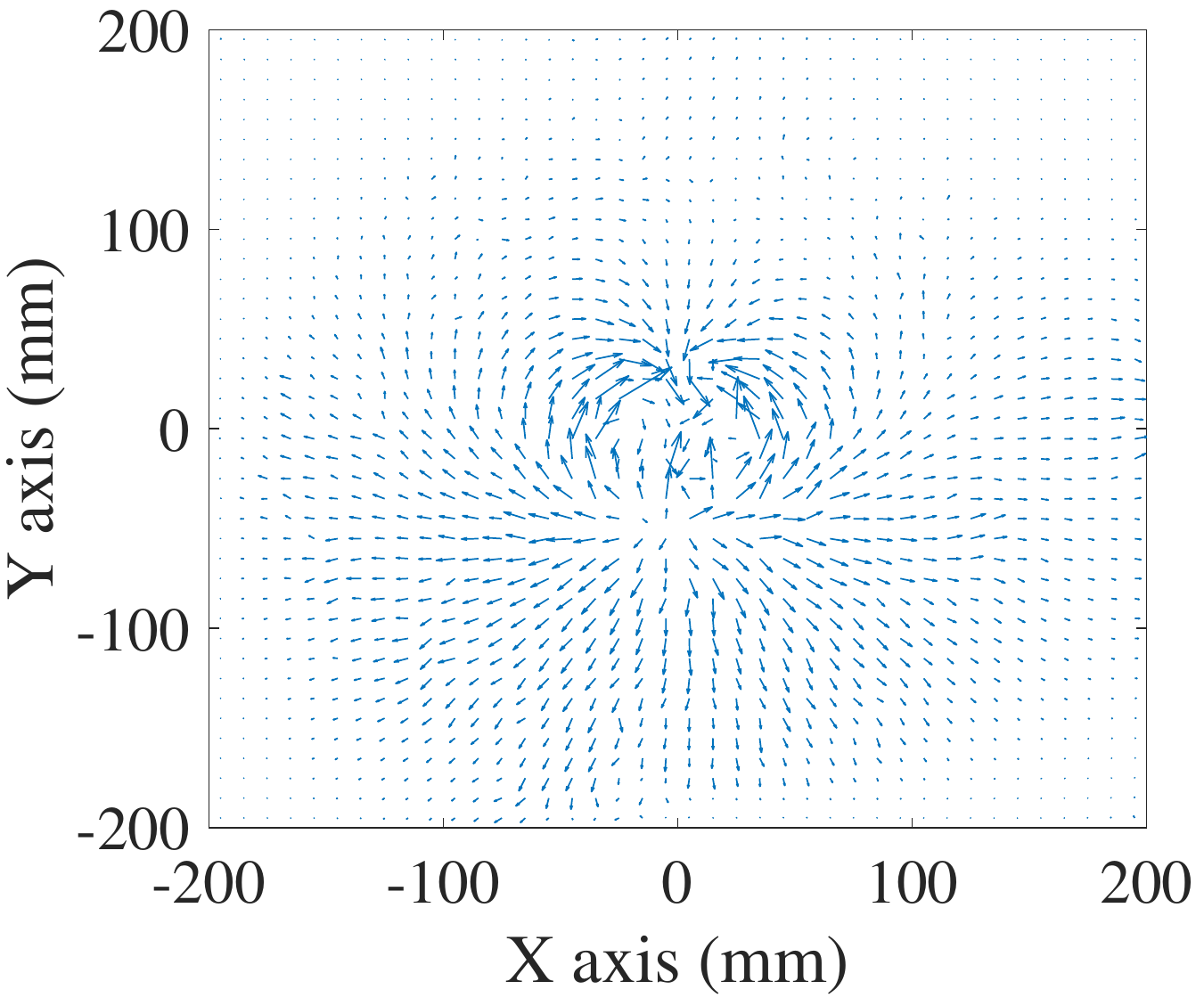}
		\label{fig:pencilMap_after_5}
	\end{subfigure}
	\caption{Pencil magnetic map for $z =-5$mm (left) $z=0$mm (center) and $z=5$mm (right), before (top) and after (bottom) optimization.}
	\label{fig:pencilMap}
\end{figure*}

The camera acts as a bridge between the screen and the Pencil.
By finding the camera's orientation in $C_s$ and then finding the Pencil orientation in the camera's coordinate system, we can finally find the orientation of the Pencil in $C_s$.
To find the orientation of the camera in $C_s$, we use a standard camera calibration technique~\cite{zhang2000flexible}.
%
%The iPad is moved to capture several images with different distances and orientations of this pattern with respect to the fixed camera.
%
%These images are given as input to camera calibration algorithm to find the extrinsic and intrinsic parameters of the camera.
%
%These parameters are then used to find the location and orientation of the camera in $C_s$.
%For pose estimation to work, the camera needs to be calibrated first.
%%
%We calibrate the camera by showing an image of a checkerboard pattern (similar to the pattern shown in Figure~\ref{fig:coordinate_systems}) on the iPad screen.
%TODO find reference
Once the camera is calibrated, we use PnP algorithm~\cite{lu2018review} to find the orientation of the checkerboard in the camera's coordinate system.
%
%we attach a checkerboard to the Pencil end to estimate its pose and in turn the Pencil's 3D orientation.
%
%Figure~\ref{fig:coordinate_systems} also shows the 3D coordinate system of the checkerboard, $C_b$.
%
%We define the center point of the checkerboard as the origin of $C_b$ and measure the coordinates of the square corners in $C_b$ to approximate the checkerboard's 3D model.
%
%Checkerboard detection is applied to the images recorded by the camera as the Pencil moves on the screen to find the corresponding points in the images.
%
%TODO find reference
%Given the 2D points from the image and the 3D points defined in checkerboard coordinate system, we use PnP algorithm to %estimate the 3D coordinates of the camera center and its 3D rotation in $C_b$ space.
%
%This camera center and orientation is then used to transform the checkerboard axes and its origin (center) into the screen space, $C_s$.

%
Due to the noise in camera calibration and pose estimation steps, the orientation obtained is noisy.
To remove this noise, we apply Kalman filtering~\cite{zarchan2000fundamentals} and smoothing to the orientation obtained over time.  
%convert the 3D orientation of the checkerboard into quaternions and 
% 
Since the checkerboard's origin is attached to the Pencil end, the 3D axes of the checkerboard in $C_s$ obtained after smoothing are indeed the Pencil axes in $C_s$ as well.
Therefore, by tracking the checkerboard axes, we can track the orientation of the Pencil as it moves on the screen.

%Once we finish, 
%
%The magnetic impact of ambience is removed from the magnetic readings.
%
We now have the touch data from the iOS API, the orientation data from the camera, and the magnetic data from the magnetometer.
For each touch sample, we transform the $M_{loc}$ and sample's corresponding $m$ to $C_p$ using equations~\ref{eq:mloc} and \ref{eq:mr} respectively.
These transformed magnetic readings in $C_p$ are finally quantized into $5 mm$ by $5 mm$ grids based on the 3D location of the $M_{loc}$ corresponding to the reading.
The average of all the magnetic data samples is computed for each grid.
Figure~\ref{fig:pencilMap} (top) shows the $XY$ plane of this 3D map built around the Pencil for $z = -5mm$, $z = 0mm$ and $z = 5mm$ in $C_p$. 

Here, we emphasize that we remove the magnetic impact of the ambient environment from the collected magnetic data before generating this map.
As a result, the same magnetic map can be used to track the Pencil movement while the victim is using their iPad in different locations. 
Hence, this process is a \textbf{one-time effort} and does not involve any input from the victim.
\majorRev{[C6, C11]}\majorRevText{We also note that the setup shown in Figure~\ref{fig:camera_setup} for Pencil's orientation tracking is only used for offline map generation and is not needed at attack time to infer victim's handwriting.}

\textbf{Map Optimization:} As shown in Figure~\ref{fig:pencilMap} (top), some of the grids in the 3D map have missing or noisy data.
%do not have any samples from the data collected for map generation.
%
%To be able to use the 3D map for tracking the Pencil movement for free-form handwriting, we need the map to be complete i.e. we need to account for the missing data.
%
Instead of using a simple interpolation approach, we adopted a vector field reconstruction technique that considers the divergence and curl of the vector field \cite{vectorfield2012}.
%. 
It operates by minimizing the energy function $J$:
\begin{equation}
J(g;f) = \sum_{m}|Ag(m)-f[m]|^{2} + \lambda_{c}||curl (g)||_1 + \lambda_d||div (g)||_1
\end{equation} 
where $g$ is the reconstructed map and $f$ is the original magnetic map.
$A$ is an operator that only selects grids containing at least one datum. 
$\lambda_c$ and $\lambda_d$ are regularization parameters.
%
%In order to be able to use the 3D map for tracking, we need to account for the missing magnetic data and deal with the noise in our readings.
%
%One may argue that why not simply use interpolation to get the missing data.
%%
%Although we have a large number of samples in the grids corresponding to the more common orientations of the Pencil, the magnetic readings are impacted by the sensor noise.
%%
%If this noisy data is directly used for interpolation, the resulting 3D map will not be accurate.
%%
%In addition, unlike 2D magnetic map, a linear interpolation method cannot be used to determine the magnetic field of the Pencil.
%%
%Consequently, we need a mechanism to reduce the noise in the collected magnetic data and generate the missing data for the map while considering the properties of magnetic fields in mind.
%

%Keeping our requirements for the 3D map in mind, we found a reconstruction algorithm for the magnetic field. 
%
%This algorithm consider that the fact that the divergence of the magnetic field should be zero.
%

%
According to Maxwell's equations bounding the magnetic field~\cite{maxwell1865viii}, the regularization parameter $\lambda_{c}$ should be small while $\lambda_d$ should be large.
\majorRev{[MR6, C12]}\majorRevText{Considering the noise in the magnetic readings, we perform a grid search to determine the optimal values of regularization parameters, guided by the optimal values given in \cite{tafti2011vector} for different signal to noise ratios.}
%
%we choose $\lambda_c= 0.3$ and  $\lambda_d = 1.5$ to reconstruct the magnetic map based on the best parameter selection procedure described in \cite{tafti2011vector}. 
%
%These regularization parameters ensure that the divergence and curl of the magnetic field are not impacted by the noise in the data and are minimized.
%
The resulting 3D magnetic map from this approach is continuous with reduced noise.
Figure~\ref{fig:pencilMap} (bottom) shows the $XY$ plane of this 3D map for $z = -5mm$, $z = 0mm$ and $z = 5mm$ after reconstruction.

\subsection{Pencil Tracking}
\label{sec:pencil_tracking}
With the reconstructed 3D map, we are now prepared to describe our system for tracking the Pencil's location and orientation as a user writes on the iPad screen.
Similar to the 2D case, we use particle filter to track the Pencil movement.
However, now the state vector for the particle filter also has to take into account the Pencil's orientation.

Earlier, we defined the Pencil orientation as its axes, $(X_{ps}, Y_{ps}, Z_{ps})$ , in $C_s$.
%
%These 3D axes basically describe the rotation of the Pencil. 
%
%The problem in representing the orientation as this 3D matrix is that rotations in 3D do not commute.
%
This definition is intuitive and works well for magnetic map generation but using this definition in particle filter is complicated.
%But for tracking, using this definition will require many parameters.
%This means that even though this way we can easily describe one orientation of the Pencil, describing the change in orientation over time in terms of 3D matrices is complicated and not continuous always.
%
Another common way to represent the orientation of a rigid body is to use the Euler angles, which represent the 3D orientation of an object using a combination of three rotations about different axes~\cite{diebel2006representing}. %TODO find reference
Although intuitive, Euler angles suffer from singularities and give erroneous orientation representation when used to track an object's successive rotations.
% description can sometimes lead to gimbal lock such that one of the rotation references is canceled. %TODO find reference
%
Fortunately, there is another formal representation for 3D rotation, which is the quaternions~\cite{diebel2006representing}, which allows easier interpolation between different orientations.
Quaternions encode any rotation in a 3D coordinate system as a four-element vector, where the squared sum of the four elements is equal to $1$.
%that is more efficient to compute and allows smooth tracking of sequential rotations.
%
Using quaternions, we can describe how the orientation changes sequentially more efficiently and smoothly ~\cite{mukundan2002}.
Hence, for using the quaternion representation to describe the Pencil's orientation, our particle filter has to consider $6$ parameters ($\langle x, y \rangle$ location of the Pencil tip and the Pencil orientation as 4-element quaternion vector) to track the Pencil movement, instead of $5$. 

To consider the orientation, we define the state vector for each particle at time $t$ as:

\begin{equation}
s_t = \ <x_t, y_t, q_{1t}, q_{2t}, q_{3t}, q_{4t}>
\end{equation}
Here, $ q_{1t}, q_{2t}, q_{3t}$ and  $q_{4t}$ represent the four scalar values for the quaternion representation of the Pencil's orientation at time $t$.
%
%The squared sum of the four quaternions is equal to $1$.

So far, we have considered that the Pencil can move in any orientation. %on the screen, which means that the state space for orientation is anywhere in 3D space.
However, in practice, the movement of the Pencil is limited by the range of motion of the human hand and wrist.
Therefore, based on our observations of the human handwriting behavior, we limit the range for the Pencil's possible orientation.
We assume that the altitude range of the Pencil is $30^\circ$ to $90^\circ$. %while it cannot be less than $45^\circ$.
Similarly, for the azimuth angle, we specify the range to be within $60^\circ$ to $170^\circ$.
We do not limit the rotation of the Pencil around its $Z$ axis.
\majorRev{[MR6, C12]}\majorRevText{These ranges are determined by analyzing the minimum and maximum values of these angles observed in the data collected by the attacker and their accomplices for training the writing behavior model (described below).}
We specify the ranges for orientation in terms of the altitude and azimuth for ease of understanding, but these ranges are actually checked in the quaternion representation.
%
%Since eventually we are going to track the orientation in terms of quaternions, we have to determine the possible axes for the Pencil in $C_s$ which satisfy the conditions on altitude and azimuth angles

%We define the the model for the Pencil's movement for free-form handwriting as:
%\begin{equation}
%s_{t+1} = s_{t} + v_t + b_t
%\end{equation}
%where $v_t$ is the velocity of the particles and $b_t$ is the random perturbation added to the state.

%Before we define the model for the Pencil's movement for free-form handwriting, we want to highlight an opportunity here.
%
\begin{algorithm}[t]
	\caption{Pencil Tracking}\label{alg:seqParticleFilter}
	%	\KwData{magnetic readings from the magnetometer, $M$, and 3D magnetic map, $f$, initial number of particles, $N$, maximum number of valid particles, $K$} 
	%	\KwResult{estimated Pencil position}
	\begin{algorithmic}[1]
		\STATE Initialize particles $S_0^i$ where $i=1, 2, ...N$ using orientation constraints
		\STATE Compute the weights %$w_0^i(X_0^i)$ and $W_0^i \propto w_0^i(X_0^i)$  %W_0^i = ComputeWeight(X_0^i)$
		\STATE Select top $K$ particles with the highest weights
		
		\FOR {t $\geq 1$}
		\STATE Sample $S_t^i$ using $S^i_{max(t-3, 1):t-1}$ based on writing behavior model%\sim s_t(x_n, \overline{X^i_{0:t-1})}$
		\STATE Update particle history $S_{0:t}^i \leftarrow (\overline{S^i_{0:t-1}}, S_t^i)$
		\STATE Compute the weights %$\alpha_t(X_{0:t}^i)$ and $W_t^i \propto \alpha_t(X_{0:t}^i)$
		\STATE Resample $S_t^i$ to obtain $K_a$ updated particles \\
		\ENDFOR
		\STATE Track Pencil using $S_{0:t}^i$
		
		\STATE \textbf{function ComputeWeight(s)}
		\STATE  \ \ \ \ $P_s \leftarrow s$
		\STATE  \ \ \ \ $M_{loc}^\prime \leftarrow P_s*(M_{loc}-T_{loc})$
		\STATE  \ \ \ \ Query Pencil magnetic map using $M_{loc}^\prime$ to obtain $m^\prime$
		\STATE  \ \ \ \ $m \leftarrow P_s^T*m^\prime$
		\STATE  \ \ \ \ $w \leftarrow \exp(-\frac{(m - r)^2}{2\sigma^2})$
		\STATE  \ \ \ \ return $w$
		\STATE  \textbf{end function}
		
		%		\STATE \textbf{function Resample(W, X)}
		%		\STATE  \ \ \ \ $q \leftarrow f(s) + e$
		%		\STATE  \textbf{end function}
	\end{algorithmic}
\end{algorithm}
Algorithm~\ref{alg:seqParticleFilter} describes the algorithm for Pencil tracking. % the free-form Pencil movement.

\textbf{Particles initialization:} We begin by initializing the particles $S_0^i$ where $i=1, 2, ...N$.
Since $6$ parameters now determine the state, we need a huge number of particles to cover the entire state space.
%
%Habiba: not sure about this
\majorRev{[MR6, C12]}\majorRevText{We employ a grid search approach on data collected from a small set of users (\eg data collected for training writing behavior model which is not included in our evaluation set) to determine the optimal number of particles initially.
To detail, we generate candidates from a grid of parameter values specified from $10000$ to $10000000$ with increasing the number of particles by $10000$ in each iteration. 
We evaluate each of these particle values on the users' data and use the parameter that leads to highest accuracy at the elbow point~\cite{bishop2006pattern} to prevent overfitting.
Based on our search results, we set the number of particles, $N$, to $5000000$.}
The $x$ and $y$ location for these particles are uniformly drawn from within the range of the iPad screen.
Altitude ($\theta_1$) , azimuth ($\theta_2$) and rotation around $Z$ axis ($\theta_3$) are drawn uniformly from the range $30^\circ-90^\circ$, $60^\circ-170^\circ$  and $0^\circ-360^\circ$ respectively.
%
%We compute the axes representation 
%For the orientation, we first generate random numbers equal to the number of particles for the altitude, azimuth, and rotation about the Pencil's $Z$ axis.
%
From these values for the three angles, we compute the axes representation for the Pencil in $C_s$.
These axes are converted into quaternions and are included in the state vector.
%
%From these randomly initialized particles, we choose those that satisfy our constraints for the two angles determining the orientation described above.
%
%Having a large $N$ initially helps us ensure that we have enough particles that meet our orientation conditions.
%
Once we have found $S_0^i$ meeting our criteria, we compute their weights using the $ComputeWeights$ function.
%the sake of 
\majorRev{[MR6, C12]}\majorRevText{For efficient computation, we specify a bound on the maximum number of particles at each timestamp as $K = 50000$.
This bound is also found through the same grid search approach described above.}
Therefore, from $S_0$, we choose $K$ particles which have the highest weights.

\begin{figure}[t!]
	\centering
	\begin{subfigure}{0.45\linewidth}
		\centering
		\includegraphics[width=1.8in]{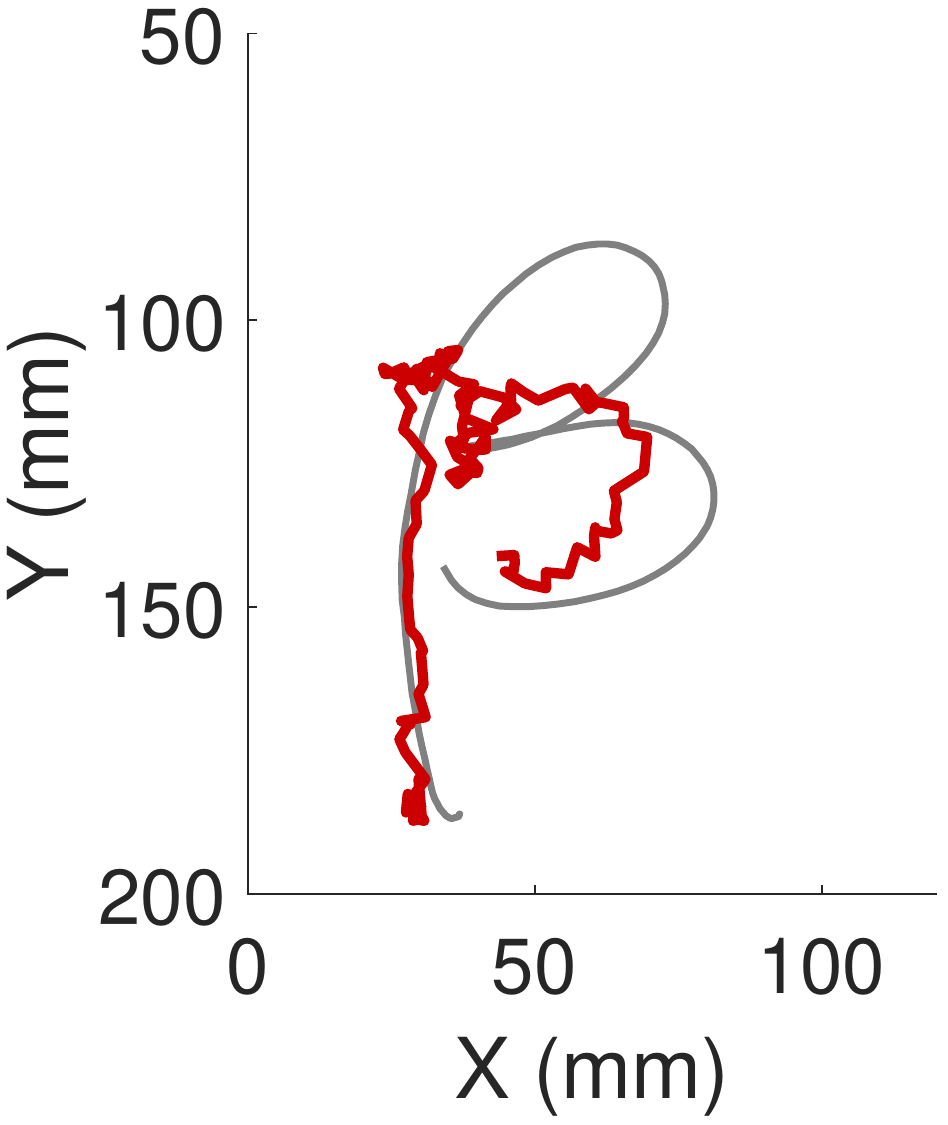}
		\caption{}
		\label{fig:particleFiler_withoutTrend1}
	\end{subfigure}
	\begin{subfigure}{0.45\linewidth}
		\centering
		\includegraphics[width=1.8in]{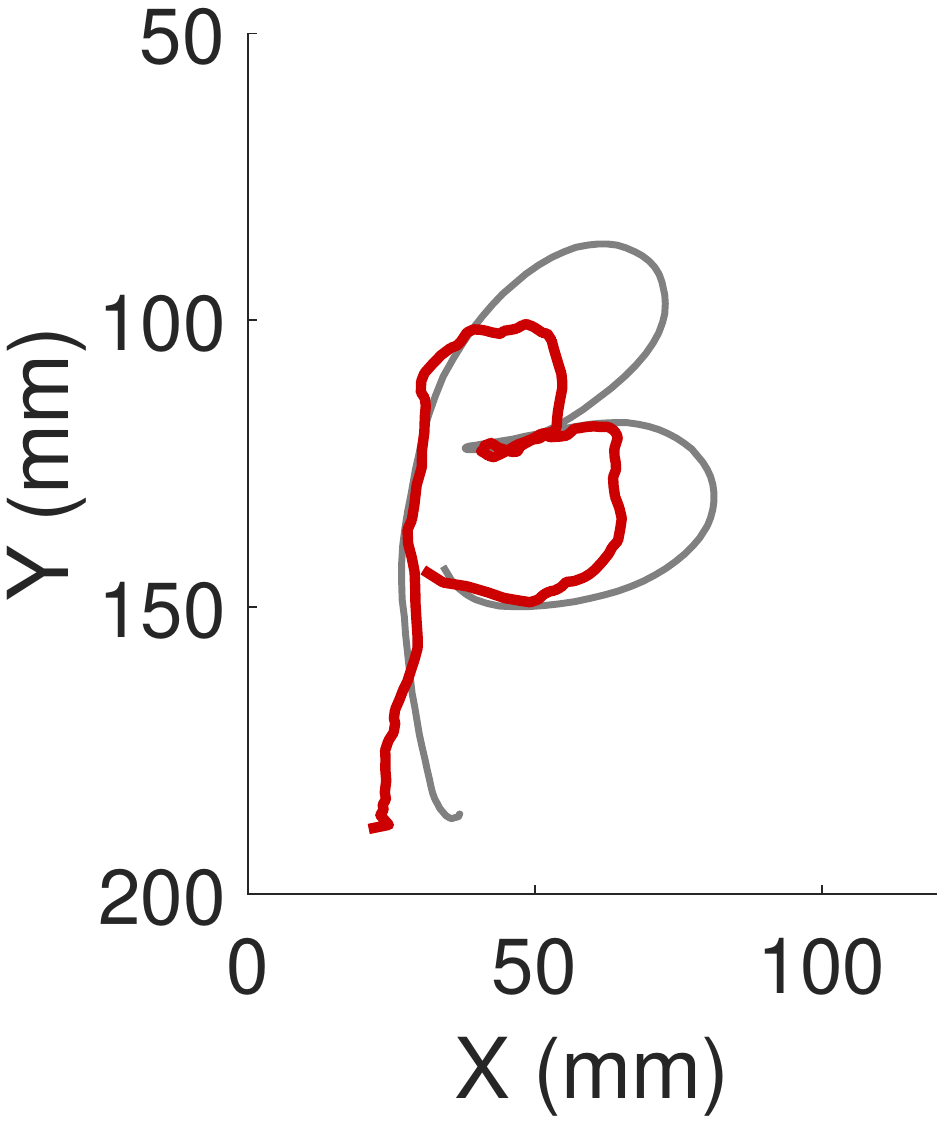}
		\caption{}
		\label{fig:particleFiler_withTrend1}
	\end{subfigure}	
	\caption{Tracking results without (a) and with (b) writing behavior model.}
	\label{fig:particleFiler_trend}
\end{figure}

\textbf{Movement model:} We define the model for Pencil's movement for natural handwriting as:
\begin{equation}
s_{t+1} = s_{t} + v_t + b_t
\end{equation}
where $v_t$ is the change in particles' states, and $b_t$ is the random perturbation added to the state.
Previously, for the fixed orientation case in Section~\ref{sec:2dtracking}, we used $v_t = 0$ since the state space was small.
For this case, we build a \textbf{writing behavior model}, which uses the changes in state from the last three timestamps to predict the state change for current timestamps using linear regression.
\majorRev{[C6, C11]}\majorRevText{To train this model, we write different letters, numbers, and shapes at different locations on the screen.
We use a camera to track the Pencil's orientation and iOS touch API to track location, as described in Section~\ref{sec:map_generation}.
This camera setup is only used while collecting training data for this model and is not used at attack time.
Data collected from a small number of users ($3$ in case of our evaluation) is sufficient for training this model.}
We capture the relationship between location and orientation change in human handwriting from this location and orientation data.
%
%We realize that prior knowledge of the Pencil's position can help estimate the Pencil's current position.
%
%In other words, keeping track of the Pencil's movement over time and using the movement history to predict the future is more efficient than completely relying on random motion of the particles.
%
%In the case of 2D particle filter, we did not use history information of the particles in tracking since the state space was much smaller than the 5D case.
% 
%However, with the orientation and location both changing, without using the history, the particles might take a long time to converge towards the correct Pencil position.
%
%Therefore, we consider using the concept of sequential particle filtering (also known as Sequential Monte Carlo method) to track the Pencil movement in this case.
%
%Keeping this in mind, we define the model for Pencil movement as:
%
%The velocity of a particle is determined at each time step using its state vector from last two time steps.
%
Figure~\ref{fig:particleFiler_withTrend1} shows the result when this model is used in state transition.
We can observe that the tracking result is more accurate and smoother than simply adding random perturbation to the last state (Figure~\ref{fig:particleFiler_withoutTrend1}).

%Since using previous state changes in the state transition lets particles consider their previous states, we can observe that the tracking result  (Figure~\ref{fig:particleFiler_withTrend1}) is much smoother than 

\textbf{Computing particle weights: }Here we transform the quaternions in the state vector to Pencil's axes, $P_s$, in $C_s$.
We use these axes to find $M_{loc}^\prime$ using equation~\ref{eq:mloc} in Section ~\ref{sec:map_generation} to query our Pencil magnetic map. 
In this equation, $T_{loc}$ is the $x$ and $y$ location in the state vector.
The result of the query is $m^\prime$, which is the magnetic reading in $C_p$. 
This reading is converted to $C_s$ using:
\begin{equation}
m = P_s^T*m^\prime
\end{equation}
Weights are assigned with the function:
\begin{equation}
w = \exp\left(\frac{(m - r)^2}{2\sigma^2}\right)
\label{eq:6dparticle_weight}
\end{equation} 
where $r$ is the magnetometer reading corresponding to the given state.
%\begin{figure}[]
%	\centering
%	\begin{subfigure}{0.9\linewidth}
%		\centering
%		\includegraphics[width=2.8in]{figures/designDetails/particleFilerNum.pdf}
%	\end{subfigure}
%	\caption{TBD}
%	\label{fig:particleFiler_sampleSize}
%\end{figure}

%The magnetic reading measurement, $q_t$, is the same as equation~\ref{eq:2dparticle_measurement} except that now $f$ is the 3D magnetic map around the Pencil.
%%
%The quaternions and Pencil tip's location in the state vector are transformed to the Pencil's coordinate system, $C_p$, to query the map and obtain the magnetic reading measurement.

\begin{figure*}[t!]
	\centering
	\begin{subfigure}{0.118\linewidth}
		\centering
		\includegraphics[width=0.7in]{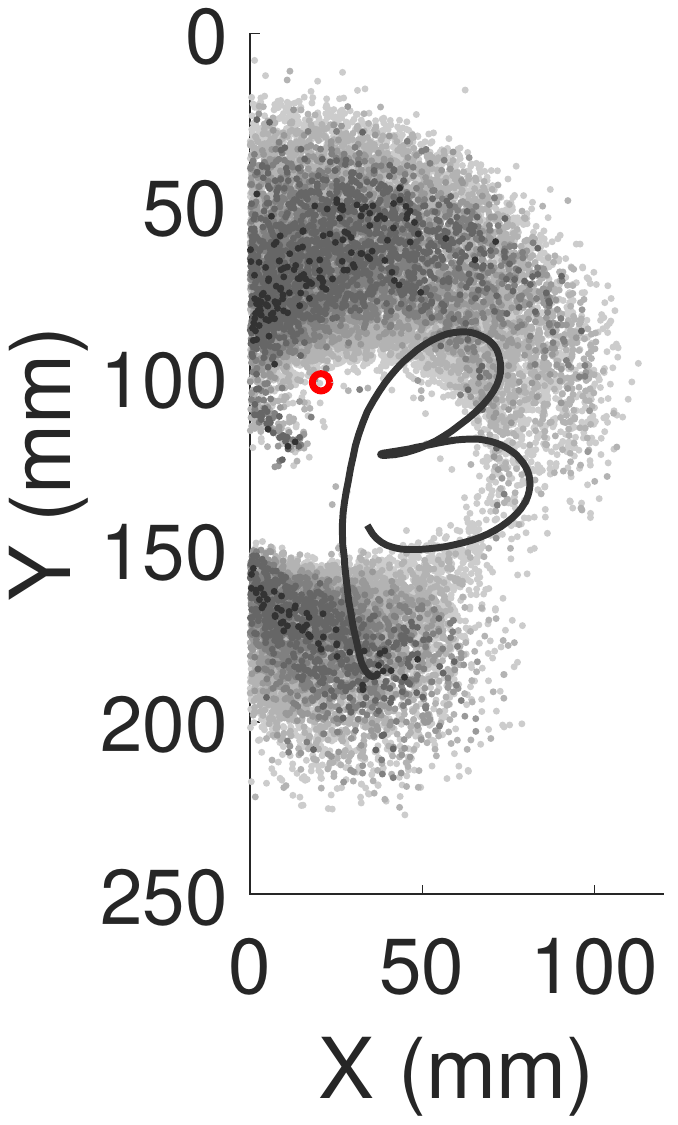}
		\caption{}
		\label{fig:particleFiler_t1}
	\end{subfigure}
	\begin{subfigure}{0.118\linewidth}
		\centering	\includegraphics[width=0.7in]{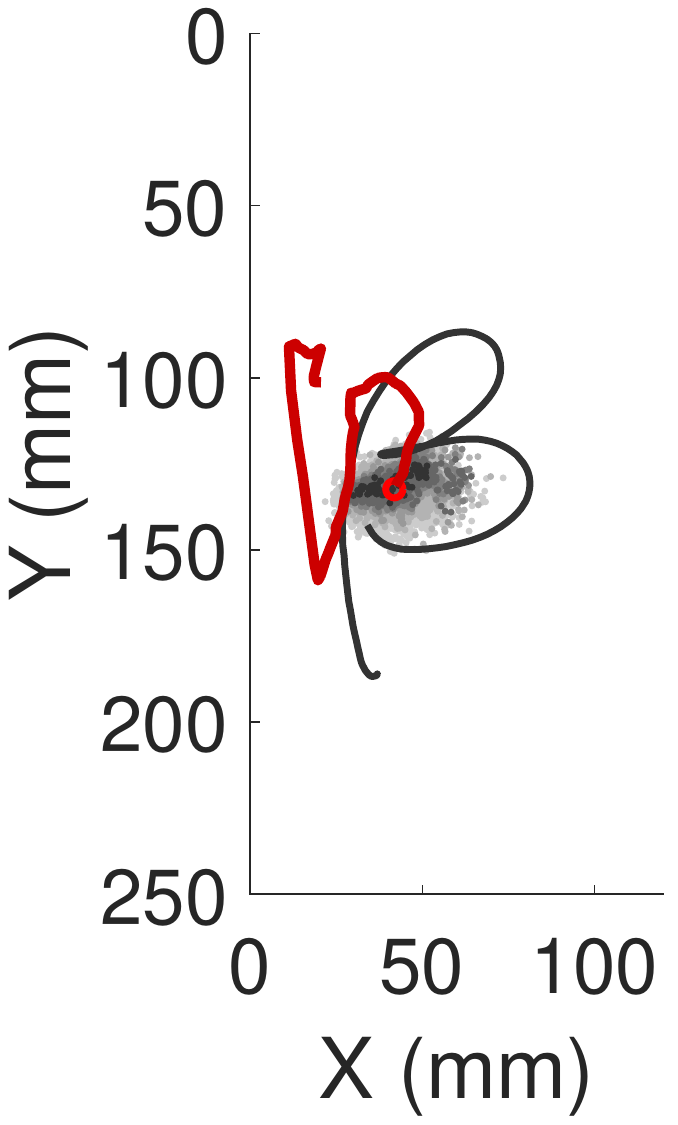}
		\caption{}
		\label{fig:particleFiler_t81}
	\end{subfigure}
	\begin{subfigure}{0.118\linewidth}
		\centering	\includegraphics[width=0.7in]{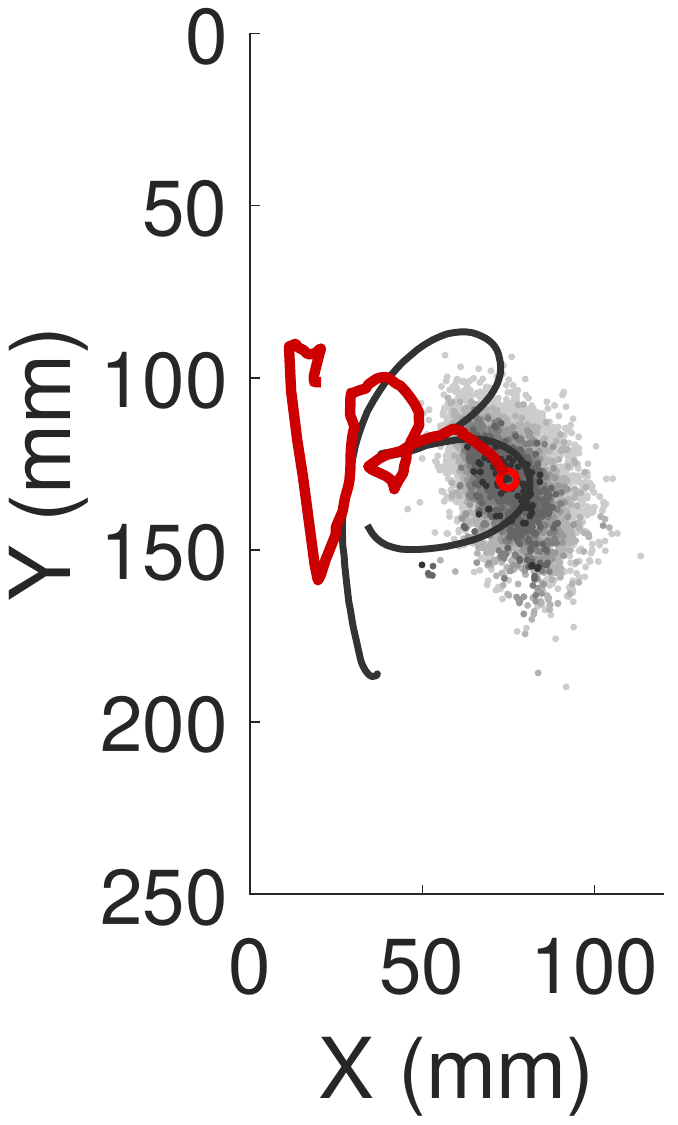}
		\caption{}
		\label{fig:particleFiler_t121}
	\end{subfigure}
	\begin{subfigure}{0.118\linewidth}
		\centering	\includegraphics[width=0.7in]{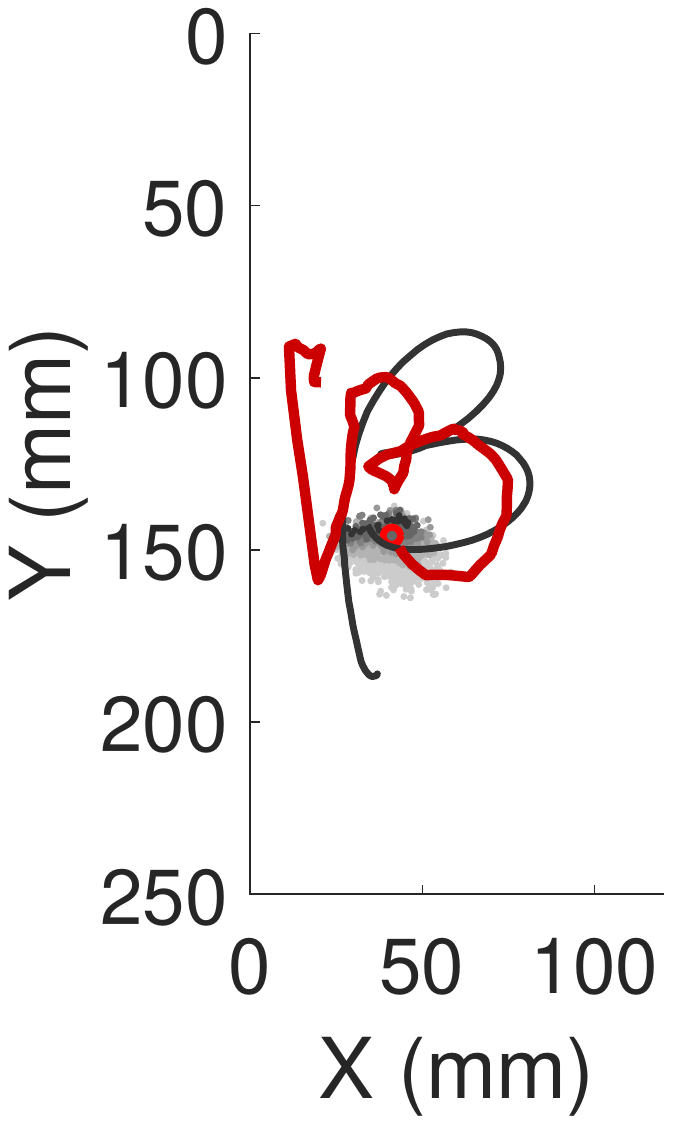}
		\caption{}
		\label{fig:particleFiler_t145}
	\end{subfigure}
	\begin{subfigure}{0.118\linewidth}
		\centering
		\includegraphics[width=0.7in]{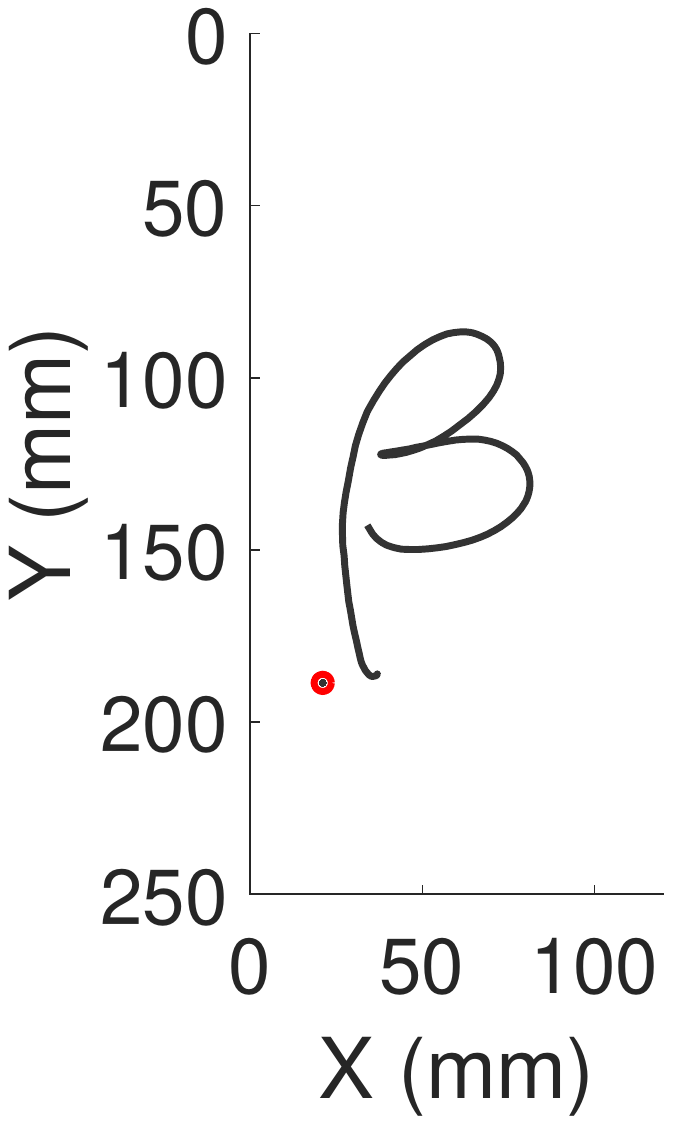}
		\caption{}
		\label{fig:particleFiler_history_t1}
	\end{subfigure}
	\begin{subfigure}{0.118\linewidth}
		\centering	\includegraphics[width=0.7in]{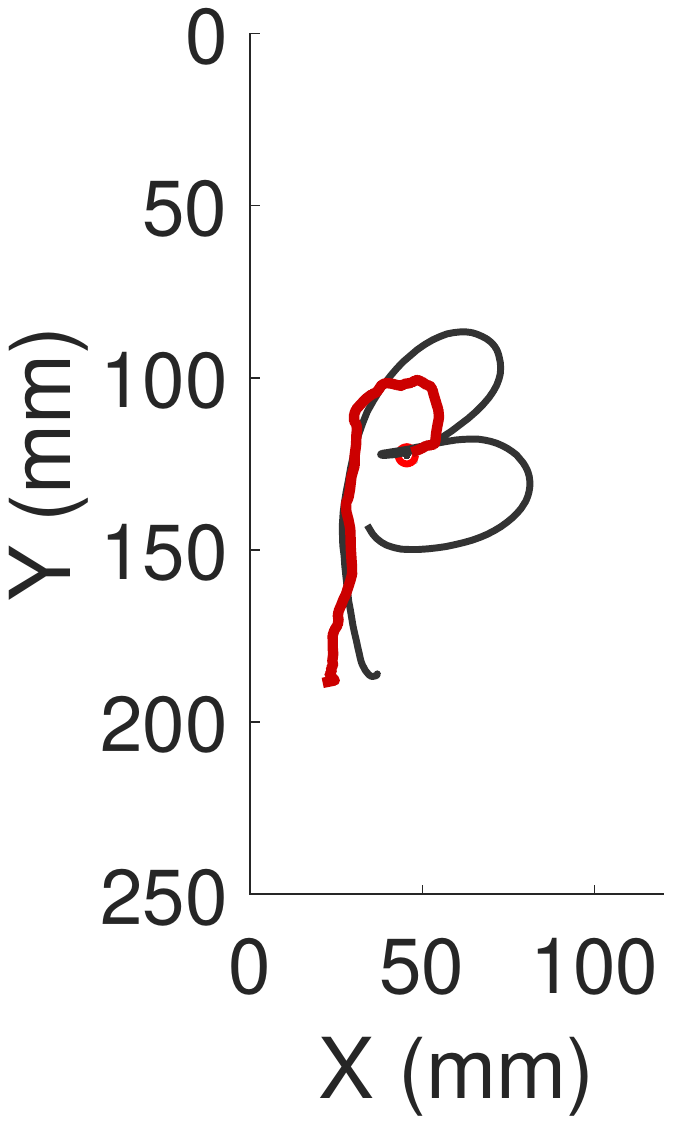}
		\caption{}		
		\label{fig:particleFiler_history_t81}
	\end{subfigure}
	\begin{subfigure}{0.118\linewidth}
		\centering	\includegraphics[width=0.7in]{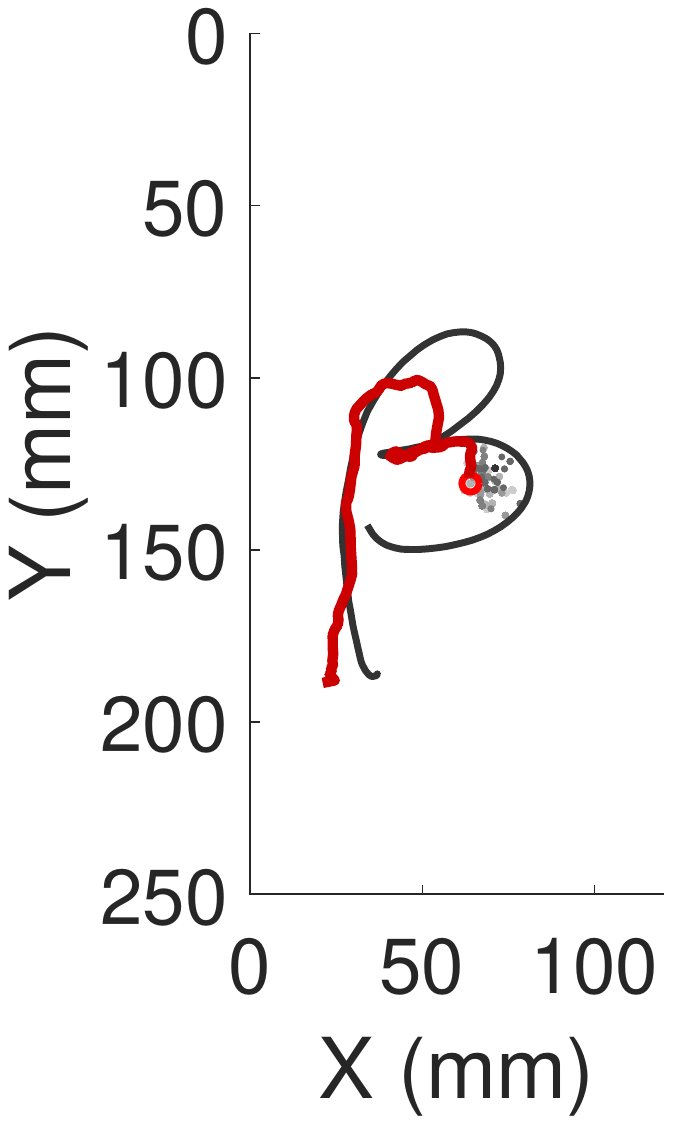}
		\caption{}
		\label{fig:particleFiler_history_t121}
	\end{subfigure}
	\begin{subfigure}{0.118\linewidth}
		\centering	\includegraphics[width=0.7in]{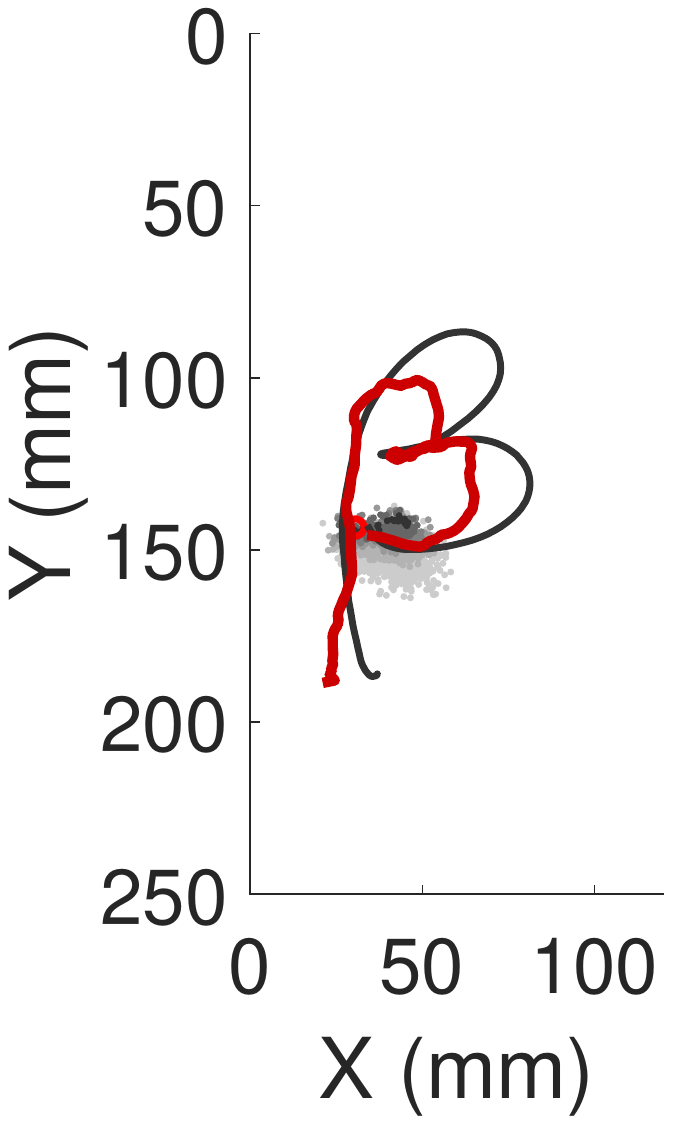}
		\caption{}
		\label{fig:particleFiler_history_t145}
	\end{subfigure}
	\caption{(a-d) shows how adaptive particle resampling chooses the number of particles over time. (e-h) show the estimation of Pencil's location when history of particles is used.}
	\label{fig:particleFiler}
\end{figure*}

\textbf{Adaptive particle resampling: }%In traditional particle filtering, the particles are resampled based on their weights, such the higher weighted particles are chosen at each time step, and the total number of particles stays the same throughout the tracking process.
While a huge number of particles, in our case $K = 50000$, are needed to determine the initial Pencil position accurately, a smaller number of particles can suffice for tracking once the particles' states start to converge.
%
%Adjusting the number of particles at each time step can greatly improve the efficiency of our algorithm.
%TODO add reference
For this purpose, we use the Kullback-Leibler distance (KLD)~\cite{KLD} resampling~\cite{fox2002kld} approach to decide the number of particles, $K_a$, on the fly for every timestamp.
KLD resampling operates by choosing the number of particles to resample such that the KLD between the particles' distribution before and after resampling does not exceed a predefined threshold.
%
%
%Based on this approach, for each time step, $t$, we first determine if the particles at time $t-1$ are spread around the state space or concentrated in a particular region.
%
%This is done by dividing the state space into bins and assigning particles to these bins based on their states at time $t-1$.
%
%We use the number of bins occupied by the particles to determine the number of particles needed at time $t$ (the number is drawn from $\chi$ distribution according to \cite{fox2002kld}).
%
If the particles are widely spread, a large number of particles are used, whereas if the particles are more concentrated, a smaller number is resampled.
We use a batched approach to increase the sample size for improving the efficiency of the resampling process.
Figures~\ref{fig:particleFiler_t1}-~\ref{fig:particleFiler_t145} show how the resampling phase operates over time.
In Figure~\ref{fig:particleFiler_t1}, the particles are scattered into two separate clusters; hence a large number of particles is resampled.
Overtime, once the particles start converging, the number of samples selected at each timestamp decreases as shown in Figure~\ref{fig:particleFiler_t81}-~\ref{fig:particleFiler_t145}.
%
%Instead of selecting and assigning the particles to bins one by one, we use a batched approach where each batch consists of $5000$ particles to increase the efficiency of the resampling process.
%
%We keep on adding particles until the number of particles selected equals the bound determined by KLD sampling or we have selected $K$ particles.
%
%Using a batched approach increases the efficiency of the resampling process. 

\textbf{Pencil location estimation: }In addition to keeping track of the state, each particle also carries its history $S_{0:t-1}$ along with it for all timestamps $t \geq 1$.
For the simple case of location tracking, we directly used the centroid of the particles to determine the Pencil's position at each timestamp.
However, in this case, given the huge state space, the particles might exist in various clusters in different regions of the space, making it difficult to track using just the centroid.
For example, in Figure~\ref{fig:particleFiler_t1}, there are two clusters of particles, and the centroid of these clusters (shown as red circle) does not match well to the ground truth Pencil location (\ie bottom part of the character $`\beta$').
In contrast, if we choose the maximum weighted particle in the last timestamp and use its history as the location estimate, we observe that all the particles converge at the beginning.
This means that using the history of the particles returns us a path for the Pencil's movement with high accuracy.
Figure~\ref{fig:particleFiler_history_t1}-\ref{fig:particleFiler_history_t81} shows the path retraced from the history of particles (red circle in Figure~\ref{fig:particleFiler_history_t81}).
%
%We can see that the particles indeed converge at the beginning as all particles are present around the red circle (Figure~\ref{fig:particleFiler_history_t1}).
%
%The trace of Pencil's position while using particles' history (red trace in Figure~\ref{fig:particleFiler_history_t145}) is much more similar to the ground truth as compared to the trace obtained from centroid of particles (Figure~\ref{fig:particleFiler_t145}).

\subsection{Stroke Detection}

\begin{figure}[t!]%
	\begin{subfigure}{1\linewidth}
		\centering
		\includegraphics[height=1in, width=3.5in]{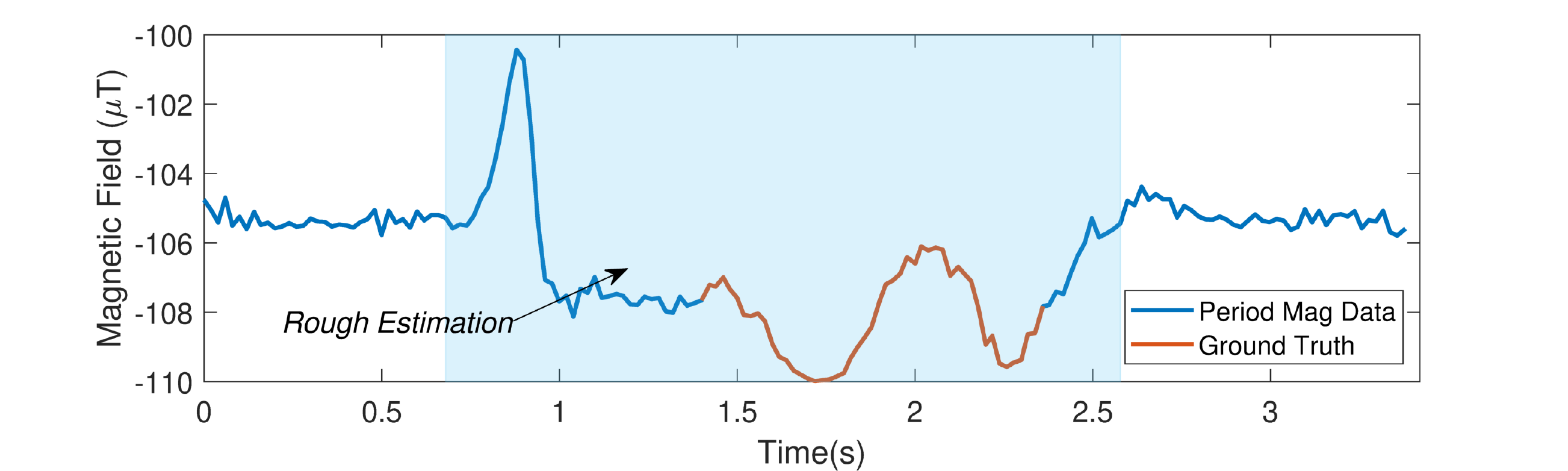}
	\end{subfigure}
		%\label{fig:basic_pattern3d}
	\begin{subfigure}{1\linewidth}
		\centering
		\includegraphics[height=1in, width=3.5in]{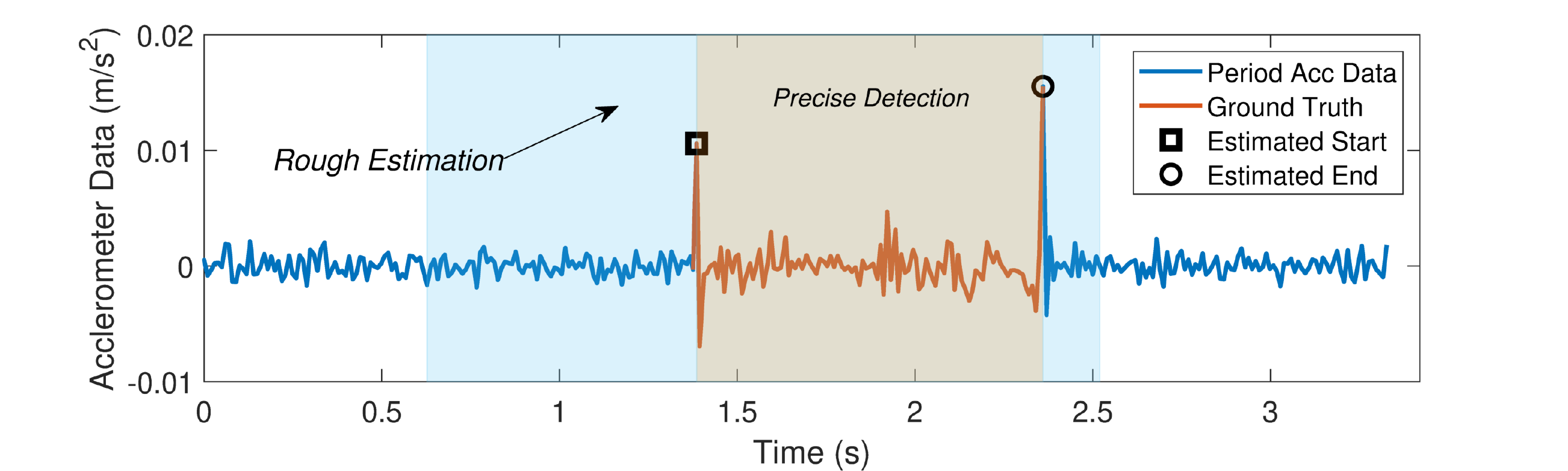}
		%\label{fig:basic_pattern3d}
	\end{subfigure}		
	\begin{subfigure}{1\linewidth}
		\centering
		\includegraphics[height=1in, width=3.5in]{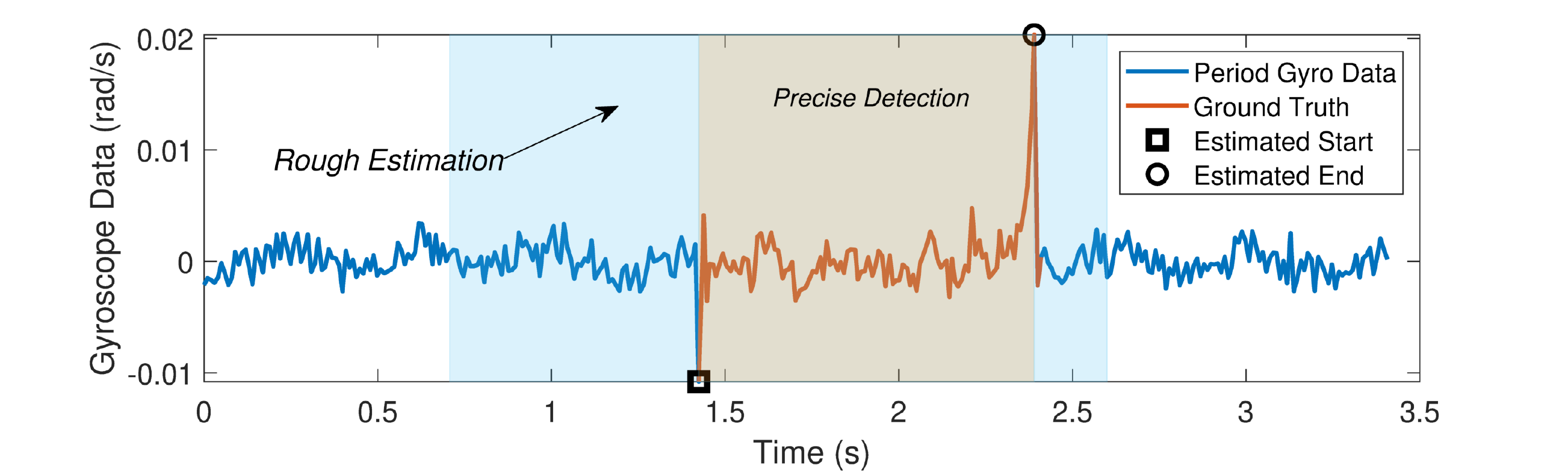}
		%\label{fig:basic_pattern3d}
	\end{subfigure}
	\caption{Stroke detection to determine the beginning and end of a stroke.}
	\label{fig:sd}
\end{figure}

To track a user's writing, we need to identify the precise period during which the user is writing from the collected motion sensors' data.
To achieve this goal, we design a \emph{two-step stroke detection} algorithm.

\emph{Step 1}: Rough estimation: 
As shown in Figure \ref{fig:sd}(a), the magnetic field around the iPad will fluctuate when a user is writing, while it is stable when the Pencil moves away from the screen. 
Using this fluctuation, we first generate a rough estimate of the period when the user is writing. 
The basic idea is to use a sliding window with an appropriate size that contains a sequence of magnetometer data and move the window along the timeline.
We mark the midpoint of this window as the starting timestamp of a stroke when the magnetic data variance is larger than a predefined threshold.
Similarly, the ending timestamp is marked as the center of the window when the variance is below the threshold.
We analyzed the data collected for training the writing behavior model for determining the optimal values for the window size and the threshold and set them to $100$ and $0.12$, respectively.

\emph{Step 2}: Precise detection:
A rough estimation is able to help us identify time periods which may contain a stroke. 
However, since the magnetic field starts fluctuating when the pencil is moving close to the iPad, it is difficult to identify the precise beginning and end of a stroke through magnetic data alone.
Therefore, we utilize accelerometer and gyroscope data to detect the strokes more precisely.
The intuition behind this step is that the iPad will vibrate slightly at both moments when the Pencil touches and is lifted from the screen.
As shown in Figure~\ref{fig:sd}(b) and (c), the beginning and end of the stroke are visible as prominent peaks in the accelerometer and gyroscope data.
Hence the same threshold-based approach can be used on this data to identify the stroke with higher precision.

%% file: evaluation.tex
\section{Evaluation}
\label{sec:evaluation}

\subsection{Data Collection and Implementation}
\label{sec:data_collection}
We evaluated \sys{} on an Apple 11" iPad Pro running iOS 12.0.
We implemented two applications for conducting our experiments.
The first application, \textit{A}, acts as the malicious application installed on the victim's iPad, mimicking a legitimate application (such as a fitness tracker) that accesses the motion sensors' data and stays alive in the background (\eg by use of location services).
Application $A$ logs the magnetometer readings at $50$Hz, and accelerometer and gyroscope data at $100$Hz in the background.
%
%The application plays an audio service to stay alive in the background; however, the sound played is inaudible to avoid detection by the user.
%
The second application, \textit{B}, mimics a chatting application with a text input region in the lower half of the iPad screen when it is used in landscape mode.
This is where the user writes using the Apple Pencil.
The input region is divided into $3$ equally sized grids.
This application acts as the application under attack from application $A$.
The sensor data collected from application \textit{A} is stored on the iPad during the data collection process and is transferred to a server for analysis. %a MATLAB
\majorRev{[C9]}\majorRevText{During our experiments, we observe that on average, with a fully charged battery, this application consumes less than $10$\% of the battery while logging the motion sensor's data, over a duration of $3$ hours.}
We use our own custom application (application $B$) for writing to record the touch API data for ground truth. 
%For ground truth, we also record the touch API data via application B while the user is writing.
%
\majorRev{[C8]}\majorRevText{We also tested these applications on the latest iOS version (14.0) to validate our assumptions regarding required permissions.}

Three authors (pretending to be attackers) collected magnetic data for the Pencil's magnetic map generation while randomly drawing on the iPad screen for a total duration of $3$ hours.
We used the setup shown in Figure~\ref{fig:camera_setup} to record the orientation of the Pencil and the touch API's data for recording the Pencil's location.
\majorRev{[MR2, C2, C6]}\majorRevText{The three authors also wrote/drew the $26$ lowercase alphabet letters ($a$-$z$), $10$ numbers ($0$-$9$), and five different shapes (square, triangle, circle, heart, and star) twice in each grid in the input region of the application \textit{B}.
This data was used to train the writing behavior model described in Section~\ref{sec:system_design} and was not a part of the evaluation set.}

To evaluate \sys, we conducted experiments with $10$ subjects, recruited by advertising on the university campus after IRB approval.
These subjects included $6$ females and $4$ males.
During the experiments, subjects were seated next to a table on which the iPad was placed.
%
%We ensured that the iPad remained static throughout the experiment duration.
%
%Before beginning an experiment, we asked the subject to detach the Pencil and hold it a few inches away from the iPad at which point we started the application $A$ to log sensor data.
%
%This allowed us to measure the ambient magnetic field without any impact from the Pencil.
%
%After a few seconds, the subject was asked to start writing.
%
We asked each subject to use the Pencil to write the $26$ lowercase alphabet letters ($a$-$z$), $10$ numbers ($0$-$9$) and five different shapes (square, triangle, circle, heart and star) twice in each grid in the input region.
\majorRev{[C7]}\majorRevText{The subjects were guided to write in their natural handwriting style and we did not impose any restriction on the size of the strokes within the input region.}
%
%If any stroke was written outside of the input region, we discarded that data. 
%
In total, we collected the motion sensor data for $1560$ letters, $600$ numbers and $300$ shapes.

\majorRev{[C15]}\majorRevText{In addition to letters, numbers, and shapes, we also tested our system's ability to track words.
We asked five subjects to write words of lengths varying between three to six letters, where each word consisted of lowercase letters randomly selected from the English alphabet.
We instructed each subject to write $30$ words for each of the four word lengths.
We also investigated our system's performance in detecting legitimate English words.
For this purpose, we asked these subjects to write $30$ words each, randomly selected from the $500$ most commonly used English words~\cite{wordFrequency}.
These words represented a diverse set of words varying in length between $3$ to $8$ letters with an average length of $5$ letters.}
%The subjects were instructed to use cursive handwriting for these experiments.

% We asked $5$ subjects to write $30$ words each using the Pencil in the input region of the application B while the iPad was placed on the table.
% %
% These words were randomly selected from the top 500 most common English words used in English conversations from the OEC text corpus~\cite{theoec}.
% %
% The subjects were instructed to use cursive handwriting for these experiments.
% %
% Apart from these commonly used words, we also evaluated the effect of word length on \sys{}'s performance by asking the volunteers to write words of varying length (between $3$ to $6$).
% %
% In this case, each word consisted of lower case letters selected randomly from the English alphabet.
% %
% Each subject wrote $30$ random words for each of the word lengths.

\majorRev{[MR3]}\majorRevText{To demonstrate the practicality of our system, we also conducted experiments with different positions of the iPad and analyzed the impact of various environment settings.}

\majorRev{[C13]}\majorRevText{We randomly selected one of the volunteers to act as an attacker and showed \sys{}'s Pencil tracking results to the attacker for each of our experiments.
The attacker was given three guesses to identify each stroke.
Hence our system's accuracy is determined as the number of strokes correctly identified by the attacker in the first three guesses.
We use this top-$k$ guess method considering the similarity in shapes of some letters in the English alphabet.}

% We also evaluated our system with different positions of the iPad.
% %
% For this purpose, we collected the data described above from $3$ subjects for scenarios where they were sitting on a couch, standing, laying down on a couch and walking inside a room, all while holding the iPad in hand.
% %

% Finally, we also investigated the impact of the environment on \sys{}'s performance.
% %
% We analyzed the affect of different human activities and metallic objects on the tracking results by collecting data from $$

\subsection{Performance Results}
\label{sec:performance_results}
We answer the following questions to evaluate \sys's performance:

\begin{enumerate}
	\item How well can \sys{} detect the letters, numbers and shapes?
	\item What is the detection accuracy for different letters, numbers, and shapes, and why?
	\item Is \sys{} able to detect words?
	\item Can \sys{} detect continuous strokes?
	\item How does the location of the Pencil tip affect the performance of \sys{}?
	\item Does the performance of \sys{} change across users?
	\item How does the positioning of the iPad affect \sys{}'s performance?
	\item How do the environmental factors impact the performance of \sys{}?
	\item How does \sys{} perform compared to other machine learning techniques?
\end{enumerate}

\begin{figure}[t!]
	\centering
		\includegraphics[height=1.8in]{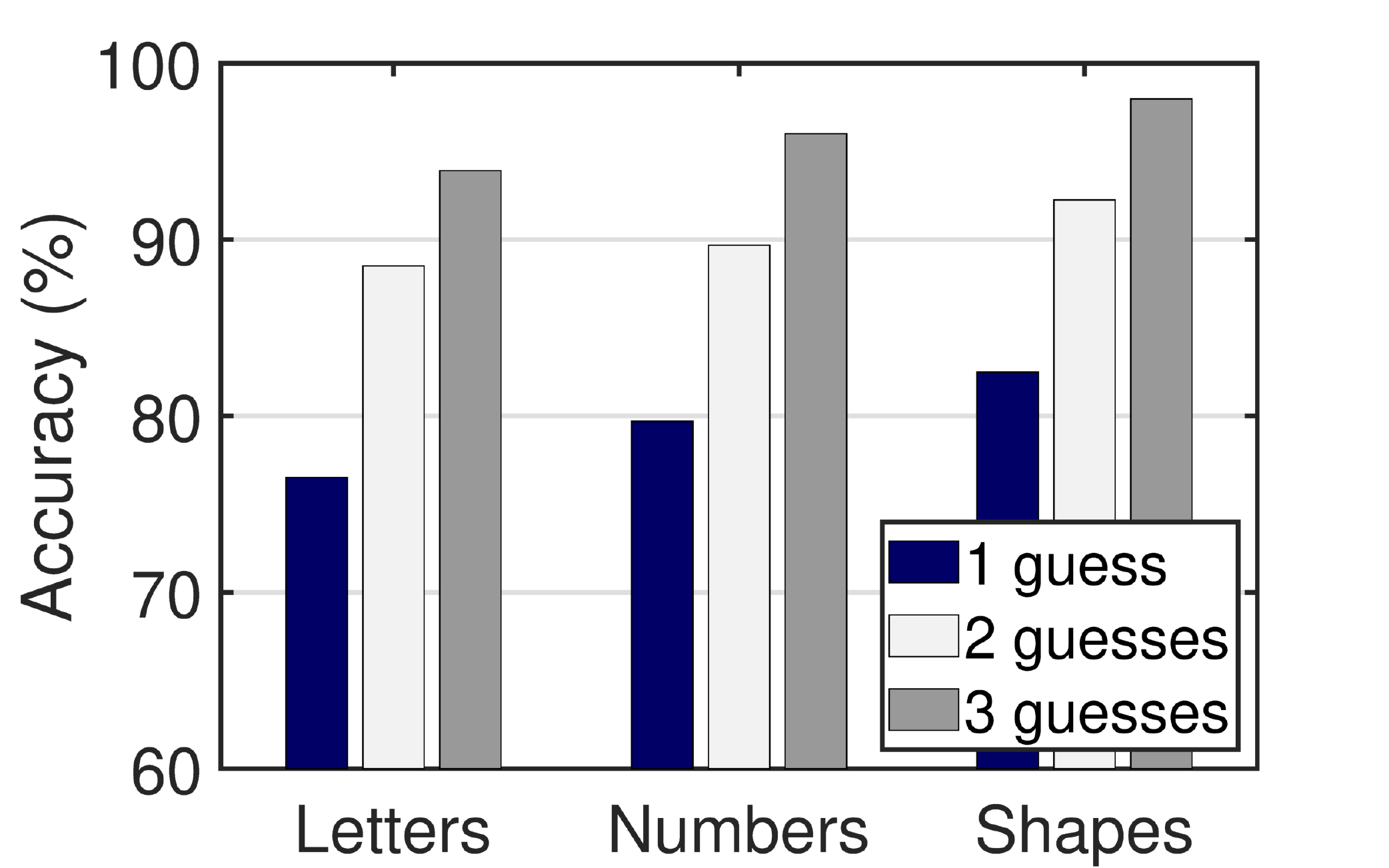}
		\caption{Overall accuracy of the correctly guessed letters, numbers and shapes in 1, 2 and 3 guesses.}
		\label{fig:overall_performance}
%	\begin{subfigure}{0.45\linewidth}
%		\includegraphics[height=1.8in]{figures/evaluation/stroke_size_distribution.pdf}
%		\caption{}
%		\label{fig:stroke_size_distribution}
%	\end{subfigure}
%	\label{fig:main_performance}
%	\caption{(a)  (b) Distribution of the sizes of the strokes drawn on the screen.}
\end{figure}

\begin{figure*}[t!]
	\centering
	\begin{subfigure}{\linewidth}
		\centering
		\includegraphics[height=1.2in]{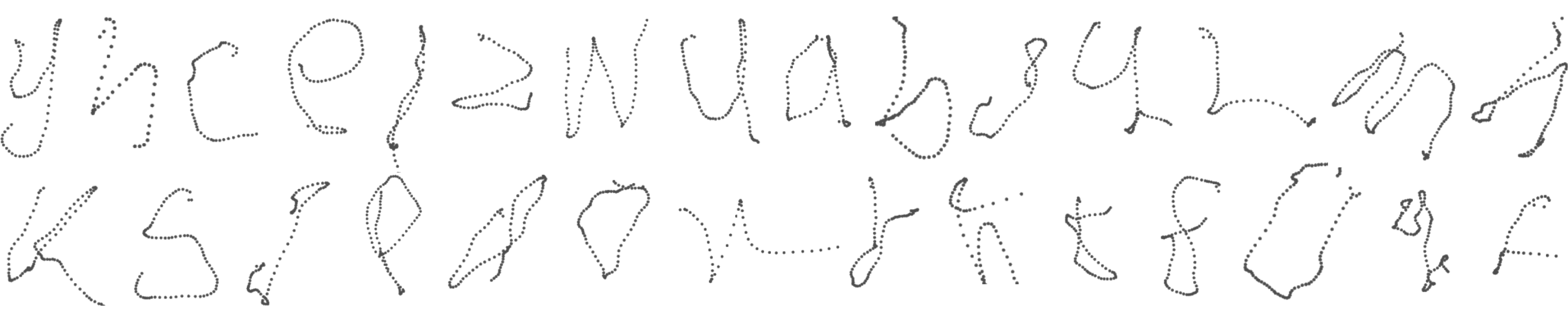}
		\caption{}
		\label{fig:eval_examples_letters}
	\end{subfigure}
	\begin{subfigure}{\linewidth}
		\centering
		\includegraphics[height=1.2in]{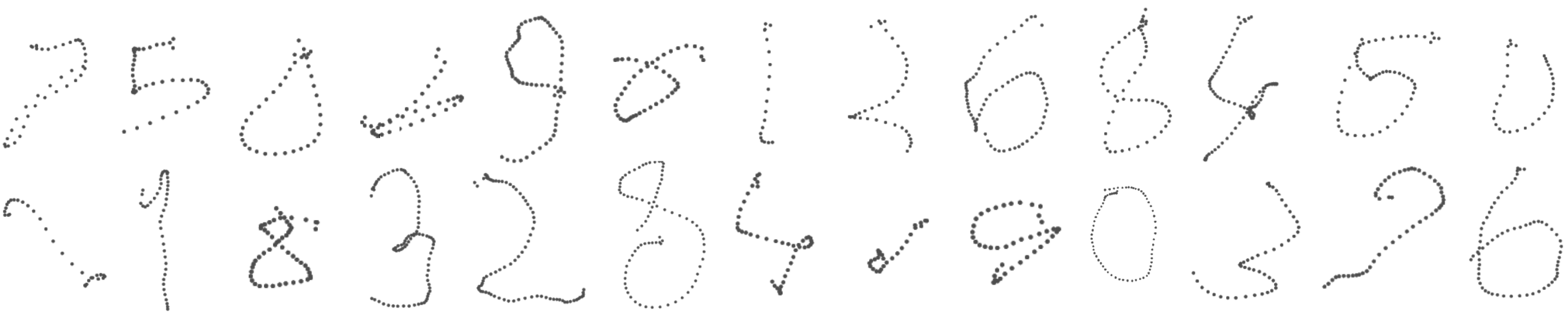}
		\caption{}		
		\label{fig:eval_examples_numbers}
	\end{subfigure}
	\begin{subfigure}{\linewidth}
		\centering
		\includegraphics[height=0.62in]{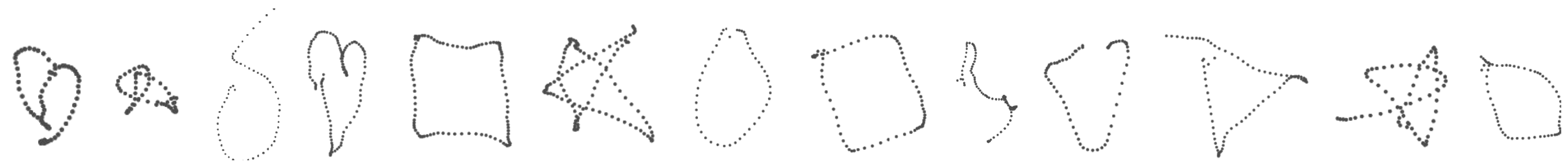}
		\caption{}
		\label{fig:eval_examples_shapes}		
	\end{subfigure}	
	\caption{Tracking result examples: (a) letters, (b) numbers, and (c) shapes.}
	\label{fig:eval_examples}
\end{figure*}

\textbf{(1) How well can \sys{} detect the letters, numbers, and shapes?}

We evaluated our system's performance by randomly selecting one of the volunteers acting as an attacker to guess what the users wrote from \sys{}'s Pencil tracking results.
The tracking results for different strokes were shuffled for this process to remove any possible bias.
We recorded the top $3$ guesses from the attacker for each stroke.
Figure~\ref{fig:overall_performance} shows the overall detection accuracy when the attacker's $1^{st}$, $2^{nd}$ or $3^{rd}$ guess is correct for different strokes.
The attacker correctly guessed $93.9$\%, $96$\%, and $97.9$\% of the letters, numbers, and shapes, respectively.
This detection accuracy is based on whether any of the $3$ guesses for a stroke matched with the actual letter, number or shape.
Figure~\ref{fig:eval_examples} shows a set of examples from the Pencil traces generated by \sys{} solely using the motion sensor data.
We encourage the reviewers to guess what letter, number, or shape is shown in each example.
We give the correct answers at the end of this paper.

% Habiba's version November 12.
\begin{comment}
\textbf{(1) How well can \sys{} detect the letters, numbers, and shapes?}

We evaluated our system's performance by randomly selecting one of the volunteers who acts as an attacker to guess what users wrote from \sys{}'s Pencil tracking traces.
%
The tracking results for different strokes were shuffled for this process to remove any possible bias.
%
%The attacker was informed about the category of the stroke before guessing.
%
We recorded the top $3$ guesses from the attacker for each stroke.
%
The attacker correctly guessed $93.9\%$, $96\%$, and $97.9\%$ of the letters, numbers, and shapes, respectively.
%
This detection accuracy is based on whether any of the $3$ guesses for a stroke matched with the actual letter, number or shape.
%
Figure~\ref{fig:overall_performance} shows the overall detection accuracy when the attacker's $1^{st}$, $2^{nd}$ or $3^{rd}$ guess were correct for different strokes.
%
%
%By analyzing the size distribution of all the strokes collected from the subjects, we observed that average size of the strokes is $30mm$
%
%Considering the size of the iPad screen is $159mm$x$229mm$, the area covered by majority of the strokes was less than $5\%$ of the screen size.
%
%%

Figure~\ref{fig:eval_examples} shows examples of the Pencil traces generated by \sys{} solely using the motion sensor data.
%
We encourage the reviewers to guess what letter, number, or shape is shown in each example.
%
We give the correct answers at the end of this paper.
\end{comment}

\begin{figure*}[t!]
	\begin{subfigure}{0.47\linewidth}
		\centering
		\includegraphics[height=1.3in]{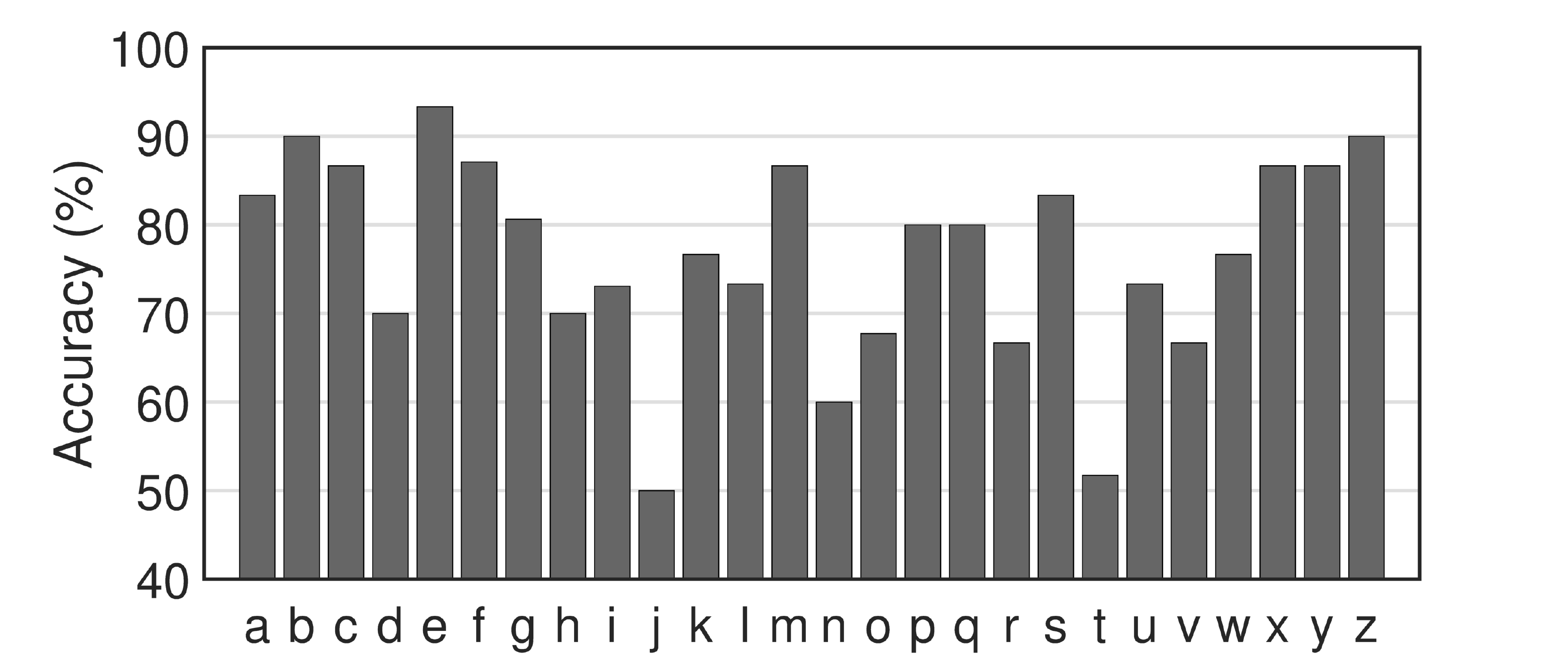}
		\caption{}
		\label{fig:individual_letters}		
	\end{subfigure}
	\begin{subfigure}{0.25\linewidth}
		\centering
		\includegraphics[height=1.3in]{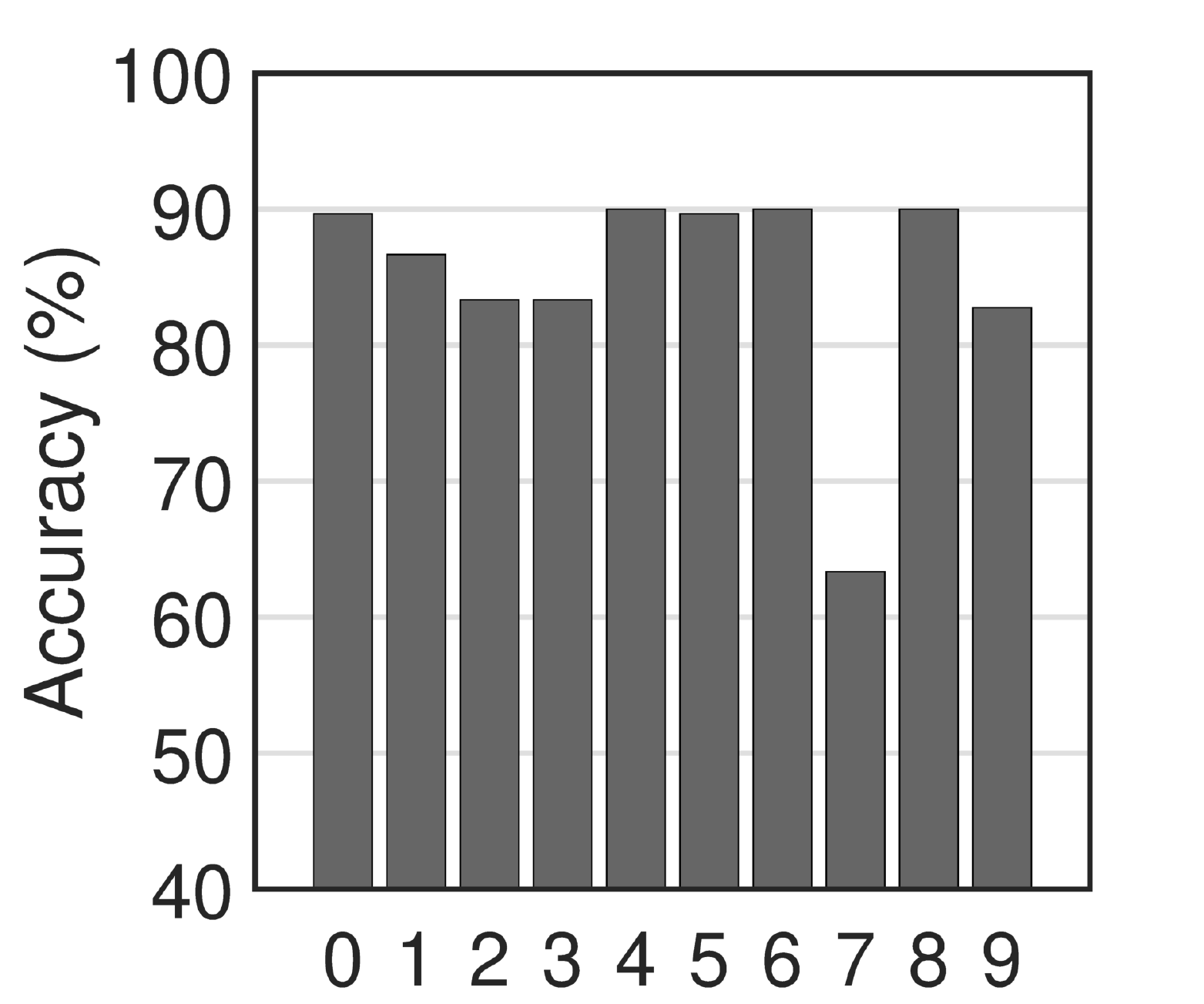}
		\caption{}
		\label{fig:individual_numbers}		
	\end{subfigure}
	\begin{subfigure}{0.25\linewidth}
		\centering
		\includegraphics[height=1.3in]{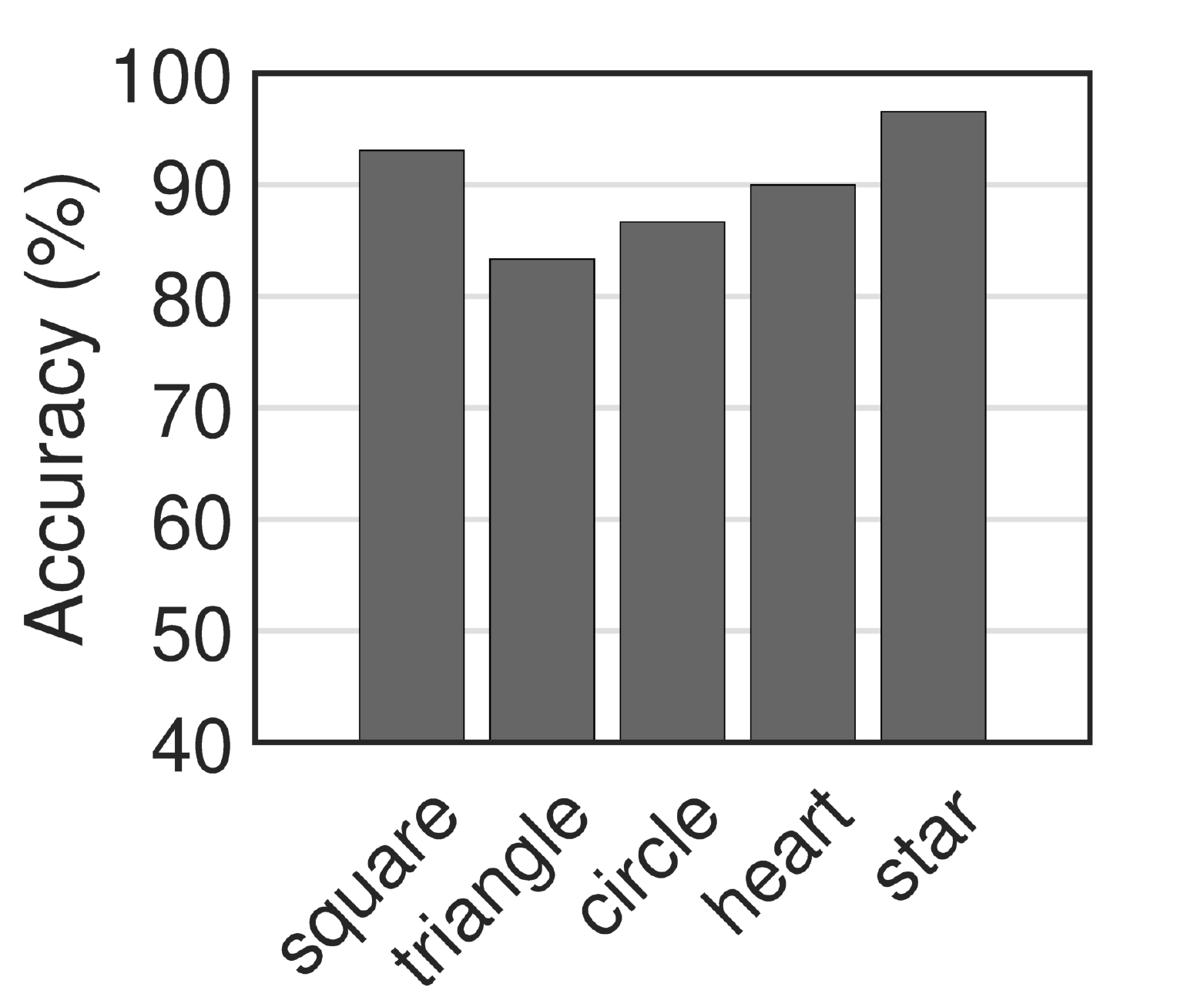}
		\caption{}
		\label{fig:individual_shapes}		
	\end{subfigure}
	\caption{Accuracy of the first guess for (a) each letter, (b) each number, and (c) each shape. }
	\label{fig:individual}
\end{figure*}
% 0.8333
% 0.9000
% 0.8667
% 0.7000
% 0.9333
% 0.8710
% 0.8065
% (h) 0.7000
% (i) 0.7308
% (j) 0.5000
% (k) 0.7667
% (l) 0.7333
% 0.8667
% (n) 0.6000
% 0.6774
% 0.8000
% 0.8000
% (r) 0.6667
% 0.8333
% (t) 0.5172
% (u) 0.7333
% (v) 0.6667
% 0.7667
% 0.8667
% 0.8667
% 0.9000

% 0.8966
% 0.8667
% 0.8333
% 0.8333
% 0.9000
% 0.8966
% 0.9000
% (7): 0.6333
% 0.9000
% 0.8276

% 0.9310
% 0.8333
% 0.8667
% 0.9000
% 0.9655

\textbf{(2) What is the detection accuracy for different letters, numbers, and shapes and why?}

We analyzed how different letters, numbers, and shapes contributed to the overall detection accuracy reported earlier.
Figure~\ref{fig:individual} shows the detection accuracy for each letter, number, and shape based on the attacker's first guess.
Here, we can clearly see that some letters, numbers, and shapes are detected better than the others in the first guess.
For example, letters like `b', `e', and `z' have a high detection accuracy.
In contrast, letters like `i', `j', and `t' have lower accuracy.
We found that these letters are usually written in two strokes. 
Since \sys{}  does not consider two strokes separately, the letter's shape is not obvious in the Pencil's trace.
This can also be seen in Figure~\ref{fig:eval_examples_letters}, where it is difficult to recognize the letter `j' (third trace in the bottom row).
Apart from this observation, some letters like `h' and `n' have a low accuracy of $70$\% and $60$\% respectively because of the similarities in their shapes.
The second letter in the top row of Figure~\ref{fig:eval_examples_letters} can be inferred as `h' although it is actually `n'.
We observe similar patterns for `r` and `v' ($67$\% accuracy for both) whose traces are similar to each other as shown in Figure~\ref{fig:eval_examples_letters} (`v' - third last letter in row 1, `r' - $8^{th}$ letter in row 2).    
Similarly, the number `7' can be confused with the number `1'.
We note that the letters, numbers, and shapes whose traces are distinct from others are detected with high accuracy in the first guess.
Examples of such cases are letters `m', number `8', and star shape.

\majorRev{[MR3, C7]}\majorRevText{Given that we did not impose any restriction on subjects' handwriting styles, we observed that the same strokes written by the same subject often varied slightly in size or style (\eg slanted).
However, since \sys{} does not rely on a specific handwriting style for accurate recognition, these variations do not affect its accuracy.}

\begin{figure}[ht!]
	\centering
	\includegraphics[height=1.5in]{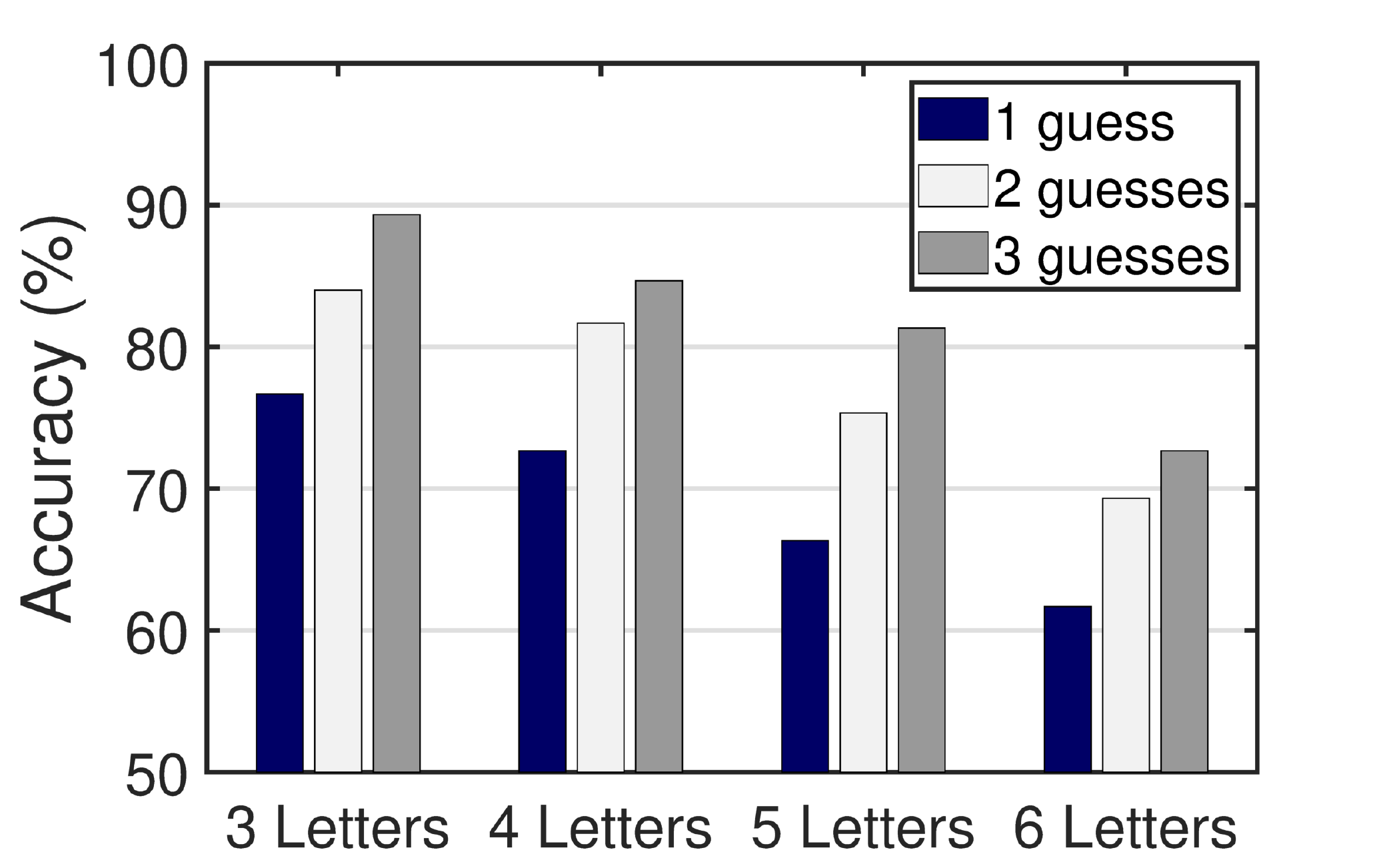}
	\caption{\sys{}'s accuracy in detecting $3$ to $6$ letter words.}
	\label{fig:performance_words}
\end{figure}

\majorRev{[MR3, MR6, C12]}\majorRevText{\textbf{(3) Is \sys{} able to detect words?}}

\majorRevText{To answer this question, we shuffled the tracking results for all the words collected from the $5$ subjects and presented them to the attacker as described previously.
Figure~\ref{fig:performance_words} shows \sys{}'s detection accuracy for the random words of varying lengths.
The attacker correctly guessed $89.3$\% and $84.67$\% of the 3 and 4 letter words.
Given the completely random nature of these words, we observe that the number of correctly guessed words decreases as the words' length increases.
This is due to the similarity in the tracking results of different letters, as described above, which becomes more prominent in longer words.
\begin{figure}[t!]
	\centering
	\includegraphics[height=0.62in]{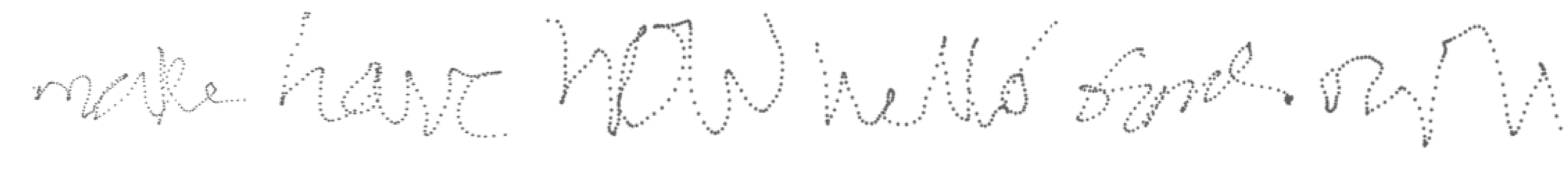}
	\caption{Examples of tracking results for the words.}
	\label{fig:words_examples}
\end{figure}
However, human handwriting in practice mainly consists of legitimate English words where little randomness is involved.
For our experiments with the commonly used English words (three 
 to eight-letter words), the attacker could guess the words with an accuracy of $93.33$\% in the first three guesses.
Figure~\ref{fig:words_examples} shows a set of examples of the tracking results for this case.
We found that the tracking results, in some cases, are noisy, making it difficult to guess all the letters in words. An example of this observation is seen in the rightmost word in Figure~\ref{fig:words_examples}.
Yet, the attacker can guess the words correctly in these cases since she is able to infer a few letters of the word from the tracking results, which allows her to figure out what the complete word is.}

\majorRevText{\textbf{(4) Can \sys{} detect continuous strokes?}}

\majorRevText{To evaluate continuous stokes, we conducted an end-to-end attack with \sys{}.
To detail, we consider a real-world scenario where the victim uses the Apple Pencil to write a text message in a chatting application. 
The attacker's goal is to infer the text message (English sentences) based on the magnetometer's readings collected by the malicious application running in the background.}

\majorRevText{In these experiments, we invited three volunteers to write $10$ different text messages on the iPad while sitting and holding it in hand.
The subjects were asked to write valid English sentences, used in common text message conversations (without punctuation and emojis), with up to $8$ words in each case.
The number of words in the sentences written by the volunteers ranged from three to eight, with a median of five words.
\sys{}'s stroke detection component was used to separate the magnetic readings for different words within each sentence and fed into the particle filter.
The attacker was able to guess $66.67$\% of all the words collected in the first three guesses through the tracking results ($42.5$\% in the first guess and $59.1$\% in two guesses).
We found that the estimated beginning and end of some strokes are not very precise when the time gap between different words while writing a sentence is small.
Therefore, the tracking results for these strokes are noisy, making it difficult for the attacker to guess the correct word.
However,  the attacker was able to guess $80$\% of the sentences correctly when the tracking output of all the words in a given sentence were shown together as part of a single sentence.
Thus, the context helps the attacker infer the complete sentence correctly even when the tracking results of individual words are noisy.}

\begin{figure}[h!]
	\centering
	\includegraphics[height=1.5in]{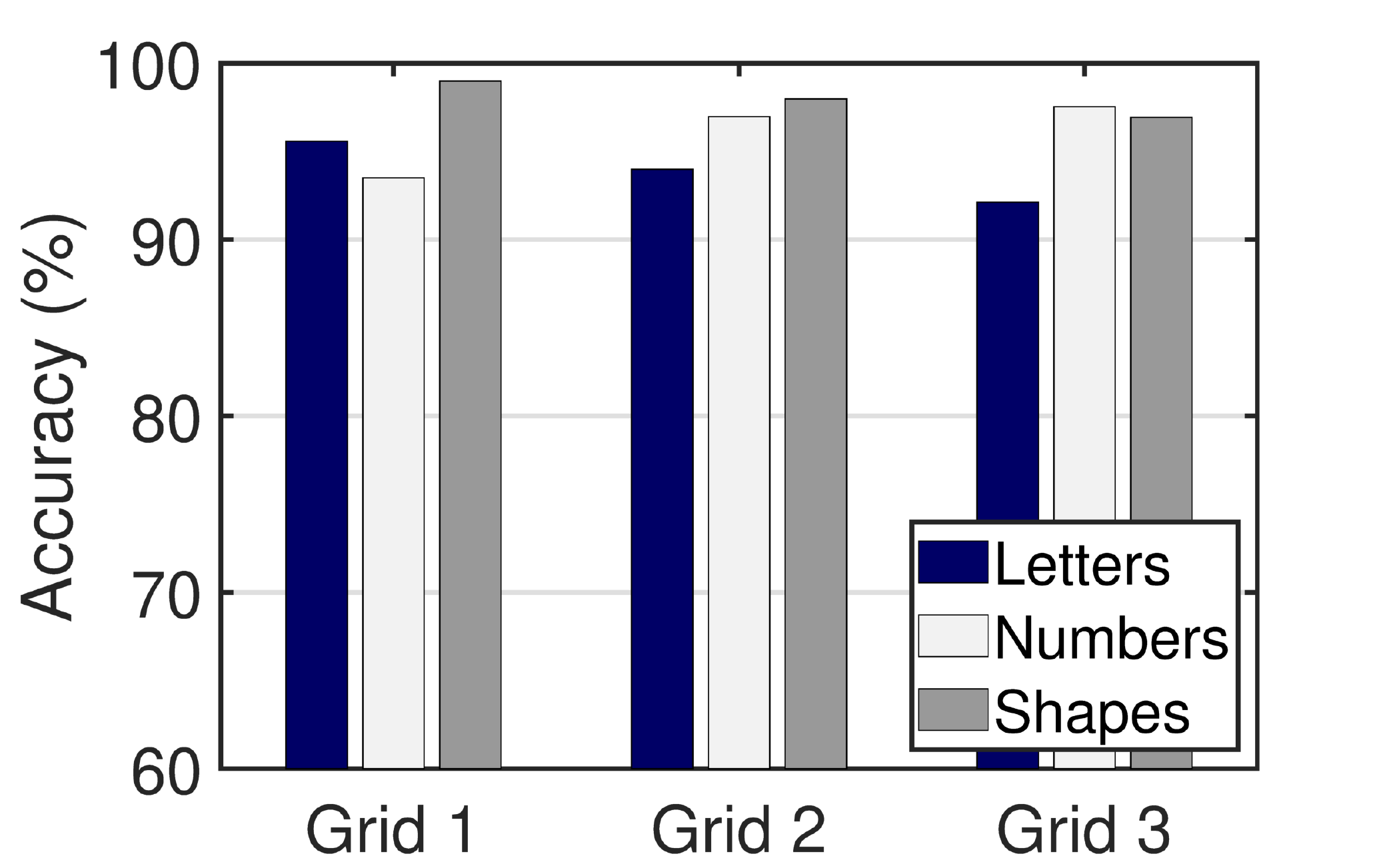}
	\caption{Accuracy of detecting letters, shapes, and numbers across different locations on the iPad screen. Grid 1, 2, and 3 are the left, center, and right part of the input region.}
	\label{fig:performance_locations}
\end{figure}

\textbf{(5) How does the location of the Pencil tip affect the performance of our system?}

As mentioned earlier, we split the input region into $3$ grids during the experiments.
This enabled us to evaluate \sys {}'s performance when a user writes on different iPad screen locations.
Figure~\ref{fig:performance_locations} shows the detection accuracy of letters, numbers, and shapes in the $3$ grids. 
Since grids $2$ and $3$ are close to the magnetometer location, we observe that the range for magnetic fluctuations caused by the Pencil is large in these grids, whereas grid $1$, which is farther from the magnetometer, is less sensitive to the Pencil's movement.
%
%Considering this, good detection accuracy can be expected for grids $2$ and $3$.
%
However,  we observe that our system performs consistently well across all grids despite detecting only small fluctuations in the grid $1$.
This observation clearly demonstrates that small changes in magnetic reading can be tracked by \sys{}.

% Habiba's version November 13.
\begin{comment}
\textbf{(5) How does the location of the Pencil tip affect the performance of our system?}

As mentioned earlier, we divided the input region into $3$ grids during the experiments.
%
This allowed us to evaluate our system's performance when a user writes on different iPad screen locations.
%
Figure~\ref{fig:performance_locations} shows the detection accuracy for letters, numbers, and shapes in the $3$ grids. 
%
Since grids $2$ and $3$ are close to the magnetometer location, we observe that the range for magnetic fluctuations caused by the Pencil is large in these grids, whereas grid $1$, which is farther from the magnetometer, is less sensitive to the Pencil's movement.
%
%Considering this, a good detection accuracy can be expected for grids $2$ and $3$.
%
However, despite detecting only small fluctuations in the grid $1$, we observe that our system performs consistently well across all grids.
%
This shows that even small changes in magnetic reading can be tracked by \sys{}.
\end{comment}

%\begin{figure}[h!]
%	\centering
%	\includegraphics[height=1.8in]{figures/evaluation/performance_users.pdf}
%	\caption{Detection accuracy of letters, shapes and numbers for different users.}
%	\label{fig:performance_users}
%\end{figure}

\begin{figure}[t!]
	\centering
	\begin{subfigure}{0.33\linewidth}
		\centering
		\includegraphics[height=1.4in]{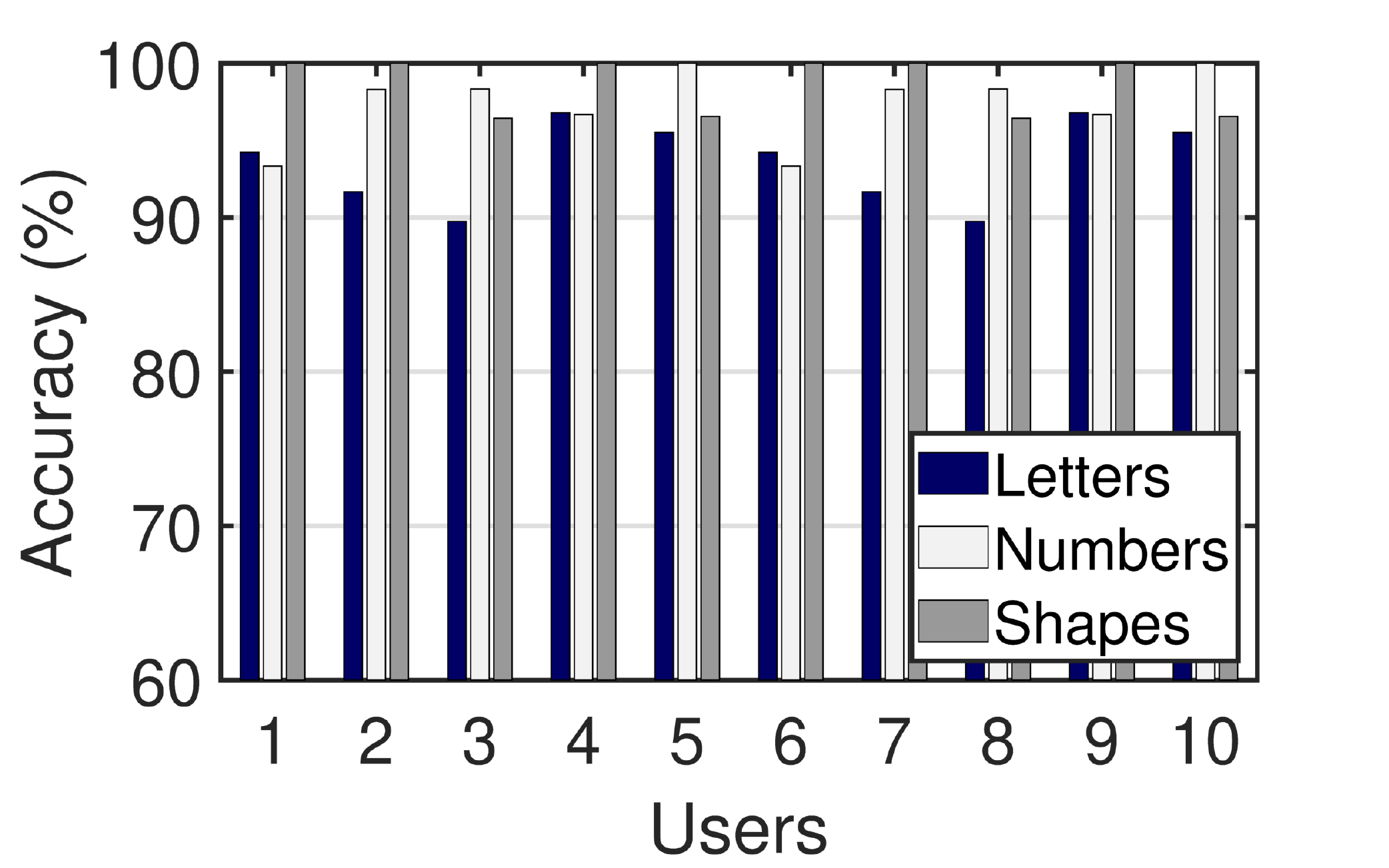}
		\caption{}
		\label{fig:performance_users}
	\end{subfigure}
	\begin{subfigure}{0.33\linewidth}
		\centering
		\includegraphics[height=1.4in]{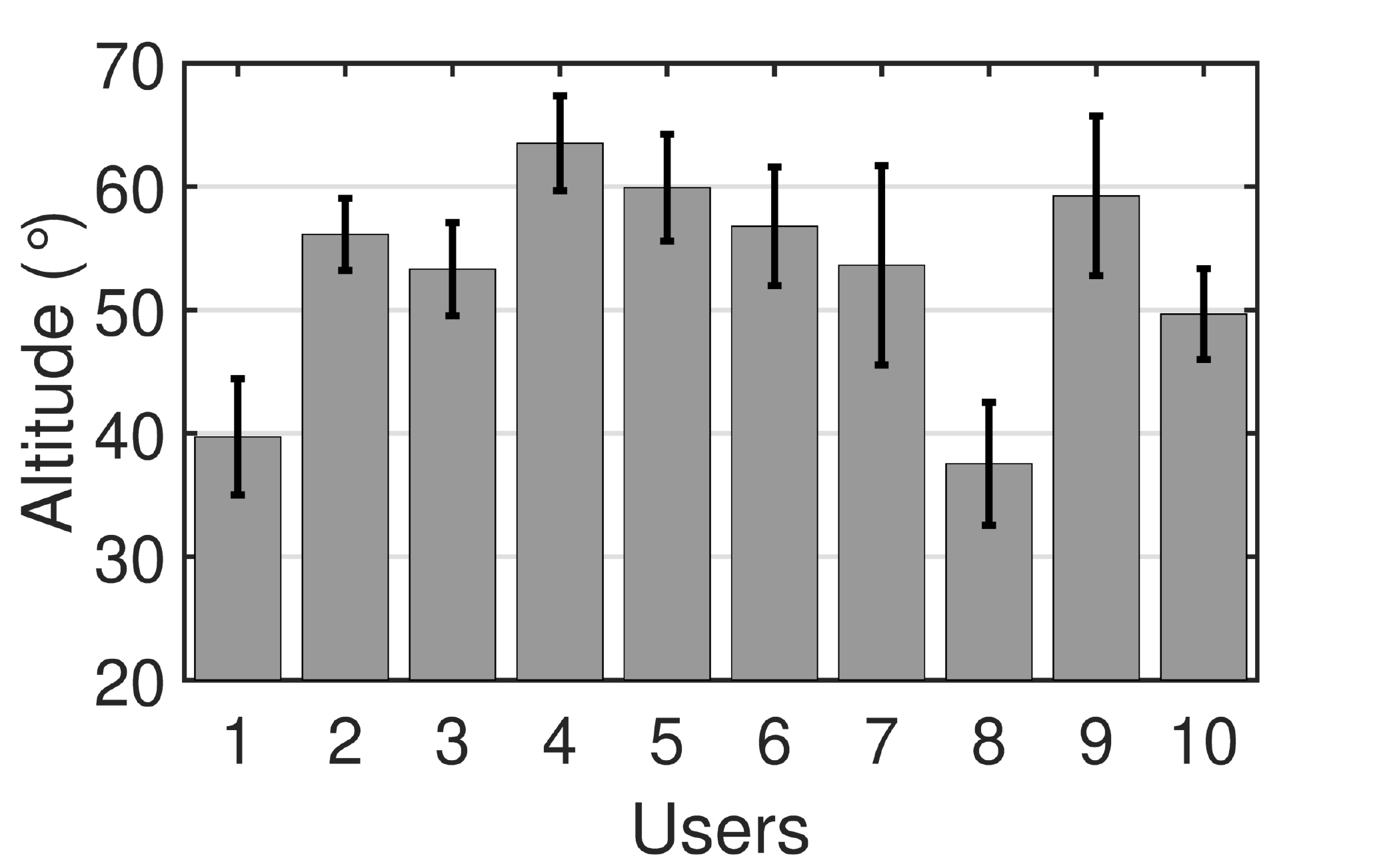}
		\caption{}
		\label{fig:altitude_distribution}
	\end{subfigure}
	\begin{subfigure}{0.33\linewidth}
		\centering
		\includegraphics[height=1.4in]{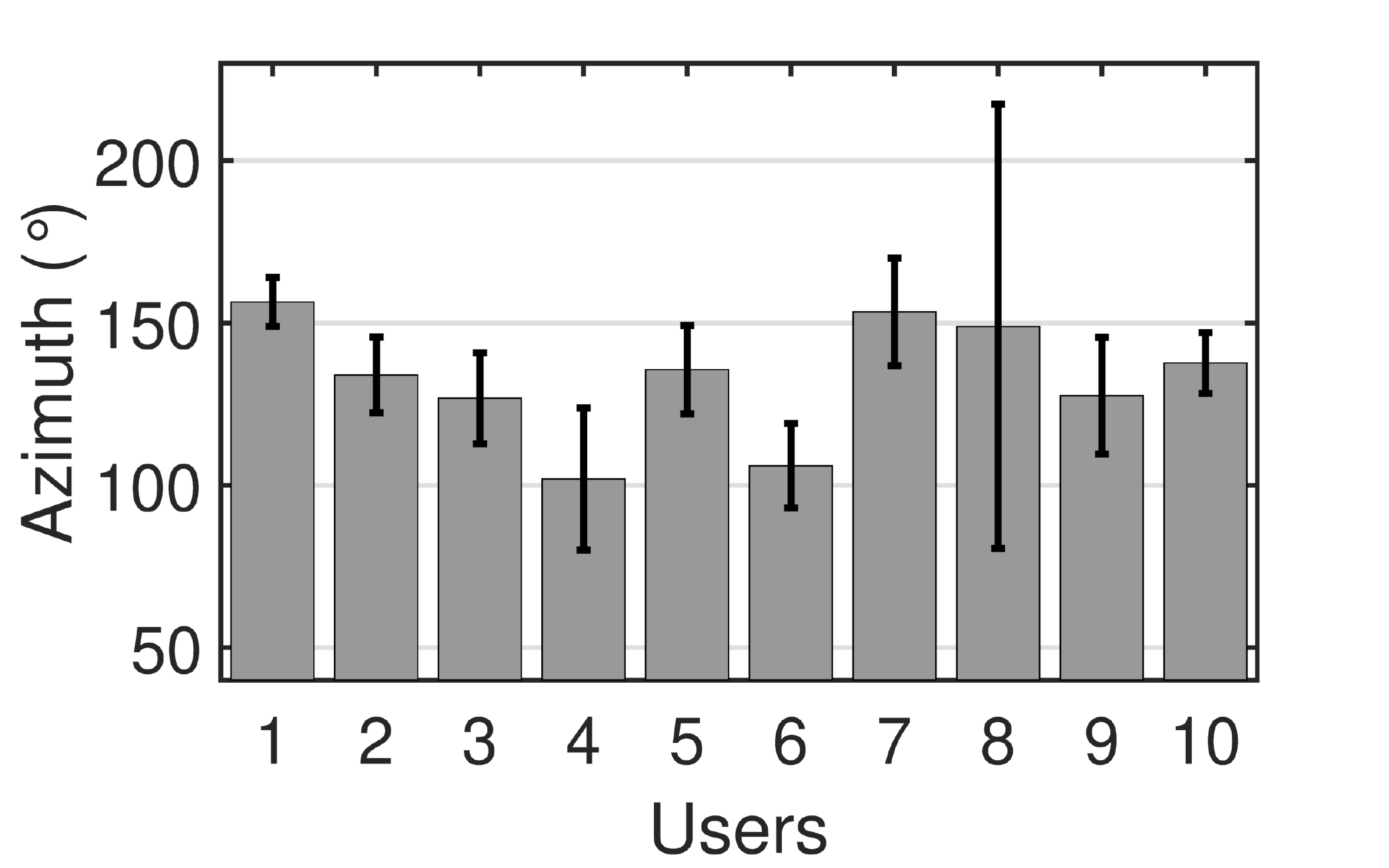}
		\caption{}
		\label{fig:azimuth_distribution}
	\end{subfigure}
	\caption{(a) Detection accuracy of letters, shapes, and numbers for different users. (b) Distribution of the altitude and (c) azimuth angles for the strokes written on the screen.}
	\label{fig:user_results}
\end{figure}

\textbf{(6) Does the performance change across users?}

We show in Figure~\ref{fig:performance_users} how well \sys{} performs across different users.
The detection accuracy is equal to or greater than $90$\% for letters, numbers, and shapes for all users.
From the touch API data recorded for ground truth, we observe that the range for altitude and azimuth angles vary broadly across $10$ users for each category. 
Figure~\ref{fig:altitude_distribution} and Figure~\ref{fig:azimuth_distribution} show the average altitude and azimuth angles along with their standard deviation for each of the $10$ users.
Though we defined a fixed range for these two angles, the consistent good accuracy for all users shows that our tracking algorithm can accommodate a broad range of Pencil orientations and is not affected by the way a user holds the Pencil while writing.

% Habiba's version November 13.
\begin{comment}
\textbf{(6) Does the performance change across users?}

Figure~\ref{fig:performance_users} shows how well our system performs across different users.
%
The detection accuracy is $90\%$ and above for letters, numbers, and shapes for all users.
%
From the touch API data recorded for ground truth, we observe that the range for altitude and azimuth angles vary broadly across the $10$ users for each category. 
%
Figure~\ref{fig:altitude_distribution} and Figure~\ref{fig:azimuth_distribution} show the average altitude and azimuth angles along with their standard deviation for each of the $10$ subjects.
%
Even though we defined a fixed range for these two angles, the consistent good accuracy for all users show that our tracking algorithm can deal with a broad range of Pencil orientations and is not impacted by how the user is holding the Pencil while writing.
\end{comment}
%
\begin{figure}[t!]
	\centering
	\begin{subfigure}{0.24\linewidth}
		\centering
		\includegraphics[height=1in]{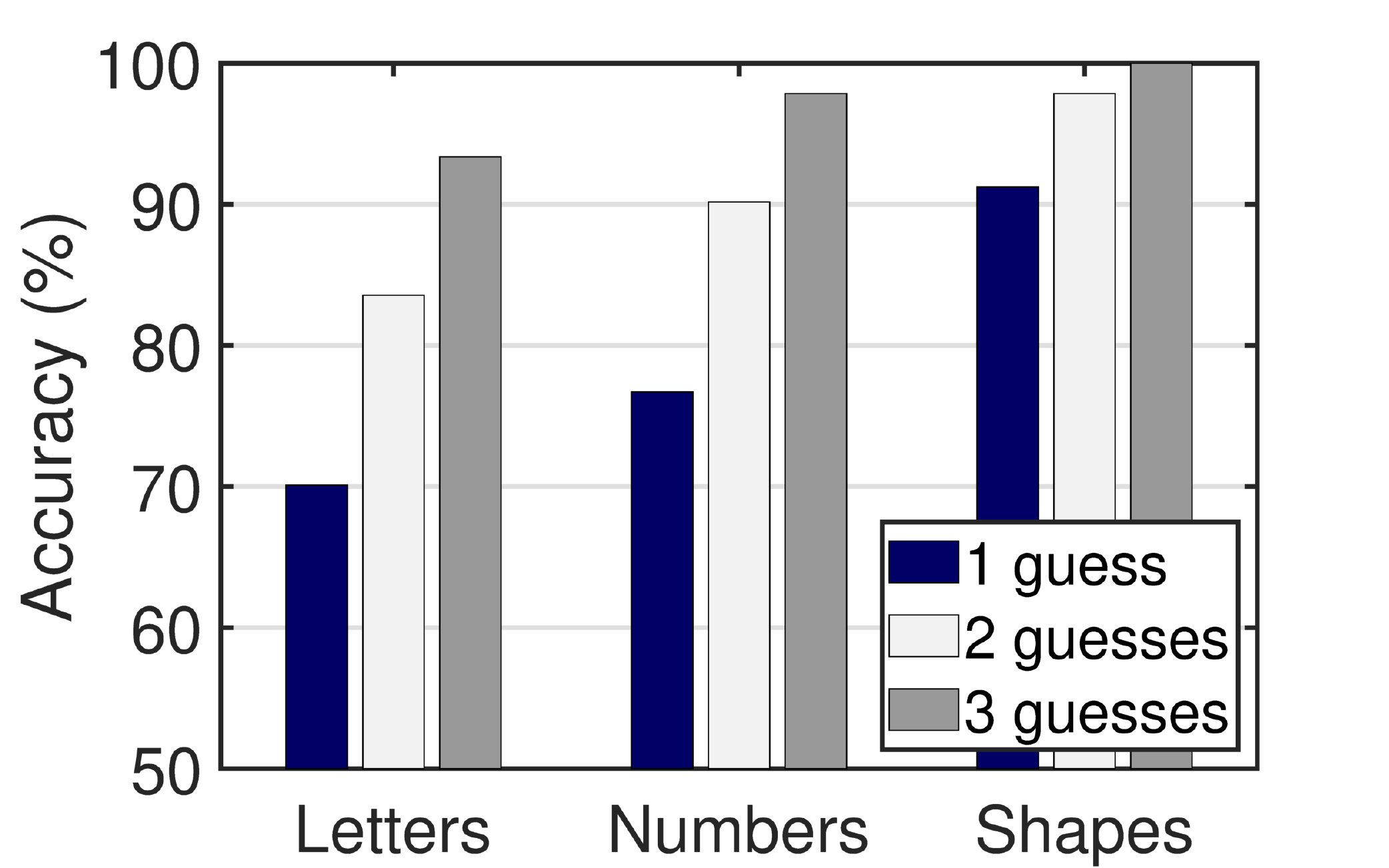}
		\caption{}
		\label{fig:performance_hand_sitting}
	\end{subfigure}
	\begin{subfigure}{0.24\linewidth}
		\centering
		\includegraphics[height=1in]{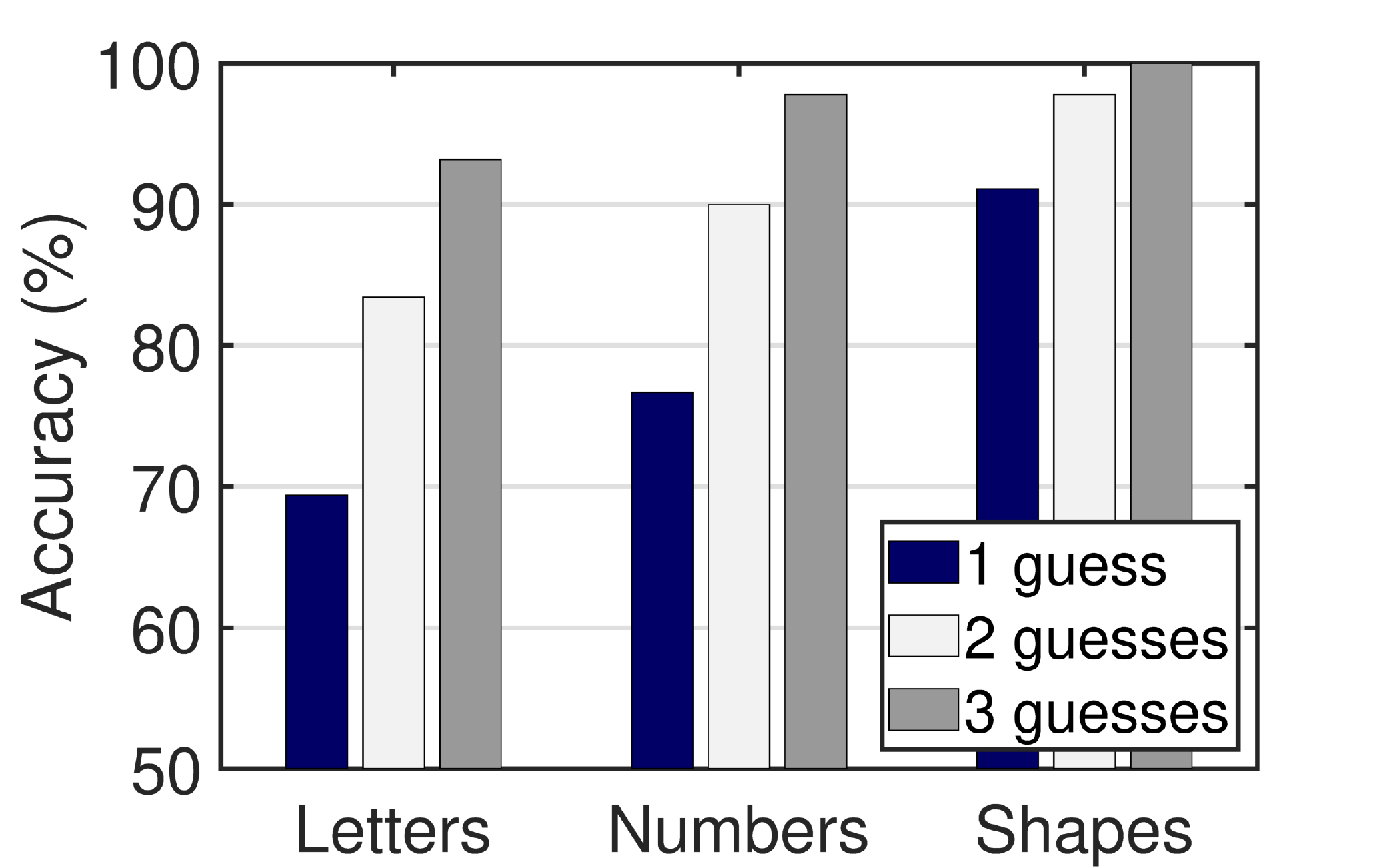}
		\caption{}
		\label{fig:performance_hand_standing}
	\end{subfigure}
	\begin{subfigure}{0.24\linewidth}
		\centering
		\includegraphics[height=1in]{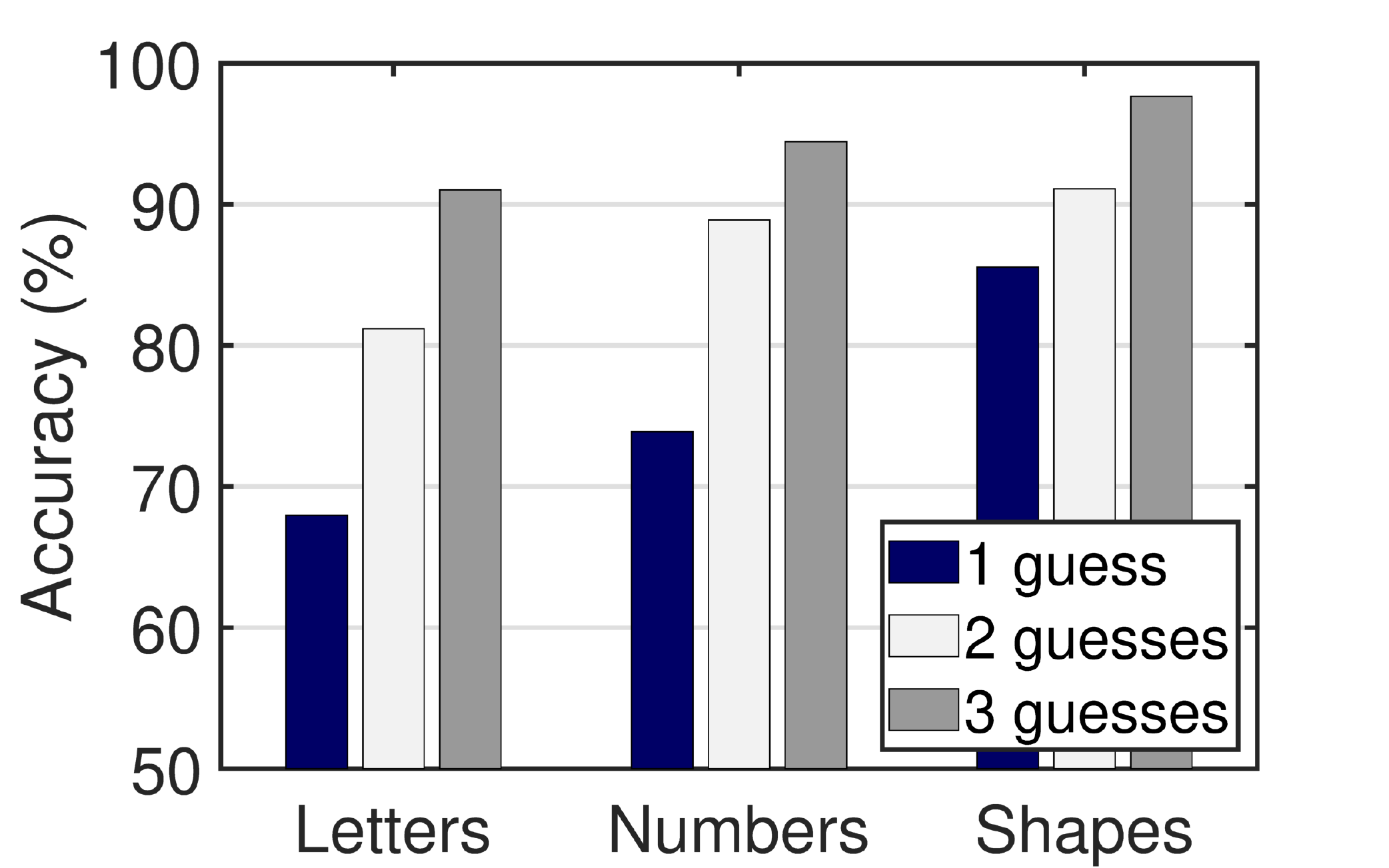}
		\caption{}
		\label{fig:performance_hand_laying}
	\end{subfigure}
	\begin{subfigure}{0.24\linewidth}
		\centering
		\includegraphics[height=1in]{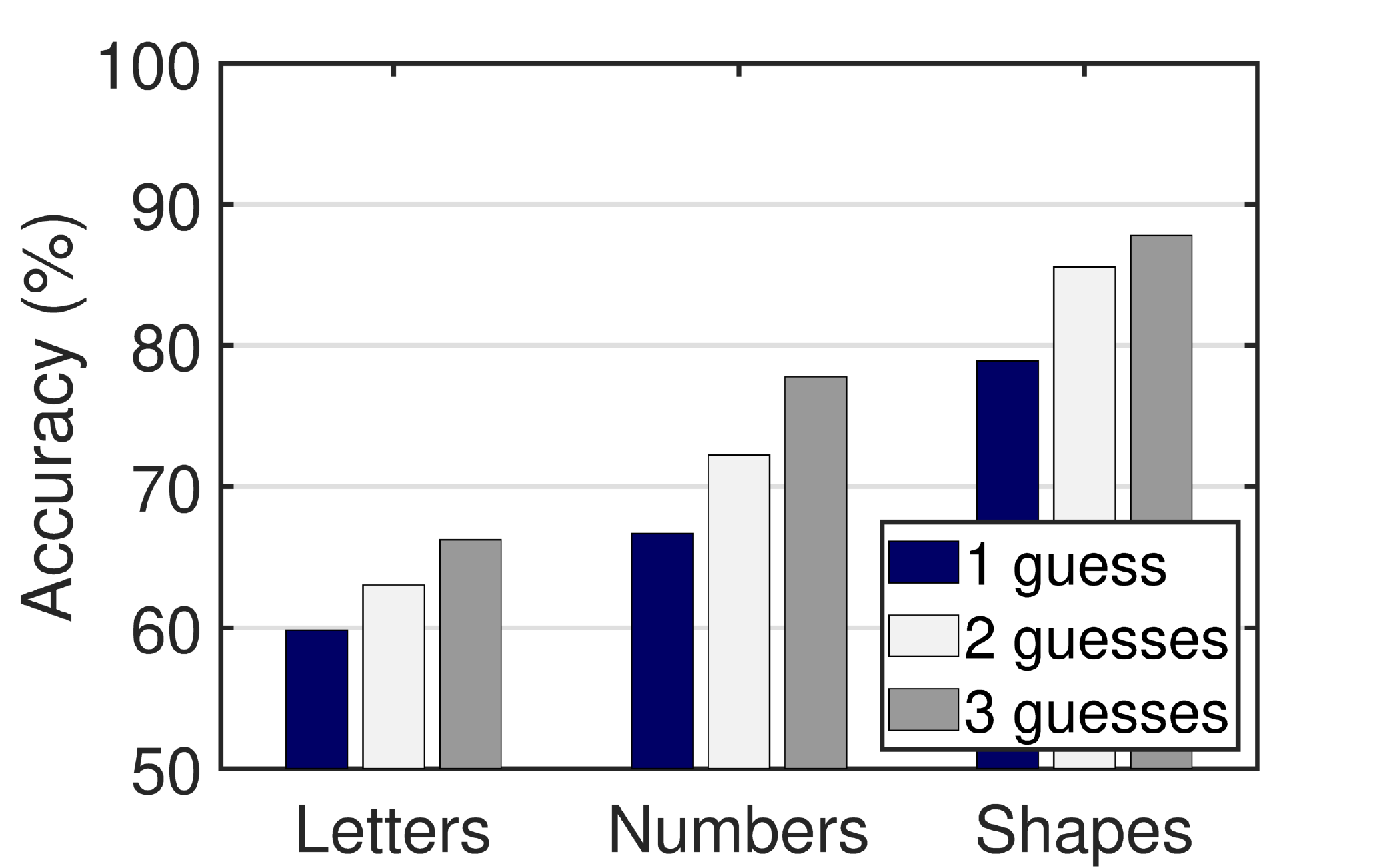}
		\caption{}
		\label{fig:performance_hand_walking}
	\end{subfigure}
	\caption{\majorRevText{Accuracy of the correctly guessed letters, numbers and shapes in 1, 2 and 3 guesses when the user is holding the iPad in hand (a) while sitting, (b) while standing, (c) while laying down, and (d) while walking.}}
	\label{fig:performance_hand}
\end{figure}
% \begin{figure}[h!]
% 	\centering
% 	\includegraphics[height=1.8in]{figures/evaluation/performance_hand_sitting.pdf}
% 	\caption{Accuracy of the correctly guessed letters, numbers and shapes in 1, 2 and 3 guesses when the user is holding the iPad in hand.}
% 	\label{fig:performance_hand}
% \end{figure}
% \begin{figure}[h!]
% 	\centering
% 	\includegraphics[height=1.8in]{figures/evaluation/performance_hand_standing.pdf}
% 	\caption{Accuracy of the correctly guessed letters, numbers and shapes in 1, 2 and 3 guesses when the user is holding the iPad in hand.}
% 	\label{fig:performance_hand}
% \end{figure}

\majorRev{[MR3, C3, C14]}\majorRevText{\textbf{(7) How does the positioning of the iPad affect \sys{}'s performance?}}

\majorRevText{We evaluated \sys{}'s performance with a variety of iPad positions. 
Apart from the experiments with the iPad placed on the table, we conducted experiments with three volunteers in the following scenarios:
1) iPad is held in the hand while sitting, 2) iPad is held in the hand while standing, 3) iPad is held in the hand while laying down, and 4) iPad is held in the hand while walking.
In each scenario, we asked the volunteers to write letters, numbers, and shapes on the iPad screen, as described earlier. 
Figure~\ref{fig:performance_hand} shows the detection accuracy for letters, numbers and shapes in the top three guesses for each scenario.
While a user is holding the iPad in hand, the writing causes small movements in the iPad body that introduce minor fluctuations in the motion sensors' data. 
However, a detection accuracy of higher than $90\%$ for all strokes shows that \sys{} is resilient to these small fluctuations and can still identify the Pencil's magnetic impact in scenarios 1, 2, and 3.
For the scenario when the user is writing on the iPad while walking, we observe that \sys{}'s detection accuracy reduces ($66.2\%$, $77.8\%$ and $87.8\%$ for letters, numbers and shapes respectively) because the continuous change in ambient magnetic readings makes it difficult to separate the magnetic impact of the Pencil and the ambient environment.}

\begin{table}[t!]
\centering
\caption{\majorRevText{\sys{}'s detection accuracy in different environmental settings.}}
\begin{tabular}{ |c | c| c| c|}
\hline
 Scenario & Letters ($\%$) & Numbers ($\%$) & Shapes ($\%$) \\ 
\hline
 Door opening and closing & 93.37 & 95.55 & 96.67 \\  
 People walking around in the room & 93.80 & 96.58 & 97.86 \\
 Watch with metallic bracelet & 91.88 & 95.55 & 96.67 \\
 Smart folio iPad case & 91.67 & 94.44 & 96.67 \\
\hline
\end{tabular}
\label{table:environment}
\end{table}

\majorRev{[MR3, C3, C7]}\majorRevText{\textbf{(8) How does the environment impact the performance of \sys{}?}}

\majorRevText{In our experiments, we observed that the magnetic field sensed by the iPad's magnetometer is not affected by regular objects such as books, clothes, etc., in the surroundings.
The experiments described above were all conducted in a lab setting where electrical devices like computers, laptops, monitors, etc., were present in the surroundings. 
Based on these results, we concluded that the existence of these devices did not impact the performance of our system. 
Similarly, events such as opening or closing the door, people walking around, and the movement of furniture in the environment had no significant effect on \sys{}'s performance.} 

\majorRevText{We also evaluated how \sys{}'s performance is affected by the presence of magnetic objects in the environment.
We invited three volunteers to perform experiments in the following scenarios: 1) while wearing a watch with a magnetic bracelet on their non-dominant wrist, and 2) while writing on an iPad with a smart folio case.
In both scenarios, the subjects were instructed to hold the iPad in hand.
Table ~\ref{table:environment} shows our system's detection accuracy in the top 3 guesses for each scenario described above.
In scenario 1, even though the magnetic bracelet is very close to the Pencil while the user is writing, its impact on the magnetic readings is much smaller than the Pencil's magnetic impact.
As a result, despite the interference, \sys{} is able to accurately track the Pencil position.
In the case of the smart folio iPad cover, although the magnets in the cover affect the magnetic readings sensed by the iPad, their magnetic impact stays consistent over time, allowing \sys{} to separate the impact of the Pencil as it moves over time. }

\majorRevText{\textbf{(9) How does \sys{} perform compared to other machine learning techniques?}}

\majorRevText{To conduct a comparative evaluation with our system, we implemented three machine learning methods to infer the users' writing from the motion sensors' data.
These methods included 1) k-nearest neighbors (k-NN) regression model,  2) Long Short Term Memory networks (LSTM) model for classification, and 3) an ensemble of LSTM models for classification trained for different locations of the iPad screen. 
We selected these algorithms as they capture the order dependence in our dataset and consider the nearest neighbors or previous states while guiding us about the pencil location over time (we detail the models in the Appendix.)
%
%The KNN regression model was trained on the data collected for Pencil's magnetic map generation, as well as the data collected from the $10$ subjects to predict the Pencil's location for each magnetic data sample and k, was set to 1.
%
The tracking results generated from the k-NN model allowed the attacker to guess only $11.54$\%, $18.67$\%, and $3$\% of the letters, numbers and shapes, respectively.
The low accuracy is due to the fact that the magnetic readings can be similar for different Pencil locations and orientations and the k-NN model fails to capture the relationship between Pencil's previous states and its current location and orientation.
Hence, the tracking results generated from this model are hardly legible.
%
% For the LSTM model, we used the data collected from the $10$ subjects for training with a stratified cross-validation approach.
% %
% The model takes the magnetic readings corresponding to a stroke (number of samples fixed to $200$) and predicts the letter, number, or shape.
% %
The LSTM model similarly achieved an accuracy of $7.69$\%, $10$\%, and $21.67$\% for letters, numbers, and shapes.
This model fails to precisely infer useful information about the users' handwriting due to the variations in magnetic readings caused by changes in the Pencil location and orientation, even when a user is writing the same character (as illustrated in Figure~\ref{fig:different_locations_orientations}),
To guide the LSTM about the Pencil location, we also trained $3$ separate LSTM models on data collected from each of the $3$ grids in the input region.
%
% For inference, we used an ensemble of these three models by selecting the model's prediction output with maximum confidence.
%
We observe that this approach slightly improves the LSTM model's accuracy and yields $11.54$\%, $16.67$\%, and $29.67$\% for letters, numbers, and shapes, respectively.
However, this model also yields less accurate results than \sys{}'s particle filter algorithm despite having a smaller search space for the Pencil's location.
%
% Habiba: need to check this
This is because this model fails to capture the combined impact of the changes in Pencil's location and orientation on the magnetic readings from the given data.
In contrast to these models, \sys{} can accurately track the Pencil's movement over time solely through the magnetic readings without requiring any large training dataset.
}

%% file: discussion.tex
\section{Limitations and Discussion}
\label{sec:discussion}
In this section, we discuss limitations and opportunities for the improvement of our system.

\vspace{1pt}\noindent\textbf{Limitations}: 
	%Our $S^3$ system needs some improvement to be a real-world attack. 	
\sys{} currently can detect individual letters, numbers, and words. 
As we have mentioned in the evaluation, \sys{} can infer sentences with a good accuracy when the tracking results of its words are analyzed together.
Although this approach allows an adversary to infer commonly used English sentences through their semantic content, it might not be feasible for complex sentences when the semantic connections among words are lost or difficult to infer.
\minorRevText{A natural extension to our work would be to incorporate language models~\cite{howard2018universal} for improving the detection accuracy of complex words and sentences.
Since language models capture long-term dependencies, hierarchical relations and sentiment of the language, a sentence drawn from the modelled language, regardless of its complexity, can be inferred from an initial imprecise guess with a high accuracy.}
% In order to continuously track sentences, with short interval in between different strokes, it is difficult for our stroke detection to work. 
	%In this case, a long period of magnetometer data will be regarded as one ``stroke'', which will reduce the possibility of tracking correctly. The second limitation of writing is, the strokes written by users should not be too small. This is because when a user writes a small character, the fluctuations of magnetometer data will not be large enough, which makes the tracking process much harder.
	%\item \textbf{Stability requirement of environment}.

We also demonstrated that \sys{}'s performance is resilient to subtle changes in the environment.
% 
%This means that our system assumes that the user's Pencil interaction with the iPad is the main source of fluctuations in magnetometer data. 
%
However, continuous changes in the ambient magnetic field, such as when a user travels in a vehicle, might interfere with Pencil's magnetic impact, making it difficult for \sys{} to infer the users' writing.
%However, for cases where the attack environment is changing rapidly, such as when a user is traveling in a vehicle, the magnetic variations from the Pencil will be mixed with the changes in the ambiance, which are difficult to separate from our current approach. \\

%\noindent\textbf{Opportunities to improve}: 
In \sys{}'s stroke detection process, more complex models through Recurrent neural networks (RNN) and Long-short term memory (LSTM) can be learned to capture the underlying sequence patterns in sensor data for detecting the beginning and ending timestamps of each stroke more accurately.
%
%We believe that our tracking performance can be improved if the stroke estimation results are more precise.
%
Furthermore, our attack has a human in the loop for guessing what the victim is writing.
Future work will expand our analysis to support models built on computer vision techniques to recognize letters, and natural language processing techniques (NLP) to infer words and sentences to infer user writings without human interaction.

\vspace{1pt}\noindent\textbf{Defense Techniques}:
We now discuss possible defenses for the side-channel attack presented in this paper.
First, we observed in our preliminary experiments that if the sampling rate for accelerometer and gyroscope data is decreased, say to $50$Hz, detecting the beginning and end of the strokes becomes very difficult.
Similarly, if the magnetic data is sampled at a frequency of $5$Hz, the tracking results become less tangible.
Hence, a potential defense against our attack would require iOS to reduce the available sampling rate for accelerometer, gyroscope, and magnetic data when a user interacts with the iPad using Apple Pencil.
Another potential defense would be to pause the motion sensors when a user interacts with the Pencil.  
However, both solutions heavily affect the operation of legitimate applications running in the background, which use motion sensors for activity, context, and gesture recognition.
Another possible defense is to apply magnetic shielding~\cite{Scott2007} to the Apple Pencil.
However, this will greatly impact the weight and hence the usability of the Pencil.
Future work will analyze the trade-offs between these defense techniques and applications' access patterns to motion sensors.
%
%An alternative approach could be having special agreements between application developers and vendors to hold them accountable for any malicious activities.
%
%We understand that these solutions have their pros and cons and require further investigation by the security community.\\

% \noindent\textbf{Other possible attacks}:
% %
% \sys{} shows that motion sensors' data can be exploited to infer what a user is writing with high accuracy. 
% %
% We believe that the vulnerability in modern stylus pencils unveiled by \sys{} will promote the investigation of other possible attacks like inferring a user's PIN code or password and guessing what a user is typing with the Pencil.
% %
% Intuitively, these attacks bear great similarities with our proposed attack. However, the search space in these cases is discrete in contrast to the free-form handwriting.
% In addition, these attacks would require accurate and continuous tracking of whether the Pencil tip is on or off the screen.
% %
% We will investigate these attacks as our future work.
%As mentioned in 1), users can write small characters continuously, which will make both the stroke detection and tracking processes harder. Another way is from the producer side. When applications are in the backend, device producers (including but not limited to Apple) can reduce the frequency that applications can use to collect sensors data. In this case, important information such as peaks that are used to detect strokes may be lost.

%% file: relatedwork.tex
\section{Related Work}
\label{sec:related_work}
\textbf{Side-channel attacks through motion sensors}:
Several recent works have demonstrated potential privacy
leaks from mobile sensors.
Wang et al. explored how motion sensors' data collected from smartwatches are used to infer what a user is typing on a keyboard~\cite{wang2015mole}.
Another work demonstrated that the changes in the accelerometer readings are powerful enough to help extract passwords typed on smartphone touchscreens~\cite{owusu2012accessory}.
Michalevsky et al. showed the MEMS gyroscopes in modern smartphones could sense acoustic signals, which can be used to identify and parse speech~\cite{michalevsky2014gyrophone}.
A recent work unveiled a side-channel attack leveraging smartphone accelerometers to eavesdrop on the smartphone speaker and reconstruct the audio signals~\cite{ba2020learning}.
Das et al. \cite{das2016tracking} combined multiple motion sensors and used inaudible audio stimulation to fingerprint different users by measuring anomalies in the signals.
Another line of work leveraged magnetometers to infer private information.
It is shown that the magnetometer of a smartphone placed next to the hard drive of a computer allows an attacker to infer patterns about the system details, such as the type of operating system and applications used~\cite{biedermann2015hard}.
Block et al. leveraged magnetometers' ability to detect locations of users' devices within commercial GPS accuracy~\cite{block2018my}.
Researchers also demonstrated that electromagnetic field measured by smartphone magnetometers could be exploited for webpage~\cite{matyunin2019magneticspy}, and device~\cite{perez2019fatal} fingerprinting.
Compared with existing side channels, we introduce a new side-channel attack 
to identify users' handwriting by analyzing the magnetic impact of the embedded magnets in modern stylus pencils sensed by device magnetometers.

\vspace{1pt}\noindent\textbf{Handwriting tracking through motion sensors:}
There have also been efforts that explored the use of motion sensors for eavesdropping on users' handwriting.
Motion sensor readings collected from a smartwatch are used to infer what the user is writing~\cite{hao19motioneaves, xia18hacker}.
However, these approaches require a compromised smartwatch to be worn on the victim's writing hand. 
In contrast, \sys{} does not require any extra device, and tracks the Pencil movement with high accuracy without restricting the victim's handwriting style.
Another work introduced \emph{Finexus} to track fingertip movements in 3D space by instrumenting the fingertips with electromagnets and measuring the corresponding magnetic field changes using four magnetometers~\cite{chen2016finexus}.
Although \textit{Finexus} tracks the fingertip movements with millimeter level accuracy, it requires multiple magnetometers for precision.
Our system tracks the pencil tip and achieves high accuracy in inferring handwriting with a single magnetometer.
Similarly, \textit{TMotion} presented a self-contained 3D input device that enables interactions in 3D space around smartphones by embedding a permanent magnet and an inertial measurement unit (IMU) in the stylus pen~\cite{yoon2016tmotion}. 
While this work is effective at tracking the stylus, it requires attaching an extra wand on top of the stylus pencil to accommodate a magnet and an extra IMU sensor, impacting the stylus pen's overall usability.
In contrast, our system requires no such hardware for accurate pencil tracking.
Lastly,  a recent work introduced a sensing system for eavesdropping on handwriting by analyzing the magnetic field changes produced by stylus pens~\cite{liu2020maghacker}.
However, a commodity smartphone with a magnetometer must be placed within $20cm$ of the victim's device to sense the magnetic field.
In contrast, \sys{} uses the magnetometer on the victim's device, resulting in a higher detection accuracy, and does not require the attacker to be present in close proximity of the victim.

%% file: conclusion.tex
\section{Conclusions}
\label{sec:conclusion}
We introduced a side-channel attack through motion sensors by exploiting the vulnerability introduced due to the introduction of embedded magnets in stylus pencils.
We presented \sys{}, a novel system that infers what a user is writing from motion sensors' data on an iPad Pro using the latest version of the Apple Pencil.
To track the Pencil movement, we developed a novel multi-dimensional particle filtering algorithm using a 3D magnetic map of the Pencil to identify the magnetic impact of different locations and orientations of the Pencil. 
We evaluated \sys{} with $10$ subjects and demonstrated that an attacker could  identify $93.9$\%, $96$\%, $97.9$\%, and $93.33$\% of the letters, numbers, shapes, and words correctly by only using the motion sensors' data.
In future work, we will expand our analysis to support more devices that use stylus pencils with embedded magnets to find potential privacy leaks.

%% file: acknowledgements.tex
We sincerely thank the anonymous reviewers for their insightful comments and valuable suggestions.
This work has been partially supported by the Purdue Research Foundation (PRF) Research Grant.

%% file: groundtruth.tex
\vspace{0.2in}
\textbf{Ground truth for examples in Figure~\ref{fig:eval_examples} in order:} Letters are y, n, c, e, l, z, w, u, a, b, g, q, v, m, h, k, s, i, p, d, o, x, r, h, t, f, j, q, and f.
Numbers are 7, 5, 0, 6, 9, 6, 1, 2, 5, 8, 4, 5, 0, 3, 1, 8, 3, 2, 8, 4, 6, 9, 0, 3, 7, and 6. 
Shapes are heart, star, circle, heart, square, star, circle, square, heart, heart, triangle, star, and square.

\textbf{Ground truth for examples in Figure~\ref{fig:words_examples} in order:} make, have, now, hello, time, own.

%% file: appendix.tex
%\appendix  %interestingly does not work.

\majorRevText{\section{Details of the ML Models}}

\majorRevText{\noindent \textbf{K-Nearest Neighbors (k-NN) regression:}}
\majorRevText{We trained a k-NN model on the data collected for generating the magnetic map for the Pencil and evaluated it on the data collected from the subjects in our user study.
The model was designed to predict the Pencil's location and orientation at time $t$ when given the magnetic readings for the previous three timestamps.
We used a grid search to find the best value for $k$ within the range of $1$ to $5$ on our training data and chose the value which resulted in the highest accuracy.
The attacker was shown the tracking results generated from predicted Pencil location trace for each stroke.}

\vspace{1mm}\majorRevText{\noindent \textbf{Long Short-term Memory Network (LSTM):}}
\majorRevText{We implemented a LSTM model for predicting which letter, number or shape corresponds to magnetic readings collected for a stroke.
We used TensorFlow's Keras library~\cite{chollet2015keras} for implementing the LSTM network.
The model required a three-dimensional input of the shape [samples, time steps, features] where samples is the number of strokes, time steps is the number of magnetic samples for each stroke (fixed to $200$ samples) and the number of features is $3$ representing the $x$, $y$ and $z$ axis of the magnetometer readings.
The output of the LSTM model was a $26$ element vector ($10$ and $5$ in the case of numbers and shapes respectively) representing the probability of a given stroke being any of the $26$ letters ($10$ numbers or $5$ shapes).
We defined the model as sequential model with two LSTM hidden layers followed by a dropout layer to reduce over-fitting on training data.
The features extracted from the LSTM layers were fed into a dense fully connected layer followed by a final output layer used to make predictions.
We trained this model on the data collected from the $10$ subjects in our user study using a leave-one-out cross validation approach~\cite{ref1}.}

\majorRevText{In an attempt to improve the LSTM network's performance, we limited the search space for the Pencil's location by implementing separate models for each of the three grids in the input region of the screen.
The same network architecture was used for the three models. 
For final inference, we used the output of the model with the highest confidence score. 
}